\documentclass[prb,aps,twocolumn,superscriptaddress,showpacs,footinbib,longbibliographyx]{revtex4-2}
\usepackage[colorlinks=true,linkcolor=black, citecolor=blue, urlcolor=blue, 
unicode=true]{hyperref}

\usepackage{xcolor,amsmath,braket,amssymb}
\usepackage[version=4]{mhchem}
\usepackage{graphicx}
\usepackage{tabularx}
\usepackage{url}
\usepackage{comment}

\begin{document}
	
	\title{Anisotropic Band-Split Magnetism in Magnetostrictive CoFe$_2$O$_4$}
	
	\author{Harry Lane}
	\affiliation{Department of Physics and Astronomy, School of Natural Sciences, The University of Manchester, Oxford Road, Manchester M13 9PL, UK}
	\affiliation{The University of Manchester at Harwell, Diamond Light Source, Didcot, Oxfordshire OX11 0DE, UK}
	\affiliation{School of Physics and Astronomy, University of St Andrews, N Haugh, St Andrews KY16 9SS, UK}
	\affiliation{School of Physics and Astronomy, University of Edinburgh, James Clerk Maxwell Building, Peter Guthrie Tait Road, Edinburgh, EH9 3FD, UK}

	\author{Guratinder Kaur}
	\affiliation{School of Physics and Astronomy, University of Edinburgh, James Clerk Maxwell Building, Peter Guthrie Tait Road, Edinburgh, EH9 3FD, UK}

	\author{Masahiro Kawamata}
	\affiliation{Department of Physics, Tohoku University, Sendai, Miyagi 980-8578, Japan}
	\affiliation{Department of Physics, Tokyo Metropolitan University, Hachioji, Tokyo 192-0397, Japan}

	\author{Yusuke Nambu}
	\affiliation{Institute for Integrated Radiation and Nuclear Science, Kyoto University, 2-1010 Asashiro-nishi, Kumatori, Osaka 590-0451, Japan}

	\author{Lukas Keller}
	\affiliation{Laboratory for Neutron Scattering and Imaging, Paul Scherrer Institut, Villigen, CH-5232, Switzerland}

	\author{Russell A. Ewings}
	\affiliation{ISIS Pulsed Neutron and Muon Source, STFC Rutherford Appleton Laboratory, Harwell Campus, Didcot, OX11 0QX, United Kingdom}

	\author{David J. Voneshen}
	\affiliation{ISIS Pulsed Neutron and Muon Source, STFC Rutherford Appleton Laboratory, Harwell Campus, Didcot, OX11 0QX, United Kingdom}
	\affiliation{Department of Physics, Royal Holloway University of London, Egham, TW20 0EX, UK}

	\author{Travis J. Williams}
	\affiliation{ISIS Pulsed Neutron and Muon Source, STFC Rutherford Appleton Laboratory, Harwell Campus, Didcot, OX11 0QX, United Kingdom}

	\author{Helen C. Walker}
	\affiliation{ISIS Pulsed Neutron and Muon Source, STFC Rutherford Appleton Laboratory, Harwell Campus, Didcot, OX11 0QX, United Kingdom}

	\author{Dwight Viehland}
	\affiliation{Department of Materials Science and Engineering, Virginia Polytechnic Institute and State University, Blacksburg, VA 24061, USA}

	\author{Peter M. Gehring}
	\affiliation{National Institute of Standards and Technology, Gaithersburg, MD 20899, USA}

	\author{Chris Stock}
	\affiliation{School of Physics and Astronomy, University of Edinburgh, James Clerk Maxwell Building, Peter Guthrie Tait Road, Edinburgh, EH9 3FD, UK}

	\date{\today}
	
	\begin{abstract}
		Single crystal spinel CoFe$_2$O$_4$ exhibits the largest room-temperature saturation magnetostriction among non-rare-earth compounds and a high Curie temperature ($T_c \sim 780$\,K), properties that are critical to a wide range of industrial and medical applications.  Neutron spectroscopy reveals a large band splitting ($\sim$ 60\,meV) between two ferrimagnetic magnon branches, which is driven by site mixing between Co$^{2+}$ and Fe$^{3+}$ cations, and a significantly weaker magnetocrystalline anisotropy ($\sim$ 3\,meV).  Central to this behavior is the competition between extremely large mismatched molecular fields on the tetrahedral $A$-site and octahedral $B$-site sublattices and the single-ion anisotropy on the $B$-site.  This creates a strong energetic anisotropy that locks the magnetic moment within each structural domain in place.  As a result of these differing energy scales, switching structural domains is energetically favored over a global spin reorientation under applied magnetic fields, and this is what amplifies the magnetostrictive nature of CoFe$_2$O$_4$.
	\end{abstract}
	
	
	\maketitle
	
	\section{Introduction}
	Increasing demand for materials used for chemical and biological sensing that integrate magnetism with elastic strain fields~\cite{Saiz22:7} has driven interest in applications such as magnetoelastic resonator antennas for wireless sensing and energy harvesting.~\cite{Zaeimbashi21:12} Magnetoelastic coupling is also a fundamental topic in condensed matter physics, particularly in the emerging field of altermagnets~\cite{Smejkal22:12, Smejkal22:12_2, Cheong24:9, Cheong25:10, Yuan20:102, Yuan21:5}—a newly proposed class of materials that break the product of parity and time-reversal symmetry.~\cite{Cheong24:9} These materials exhibit piezomagnetism,~\cite{Yershov24:110,Aoyama24:8} a linear coupling between strain and magnetic field, which has been recognized as a defining signature of altermagnetic behavior. The presence of a time-reversal-symmetry-breaking field in such systems can lift the degeneracy of antiferromagnetic magnons, thereby enabling control over magnon chirality and opening pathways for spintronic applications.~\cite{Ma21:12} The coupling in magnetoelastic materials arises from anisotropy, which can be directly characterized with neutron spectroscopy by measuring gapped magnetic excitations.  In this work, we apply neutron spectroscopy to investigate the crystalline and magnetic anisotropy in the highly magnetostrictive ferrimagnetic spinel CoFe$_{2}$O$_{4}$, which, in contrast to altermagnets, exhibits a quadratic coupling between magnetism and strain.  We also use neutron and x-ray powder diffraction to identify the consequences of the ferrimagnetic long-range order below $T_c$ on both the nuclear structure and the magnetic dynamics.
	
	Magnetostriction is defined as the change in the dimensions of a material in response to a change in magnetization. It is called \textit{forced} magnetostriction if the dimensions change as a function of applied external magnetic field and \textit{spontaneous} magnetostriction if the dimensions change in zero applied field below a magnetic ordering transition.  The effect is a consequence of magnetoelastic coupling that stems from either a strongly distance-dependent exchange interaction~\cite{Callen65:139} or the presence of spin-orbit coupling~\cite{Callen63:129} that, in combination with the local crystalline electric field, gives rise to single-ion anisotropy. These energetics produce an interplay between strain, shape anisotropy, and domain formation that can be tuned with a magnetic field, thus making magnetostrictive materials extremely useful for a wide variety of technical applications such as actuators, transducers, and sensors.~\cite{Rajagopal21:92,Herzer03:254,Li10_2:19,Langlois16:tech,Fletcher02:thesis,Olabi06:29,Dhilsha05:55,Ekreem07:191}
	
	Most ferromagnetic materials display a measurable magnetostrictive coefficient below their respective Curie ordering temperatures.~\cite{Hathaway93:18} These include elemental metals such as Co, Ni, and Fe,~\cite{McCorkle23:22,Lee71:326,Lacheisserie83:31} rare-earth alloys such as Terfenol-D (Tb$_{x}$Dy$_{1-x}$Fe$_2$, $x \approx 0.3$),~\cite{Clark76:29,Jiles94:27} and transition metal oxides such as FeTi$_2$O$_4$.~\cite{Ishikawa71:26}  Magnetostriction occurs through the alignment of the magnetic domains~\cite{Kittel49:21} in a sample and through magnetoelastic coupling, which provides a means of controlling the strain.  Magnetostriction therefore requires anisotropic magnetic moments that are coupled to the underlying crystal lattice.
	
	Rare-earth magnets are an ideal platform for studies of magnetostriction given their large spin-orbit coupling constant $\lambda_{SO}$, which scales with atomic number as $\sim Z^{2}$,~\cite{Landau:book} and highly anisotropic $f$-orbitals.  However, many rare-earth compounds order magnetically at cryogenic temperatures, which prevents their use in devices that must operate at ambient conditions.  Among the many known magnetostrictive materials, CoFe$_2$O$_4$ is of particular interest due to its stability, chemical flexibility,~\cite{Turtelli14:60,McCallum01:27,Bozorth55:99} and high Curie temperature $T_c$.  It is also composed of earth-abundant transition metal ions, which makes it cheaper than the rare-earth counterparts.  CoFe$_2$O$_4$ crystallizes in a high-temperature cubic structure belonging to the $Fd\overline{3}m$ space group ($a=8.385$\,\AA) and transforms to a tetragonal structure belonging to the $I4_{1}/amd$ space group ($a=b=8.388$\,\AA\ and $c=8.378$\,\AA) below $T_c \sim 780$\,K.~\cite{Abes16:93,Yang08:77}  Complementary diffraction and microscopy studies on CoFe$_2$O$_4$ report that the magnetostrictive properties result from a switching of the tetragonal structural domains.~\cite{Abes16:93}  As CoFe$_2$O$_4$ is used in applications as far ranging as drug delivery and sensors, most studies of this material have been devoted to nanoparticle~\cite{Song04:126} and thin-film samples; very few pertain to bulk single crystals.  We apply neutron spectroscopy to characterize the bulk magnetic dynamics of CoFe$_{2}$O$_{4}$ and relate the results to the favorable magnetostrictive properties.  In particular, we discuss the interplay between the energetics associated with anisotropy and magnetic exchange.
	
	This paper is structured as follows: Sect.~\ref{sect:experimental} provides details about the \ce{CoFe2O4} sample growth and the methods used in our experiments; Sect.~\ref{sect:structure} discusses the crystallographic and magnetic structures of \ce{CoFe2O4}; Sect.~\ref{sect:spin_dynamics} summarizes the neutron inelastic scattering measurements of the spin dynamics; and Sect.~\ref{sect:theory} describes the theoretical modeling.  The paper ends with a discussion of our results and a conclusion in Sects.~\ref{sect:discussion} and~\ref{sect:conclusion}, respectively.
	
	\section{Experimental details}
	\label{sect:experimental}
	
	Powder samples of \ce{CoFe2O4} were prepared by mixing stoichiometric ratios of cobalt oxide and iron oxide. The mixture was sintered multiple times before undergoing a final calcination at 1073\,K in a box furnace.  To determine the structural changes associated with the magnetic transition, x-ray diffraction measurements were made from 300\,K to 1000\,K using a Bragg-Brentano geometry with a Rigaku SmartLab diffractometer equipped with a high-temperature furnace. An incident x-ray wavelength of $\lambda = 1.54$\,\,\AA\ was selected using a Johansson monochromator and then focused on the sample position using Cross Beam Optics (CBO) with parabolic mirrors. Neutron diffraction was used to determine the magnetic structure and the associated statically ordered magnetic moment using the DMC cold-neutron diffractometer located at the SINQ spallation source (Paul Scherrer Institute). Measurements were conducted with an incident neutron wavelength of $\lambda$ = 2.46\,\AA. A high-temperature furnace and a helium cryostat were used to characterize the magnetic structure between 900\,K and 1.6\,K.
	
	A large single crystal ($\approx 10$\,g) of CoFe$_{2}$O$_{4}$ was grown in a floating-zone furnace under high oxygen pressure ($\approx 50$\,atm). Neutron scattering measurements were carried out on this crystal using the MERLIN time-of-flight spectrometer located at the ISIS Neutron and Muon Source.~\cite{Bewley06:385}  The crystal was aligned such that Bragg reflections of the form $(HHL)$ lay within the horizontal scattering plane.  The sample was cooled to $\approx$ 10\,K using a closed-cycle cryostat.  Data were collected with an incident neutron energy $E_i =75$\,meV and the Gd neutron chopper set to 200\,Hz, which provided an energy resolution of 5.8\,meV (FWHM) at the elastic line. The repetition-rate multiplication mode available on MERLIN provided additional data (with better energy resolution) at incident neutron energies of 20\,meV  (elastic resolution: 0.80\,meV) and 9\,meV (elastic resolution: 0.27\,meV). During the experiment, the crystal was rotated through $170^\circ$ in $0.5^\circ$ steps.  
	
	Because the magnetic excitations in CoFe$_{2}$O$_{4}$ extend to high energies, further neutron measurements were conducted on the MAPS time-of-flight spectrometer (ISIS, UK).~\cite{Ewings19:90} The ``Sloppy" chopper package on MAPS combined with the longer secondary flight path, which allows access to smaller scattering angles, facilitates measurements at high energy transfers. The same crystal was used, still aligned in the $(HHL)$ zone and cooled to $\approx 10$\,K using a closed-cycle cryostat.  Data were measured with $E_i=250$\,meV and the Fermi chopper frequency set at 400\,Hz, offering a resolution at the elastic line of 12.3\,meV. All data from MERLIN and MAPS were reduced using the Mantid data analysis framework~\cite{Arnold14:764} and analyzed using the Horace package.~\cite{Ewings16:834} Given the small size of the tetragonal distortion defined by the resolution of our diffraction experiments, the data were reduced in the cubic $Fd\overline{3}m$ space group. Except where otherwise stated, the data are symmetrized according to the $m\overline{3}m$ point group to improve statistics.
	
	\section{Magnetic and Nuclear Structures}
	\label{sect:structure}
	
	\begin{figure*}
		\centering
		\includegraphics[width=0.95\textwidth]{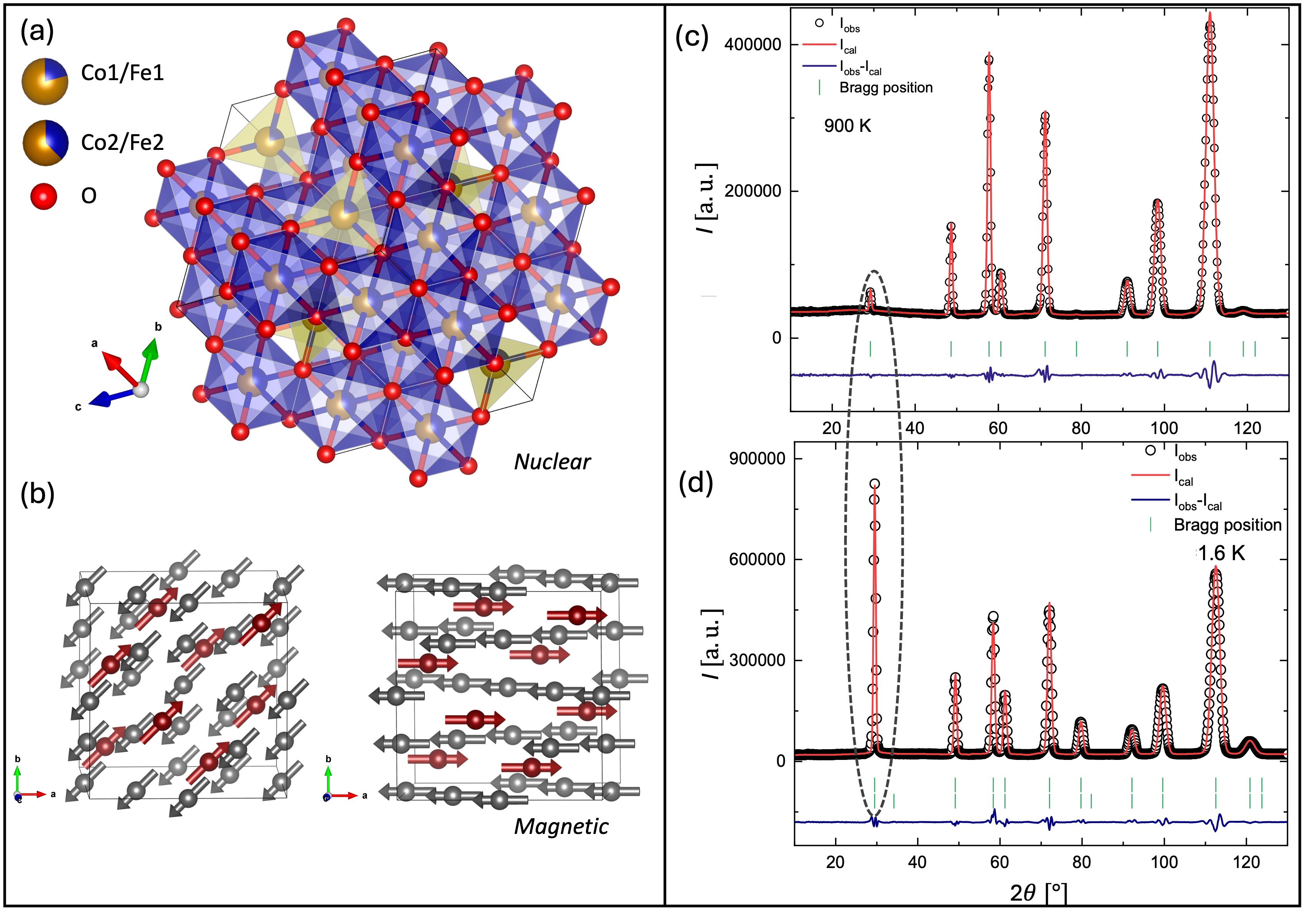}
		\caption{(a) View along [111] of the \ce{CoFe2O4} inverse spinel structure: space group $Fd\overline{3}m$ (227).  The Co and Fe cations occupy tetrahedral $A$ and octahedral $B$ sites. (b) Two possible magnetic structures consistent with the neutron diffraction data. The red and grey colors represent the two distinct magnetic sublattices. (c-d) Neutron diffraction (DMC,PSI) patterns measured above (900\,K) and below (1.6\,K) the transition temperature $T_c \sim 780$\,K.  Solid red lines represent the Rietveld refinement using Fullprof software.~\cite{rodriguez2001} The large dotted oval highlights the large change in the (111) Bragg peak intensity across $T_c$.}
		\label{fig:neutron}
	\end{figure*}
	
	\begin{figure*}
		\centering
		\includegraphics[width=1\textwidth]{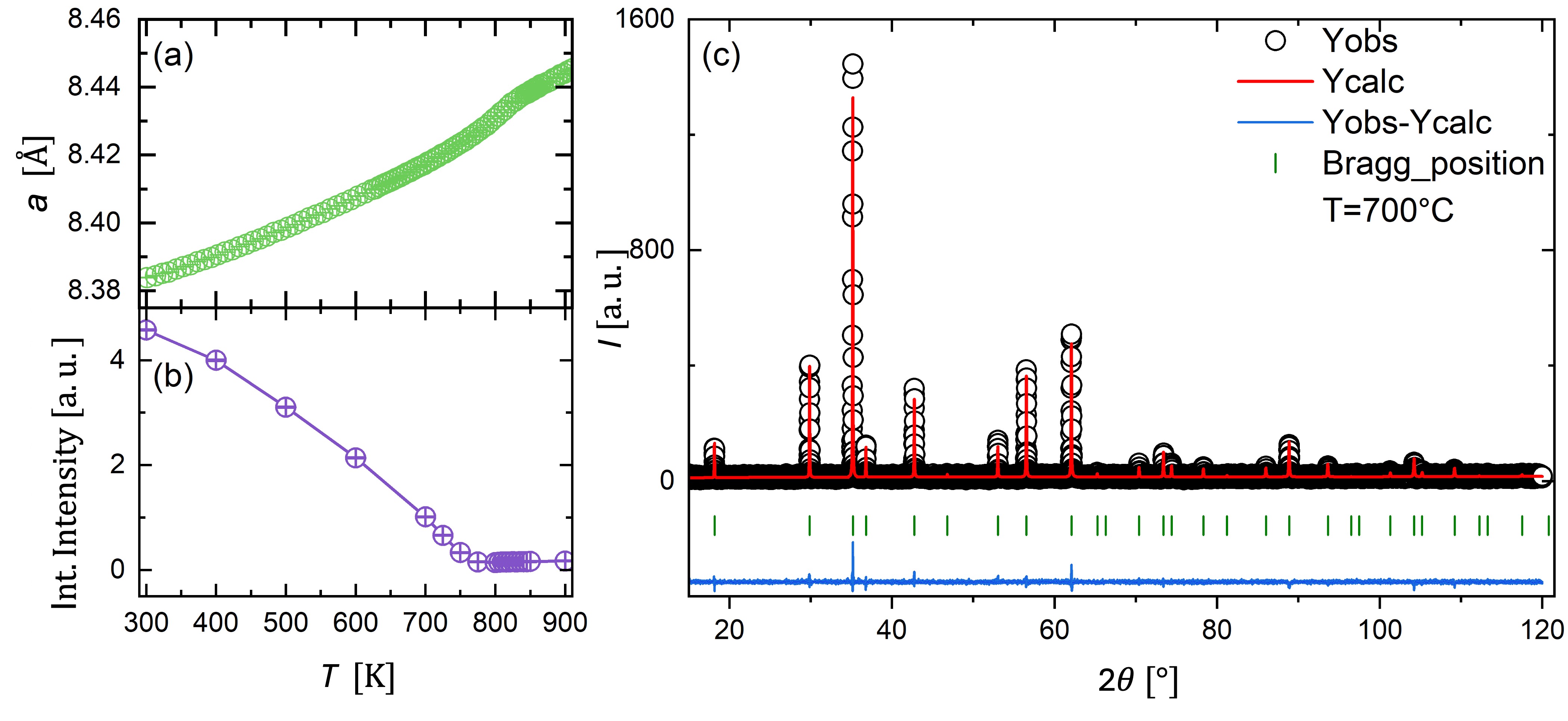}
		\caption{(a) Temperature dependence (Smartlab, Edinburgh) of the lattice constant measured with x-ray diffraction.  Note that all refinements were done using the cubic $Fd\overline{3}m$ space group.  (b) Temperature dependence of the (111) Bragg peak integrated intensity, which is almost entirely magnetic, obtained from the DMC (PSI) neutron diffractometer (see Figs. 1 $c,d$). The onset of magnetic order coincides with a change in the curvature of the thermal expansion. (c) Rietveld refinement of the X-ray SmartLab data measured at 973\,K.}
		\label{fig:lattice_constants}
	\end{figure*}
	
	The nuclear structure shown in Fig. \ref{fig:neutron} $(a)$ depicts the high-temperature cubic phase of CoFe$_2$O$_4$, which belongs to the cubic $Fd\overline{3}m$ space group (point group symmetry $m\bar{3}m$).  The parameters obtained from a Rietveld refinement of the neutron diffraction data measured at 900\,K are listed in Table \ref{Table:neutron_param}.  Of particular importance is the presence of two distinct crystallographic sites for the magnetic Fe$^{3+}$ and Co$^{2+}$ ions with differing Wyckoff symmetry (denoted as $8a$ and $16d$ in Table \ref{Table:neutron_param}). CoFe$_{2}$O$_{4}$ belongs to the family of inverse spinels $AB_2$O$_4$ in which the tetrahedral $A$ sites (site symmetry $8a$ with eight equivalent positions in the unit cell) are occupied by trivalent cations and the octahedral $B$ sites (site symmetry $16d$ with sixteen equivalent positions in the unit cell) are occupied preferentially by divalent ions. But when the two cations are similar in size, as is the case for Fe$^{3+}$ and Co$^{2+}$, significant site mixing can occur between the $A$ and $B$ sites. The greater stability of Fe$^{3+}$ compared with that of Fe$^{2+}$, together with the requirement of charge neutrality, dictates that the Fe and Co valences are unchanged on mixing~\cite{Moyer11:83,Sharma22:6}. The local environment of the $B$ site is trigonally distorted from a perfect octahedral environment. This provides a potential source of magnetic anisotropy for orbitally active magnetic ions. We discuss the implications of these two different local crystalline electric field environment on the spin magnetism below. 
	
	We used neutrons to determine the degree of cation mixing in CoFe$_2$O$_4$ because the x-ray scattering cross section varies as the square of the atomic number $Z$, and there is little contrast between Fe ($Z = 26$) and Co ($Z=27$).  By comparison, the difference between the neutron nuclear coherent scattering lengths of Fe (9.45\,fm) and Co (2.49\,fm) is quite large.  The neutron diffraction results at 900\,K are summarized in Table \ref{Table:neutron_param}, and they clearly indicate significant site mixing between the $A$ and $B$ sites.  A M$\Ddot{o}$ssbauer study of CoFe$_2$O$_4$ shows that the degree of site mixing is highly sensitive to cooling conditions during sample preparation whereas the magnetic transition temperature is not.~\cite{Sawatzky68:39}  This suggests that the underlying magnetism is robust against structural disorder from site mixing. 
	
	\begin{table*}[h]
		\centering
		\caption{Refined structural parameters for CoFe$_{2}$O$_{4}$ at 900~K extracted from the Rietveld refinement of neutron powder diffraction data from DMC with $\lambda$ =2.460~\AA. Space group, $Fd\overline{3}m$, No.~227, ($a$ = 8.44434(18)~\AA, atomic displacement parameters, B$_{iso}$ (~\AA$^{2}$), $R_p$ = 5.80 \%, and $wR_p$ = 5.78\%). The implicit error in the site occupancy is $\sim$ 4-5 \%.}
		\label{Table:neutron_param}
		\begin{tabular*}{\columnwidth}{@{\extracolsep{\fill}}ccccccc}
			\hline
			Atom & $x$ & $y$ & $z$ & Occ. & B$_{iso}$ & Site \\ \hline
			Fe1 & 1/8  & 1/8  & 1/8  & 76\% & 0.750(117) & 8$a$ \\
			Co1 & 1/8  & 1/8  & 1/8  & 24\% & 0.750(117) & 8$a$ \\ 
			Fe2 & 1/2 & 1/2  & 1/2 & 38\% & 1.077(100) & 16$d$ \\
			Co2 & 1/2 & 1/2  & 1/2 & 62\% & 1.077(100) & 16$d$ \\ 
			O & 0.25584(14) & 0.25584(14) & 0.25584(14) & 100\% & 1.654(51) & 32$e$ \\ 
			\hline
		\end{tabular*}
	\end{table*}
	
	Neutrons carry a magnetic moment, which makes them an excellent probe of spin magnetism.  The  neutron powder diffraction patterns shown in Figs. \ref{fig:neutron} $(c,d)$ reveal the presence of additional Bragg intensity at 1.6\,K that is absent at 900\,K (see, for example, the large change in intensity at $2\theta \sim 30^\circ$ and 80$^\circ$). This extra intensity is consistent with ferroic magnetic order, as it appears only at allowed nuclear Bragg peak positions.  Furthermore, all magnetic peaks at low temperatures exhibit wave-vector-resolution-limited linewidths, which is the hallmark of spatially long-range magnetic order.  Therefore, we assign the propagation vector $\boldsymbol{k} = (0,0,0)$ to the magnetic ground state, in agreement with Ref.~\cite{Teillet93:123}.  The refined ferrimagnetic structure, discussed below, is shown in Fig. \ref{fig:neutron} $(b)$ and involves ordering of the Co$^{2+}$ and Fe$^{3+}$ moments on both $A$ and $B$ sites.
	
	The magnetic structures compatible with $\boldsymbol{k} = (0,0,0)$ and the nuclear symmetry were analyzed with representation analysis. For generality, we refined the magnetic structure while allowing the direction of the spin moment on the $A$ and $B$ sites to vary without constraints.  To do this, we refined the magnetic structure in the magnetic space group $F\overline{1}$.  Given the presence of two sites with distinct Wyckoff site symmetries, there are two different sets of magnetic basis vectors that are refined: one for the $A$ site (symmetry $8a$) and one for the $B$ site (symmetry $16d$).  Basis vectors $\psi_{i}$ for the magnetic $A$-site Co1/Fe1 and $B$-site Co2/Fe2 are tabulated, respectively, in Table \ref{Table:basis_A} and Table \ref{Table:basis_B}.  Only two sites are needed for the basis vectors on the $A$ site and four for those on the $B$ site because each is related via translation symmetry to an additional four sites in the unit cell, thereby returning the number of equivalent sites expected based on Wyckoff notation.  While non-collinear structures are possible solutions to the magnetic refinement, given the constraints of powder averaging, and to simplify evaluation of the magnetic moments, we have fixed the structure to be purely collinear and therefore only consider the real components of these basis vectors.  In support of this simplifying assumption are M$\Ddot{o}$ssbauer studies on the inverse spinel Fe$_{2}$O$_{3}$,~\cite{Tucek11:99} which are highly suggestive of collinear ferrimagnetism, and refined magnetic structures in other magnetic spinels that display collinear magnetism. ~\cite{Reehuis03:35,Zhang21:104,Guratinder19:100} Tables \ref{Table:basis_A} and \ref{Table:basis_B} list the irreducible representations $\Gamma_{\nu}$ (IRs) that characterize the transformation properties.
	
	\begin{table*}[h]
		\centering
		\caption{Magnetic basis vectors $\psi_{j}$ and irreducible representations $\Gamma_{\nu}$ for the $A$-site Co1/Fe1:(0.125,  0.125,  0.125) (Wyckoff symmetry $8a$). There are two equivalent sites with coordinates $A1$: (0.125, 0.125, 0.125) and $A2$: (0.875, 0.875, 0.875). The decomposition of the magnetic representation into irreducible representations of G$_{k}$:$\Gamma_{8}\oplus\Gamma_{9}$ is given (taken from the program SARA$h$.~\cite{Wills00:276,Wills09:book}) Six basis vectors with purely real components are presented.}
		\label{Table:basis_A}
		\begin{tabular*}{\columnwidth}{@{\extracolsep{\fill}}cccc}
			\hline
			\textbf{IRs} & \textbf{mode} & \textbf{$A1$} & \textbf{$A2$}  \\
			\hline
			$\Gamma_8$ & $\psi_{1}$ & (8 0 0) & (-8 0 0)\\
			& $\psi_{2}$ & (0 8 0) & (0 -8 0) \\
			& $\psi_{3}$ & (0 0 8) & (0 0 -8)\\
			$\Gamma_9$ & $\psi_{4}$ & (8 0 0) & (8 0 0)\\
			& $\psi_{5}$ & (0 8 0) & (0 8 0) \\
			& $\psi_{6}$ & (0 0 8) & (0 0 8)\\
			\hline
		\end{tabular*}
	\end{table*}
	
	\begin{table*}[h]
		\centering
		\caption{Magnetic basis vectors $\psi_{j}$ and irreducible representations $\Gamma_{\nu}$ for the $B$-site  Co2/Fe2:(0.5,  0.5,  0.5) (Wyckoff symmetry $16d$). The coordinates of the equivalent positions are $B1$: (0.5, 0.5, 0.5), $B2$: (0.5, 0.25, 0.25), $B3$: (0.25, 0.5, 0.25), and $B4$: (0.25, 0.25, 0.5). The decomposition of the magnetic representation into irreducible representations of G$_{k}$:$\Gamma_{3}\oplus\Gamma_{5}\oplus\Gamma_{7}\oplus2\Gamma_{9}$. There are twelve basis vectors, and the respective components for each site are given below.  Only real components are presented.  The table was taken from the program SARA$h$.~\cite{Wills00:276,Wills09:book}}
		\label{Table:basis_B}
		\begin{tabular*}{\textwidth}{@{\extracolsep{\fill}}cccccc}
			\hline
			\textbf{IRs} & \textbf{mode} & \textbf{$B1$} & \textbf{$B2$} & \textbf{$B3$} & \textbf{$B4$} \\
			\hline
			$\Gamma_3$ & $\psi_{1}$ & (12 12 12) & (12 -12 -12) & (-12 12 -12) & (-12 -12 12) \\
			$\Gamma_5$ & $\psi_{2}$ & (6 -6 0) & (6 6 0) & (-6 -6 0) & (-6 6 0) \\
			& $\psi_{3}$ & $(2\sqrt{3}\ 2\sqrt{3}\ -4\sqrt{3})$ & $(2\sqrt{3}\ -2\sqrt{3}\ 4\sqrt{3})$ & $(-2\sqrt{3}\ 2\sqrt{3}\ 4\sqrt{3})$ & $(-2\sqrt{3}\ -2\sqrt{3}\ -4\sqrt{3})$ \\
			$\Gamma_7$ & $\psi_{4}$ & (0 -2 2) & (0 2 -2) & (0 2 2) & (0 -2 -2) \\
			& $\psi_{5}$ & (2 0 -2) & (-2 0 -2) & (-2 0 2) & (2 0 2) \\
			& $\psi_{6}$ & (-2 2 0) & (2 2 0) & (-2 -2 0) & (2 -2 0) \\
			$\Gamma_9$ & $\psi_{7}$ & (4 4 4) & (4 -4 -4) & (4 -4 4) & (4 4 -4) \\
			& $\psi_{8}$ & (4 -2 -2) & (4 2 2) & (4 2 -2) & (4 -2 2) \\
			& $\psi_{9}$ & (4 4 4) & (-4 4 4) & (-4 4 -4) & (4 4 -4) \\
			& $\psi_{10}$ & (-2 4 -2) & (2 4 -2) & (2 4 2) & (-2 4 2) \\
			& $\psi_{11}$ & (4 4 4) & (-4 4 4) & (4 -4 4) & (-4 -4 4) \\
			& $\psi_{12}$ & (-2 -2 4) & (2 -2 4) & (-2 2 4) & (2 2 4) \\
			\hline
		\end{tabular*}
	\end{table*}

	\begin{table*}[h]
		\centering
		\caption{Refined magnetic moments for CoFe$_{2}$O$_{4}$ at 1.6\,K extracted from the Rietveld refinement of neutron powder diffraction data measured on DMC with $\lambda$= 2.460\AA. The space group is $F\overline{1}$ and $\vec{k}$ = (0,0,0) r.l.u.  The magnetic moment directions are given in the first column, and the values of the magnetic moments on the $A$ and $B$ sites are denoted as $m_{a}$ and $m_{b}$, respectively, in units of $\mu_{B}$.  The ``goodness of fit" is given by the $R_{mag}$ factor.}
		\label{Table:moment}
		\begin{tabular*}{\columnwidth}{@{\extracolsep{\fill}}cccc}
			\hline
			Moment & $m_{A}$ & $m_{B}$ & $R_{mag}$   \\ \hline
			//[100] & 4.11 (4)  & 3.83 (3)  & 1.36    \\
			//[110] & 4.10 (4) & 3.83 (3)  & 1.41    \\  
			\hline
		\end{tabular*}
	\end{table*}
	
	For the $A$ site listed in Table \ref{Table:basis_A}, the magnetic structure associated with the magnetic propagation vector $\boldsymbol{k} = (0,0,0)$ has six basis vectors ($\psi_{1-6}$) falling under two irreducible representations ($\Gamma_{8,9}$). For the $B$ site listed in Table \ref{Table:basis_B}, given the different site symmetry, there are twelve basis vectors ($\psi_{1-12}$) and four irreducible representations ($\Gamma_{3,5,7,9}$).  Refinement of the neutron powder data shows that the basis vectors on the $A$ site transforming as the $\Gamma_{9}$ irreducible representation provide the best description of the data.  On the $B$ site, two possibilities give equivalent fits, one transforming as $\Gamma_{9}$ and one as $\Gamma_{7}$ and $\Gamma_{9}$.  This results in two possible magnetic structures: one with the moment pointing along $[100]$ and the other with the moment pointing along $[110]$.  As shown in Table \ref{Table:moment}, both possibilities result in equivalent qualities of fit as characterized by the $R$ factors listed.  The two possible magnetic structures are illustrated in Fig. \ref{fig:neutron} $(b)$.  
	
	It is instructive to compare the refined moment values in Table \ref{Table:moment} to expectations based on full magnetic ordering.  Fe$^{3+}$ consists of five $d$ electrons across five $3d$ orbitals, and Co$^{2+}$ has seven electrons in the outer $3d$ shell.  Therefore we expect Fe$^{3+}$ sites to have $S={5\over 2}$ and Co$^{2+}$ to have $S={3\over 2}$.  If the low-temperature orbital moment is quenched ($\langle \vec{L} \rangle=0$), then the ordered moment should scale as the linear combination of these two spin values weighted by the occupancy listed in Table \ref{Table:neutron_param}, with each of the $A$ and $B$ sites having an effective spin $\tilde{S}$.  This would imply an expected average effective moment of $\tilde{\mu}=g\tilde{S}$ (with the Land\'{e} factor taken to be 2) of 4.5\,$\mu_{B}$ for site $A$ and 3.8\,$\mu_{B}$ for site $B$.  These values agree well with those listed in Table \ref{Table:moment} and demonstrate consistency between the nuclear and magnetic structures refined from the neutron powder diffraction data.
	
	We now discuss the critical properties of the magnetostructural transition at $\sim$ 780\,K.  The local coordination of a magnetic ion influences its spin and orbital properties.  However, with five $d-$electrons, assuming a weak or intermediate crystal field strength,~\cite{Abragam:book} Hund's rules combined with the Pauli exclusion principle imply that the Fe$^{3+}$ cation will always be pure $S={5 \over 2}$ with no orbital degree of freedom regardless of whether it occupies a tetrahedrally ($A$) or octahedrally ($B$) coordinated site.  The single-ion physics of Co$^{2+}$, on the other hand, is very different.~\cite{Cowley13:88}  In an octahedral environment, applying Hund's rules and the Pauli principle to the seven $d-$electrons results in an orbital degeneracy.  This can be seen by populating the strong crystal-field basis consisting of the lower-energy, triply-degenerate $|t_{2g}\rangle$ and the higher-energy, doubly-degenerate $|e_g\rangle$ orbitals with an orbital degree of freedom existing in the lower-$|t_{2g}\rangle$ manifold.  This situation maps exactly onto an orbital degree of freedom with an effective $l_{eff}$=1.~\cite{Sarte20:100,Abragam:book}   
	
	The existence of an orbital degree of freedom~\cite{Slonczewski61:32} for a Co$^{2+}$ cation in an octahedral environment implies that the high-temperature nuclear structure may be unstable against a distortion that breaks the orbital degeneracy via the Jahn-Teller Theorem.~\cite{Jahn37:161,Gehring75:38}  Indeed, the high-resolution x-ray powder diffraction data shown in Fig. \ref{fig:lattice_constants} $(a)$ reveals a discontinuity in the thermal expansion near 780\,K that is consistent with a structural distortion and the onset of spontaneous magnetostriction. That the inflection point of this feature approximately coincides with the onset of magnetic long-range order, as reflected by the appearance of intensity at the $\vec{Q}$=(111) magnetic Bragg peak shown in Fig. \ref{fig:lattice_constants} $(b)$, suggests that there is a symmetry-breaking transition driven by a lifting of a magnetoelastic degeneracy that links with the spin magnetism through the presence of spin-orbit coupling. The width of the anomaly in the lattice parameter, with deviations from the high temperature trend appearing slightly above the ordering temperature suggest that short range magnetic correlations may play a role in driving this magnetostructural transition, as recently observed in MnTe.~\cite{Baral23:33} Our refinements of the magnetic moment further suggest that the low temperature orbital degeneracy is quenched and that only spin-magnetism contributes to the static response.
	
	However, despite the discontinuity in the thermal expansion displayed in Fig. \ref{fig:lattice_constants}, Rietveld refinements of the data measured in the magnetically ordered phase (at room temperature) and in the magnetically disordered phase (at 900\,K) using the Jana software~\cite{Petricek14:229} (summarized in Tables \ref{Table:xray_1} and \ref{Table:xray_2}) yield no observable changes in nuclear symmetry from the high-temperature cubic phase; both refinements to the $Fd\overline{3}m$ space group give the same quality of fit within error.  This is surprising because two high-resolution x-ray studies of single crystal CoFe$_2$O$_4$~\cite{Abes16:93,Yang08:77} report a low-temperature tetragonal phase with space group $I4_{1}/amd$. We note that this space group is a subgroup of the space group $Fd\overline{3}m$ that we used to fit our x-ray and neutron powder diffraction data. Taken together, these results highlight the subtle nature of the structural distortion at the magnetostructural transition.  
	
	The existence of a structural transition at the magnetic transition temperature $T_{c} \sim 780$\,K is supported by the fact that the low-temperature magnetic structure is not compatible with the $Fd\overline{3}m$ space group.  This is shown in Table \ref{Table:symmetry}, which lists the magnetic space groups consistent with the irreducible representations $\Gamma_{9}$ and $\Gamma_{7}$ that best describe the magnetic structure above.  These produce either a tetragonal or rhombohedral space group, but not cubic. The (111) magnetic Bragg peak shown in Fig. \ref{fig:lattice_constants} $(b)$ exhibits a continuous change of intensity starting at $T_c$ that is consistent with a second-order phase transition.  Landau theory would then imply that only one irreducible representation for the magnetic structure exists at low temperatures, and this leaves only a magnetostructural transition to the $I4_{1}/am'd'$ space group listed in Table~\ref{Table:symmetry}, which has a tetragonal unit cell.  We therefore conclude that CoFe$_{2}$O$_{4}$ undergoes magnetoelastic transition from a high-temperature cubic phase to a low-temperature tetragonal phase at $T_c$.  This picture is consistent with a common trend in spinel compounds.~\cite{Reehuis03:35,Tchern03:93,Lane23:5,Wolin18:2,Tsurkan21:926,Radaelli05:7} Moreover, first principles calculations support a tetragonal phase being a ground state.~\cite{Fritsch12:86}
	
	The tetragonal distortion, while it must occur based on symmetry arguments, is evidently beyond the resolution of our x-ray diffractometer.  On the other hand, we should have easily observed the tetragonal splitting reported for single-crystal specimens of CoFe$_2$O$_4$ in Ref. {Abes16:93}.  We therefore conclude that the distortion in our powder samples is considerably less than that found in single crystals.  This is not unusual.  Differences in structural distortions between powders and single crystals have been reported in disordered ferroelectrics ~\cite{Xu03:67,Xu04:70,Conlon04:70,Stock04:69,Gehring04:16,Xu06:79,Donovan24:36,Kumar20:8} and disordered random field magnets.~\cite{Hill97:55} Insofar as the degree of structural disorder and mixing between $A$ and $B$ sites is highly sensitive to the sample preparation in CoFe$_{2}$O$_{4}$~\cite{Sawatzky68:39} and other spinels,~\cite{Islam12:85} it is entirely plausible that the static structural response characterized by the splitting of Bragg peaks could vary in size across differing samples and between single crystals and powders.
	
	\begin{table*}[h]
		\centering
		\caption{Refined structural parameters for CoFe$_{2}$O$_{4}$ at 300\,K extracted from a Rietveld refinement of the Smartlab Rigaku diffraction data measured with a monochromatic wavelength, Cu K$\alpha$ $\lambda =1.54$\,Å. Space group, $Fd\overline{3}m$, No.~227, \(a=8.3839(83)\)~Å,U$_{iso}$ (fixed),\(R_p = 16.46\%\), and \(GoF = 0.86\%\).}	
		\label{Table:xray_1}
		\begin{tabular*}{\textwidth}{@{\extracolsep{\fill}}cccccc}
			\hline
			Atom & $x$ & $y$ & $z$ & Occ.  & Site \\ \hline
			Fe1 & 1/8  & 1/8  & 1/8  & 0.032917 & 8$a$ \\
			Co1 & 1/8  & 1/8  & 1/8  & 0.00875  & 8$a$ \\ 
			Fe2 & 1/2 & 1/2  & 1/2 & 0.050833  & 16$d$ \\
			Co2 & 1/2 & 1/2  & 1/2 & 0.0325  & 16$d$ \\ 
			O & 0.2528 & 0.2528 & 0.2528 & 0.166667 & 32$e$ \\ 
			\hline
		\end{tabular*}
		
		\vspace{2em}
		
		\caption{Refined structural parameters for CoFe$_{2}$O$_{4}$ at 900\,K extracted from a Rietveld refinement of the Smartlab Rigaku diffraction data measured with a monochromatic wavelength, Cu K$\alpha$ $\lambda =1.54$\,Å. Space group, $Fd\overline{3}m$, No.~227, \(a = 8.4447(7)\)~Å,U$_{iso}$ (fixed), \(R_p = 15.39\%\), and \(GoF = 0.85\%\).}
		\label{Table:xray_2}
		\begin{tabular*}{\textwidth}{@{\extracolsep{\fill}}cccccc}
			\hline
			Atom & $x$ & $y$ & $z$ & Occ. & Site \\ \hline
			Fe1 & 1/8  & 1/8  & 1/8  & 0.032917  & 8$a$ \\
			Co1 & 1/8  & 1/8  & 1/8  & 0.008750 & 8$a$ \\ 
			Fe2 & 1/2 & 1/2  & 1/2 & 0.050833  & 16$d$ \\
			Co2 & 1/2 & 1/2  & 1/2 & 0.032500  & 16$d$ \\ 
			O & 0.2528 & 0.2528 & 0.2528 & 0.166667 & 32$e$ \\ 
			\hline
		\end{tabular*}
	\end{table*}
	
	\begin{table*}[h]
		\centering
		\caption{Magnetic space groups vs irreducible representations}
		\label{Table:symmetry}
		\begin{tabular*}{\columnwidth}{@{\extracolsep{\fill}}ccc}
			\hline
			& 8$a$-site & 16$d$-site  \\ \hline
			$I4_1/am^\prime d^\prime$ & $\Gamma_9$ & $\Gamma_9$ \\
			$Imm^\prime a^\prime$ & $\Gamma_9$  & $\Gamma_7,\Gamma_9$  \\  
			\hline
		\end{tabular*}
	\end{table*}
	
	\section{Spin dynamics}
	\label{sect:spin_dynamics}
	
\begin{figure*}
		\centering
		\includegraphics[width=1\textwidth]{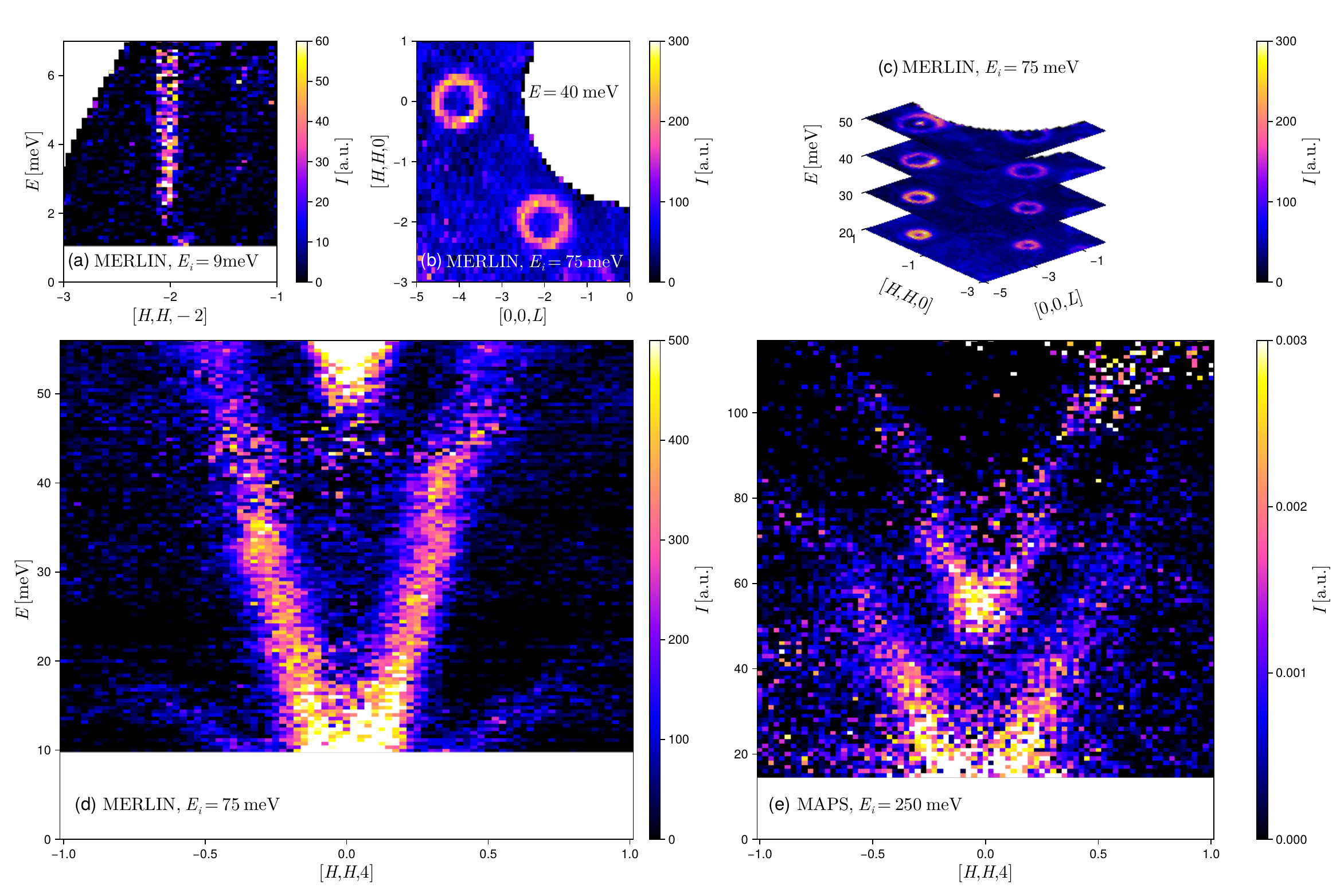}
		\caption{(a) Anisotropy gap at low energy measured on MERLIN.  Data have been integrated $\pm 0.1$\,r.l.u perpendicular to the slice direction. Intensities, $I$, given in arbitrary units. (b) Unsymmetrized constant-energy slice at $E = 40 \pm 2$\,meV. (c) Constant-energy slices (integrated $\pm 2$~meV) through the symmetrized data measured with $E_i =75$~meV. (d) Spin-wave dispersion of the lower magnon mode measured on MERLIN, integrated $\pm 0.1$\,r.l.u perpendicular to the slice direction. (e) Data from MAPS showing the upper magnon mode, integrated $\pm 0.1$\,r.l.u perpendicular to the slice direction. $(\mathbf{Q},E)$ slices have been background subtracted.}
		\label{fig:inelastic}
	\end{figure*}

	We have used neutron spectroscopy to identify the energy scales associated with the magnetic spin dynamics in CoFe$_2$O$_4$. Our neutron scattering measurements on MERLIN and MAPS were made at 10\,K, well below the magnetostructural transition temperature $T_c \sim 780$\,K, and are summarized in Fig.~\ref{fig:inelastic}. Data taken with the lowest incident energy ($E_i = 9$\,meV) provide the best elastic energy resolution (0.27\,meV) and reveal an energy gap $E_g \le 3$\,meV for wave vectors $q$ near the magnetic zone center (Fig.~\ref{fig:inelastic} $a$).  We note that the energy scale of this gap corresponds to $\sim 35$\,K, which is more than 20 times smaller than $T_c$. Constant-energy slices of the unsymmetrized data (Fig.~\ref{fig:inelastic} $b,c$) in the reciprocal-lattice plane $(HHL)$ show strongly dispersive, three-dimensional, spin-wave excitations that reflect the underlying average cubic crystal symmetry. $(\mathbf{Q},E)$ data slices taken along $[HH0]$ and centered on the $(004)$ magnetic Bragg peak show that the spin-wave excitations extend well beyond 50\,meV and weaken significantly in intensity near the zone boundary.  This weakening could indicate the presence of a strong damping mechanism that reduces the lifetime of short-wavelength magnetic excitations (Fig.~\ref{fig:inelastic} $d$).~\cite{Masuda06:96}  
	
	Panel (d) of Fig.~\ref{fig:inelastic} shows a second mode that emerges from the same magnetic zone center at higher energy, between 50 and 60\,meV. We will refer to this higher-energy mode as the ``upper mode" and the lower-energy excitation as the ``lower mode."  Below 20 meV, weak scattering from the transverse acoustic phonon branch propagating along [110] is also observed. This is due to the strong nuclear Bragg intensity at the $(004)$ position, which, combined with the $Q^{2}$ dependence of phonon scattering intensity, leads to a measurable signal at this region of momentum transfer.
	
\begin{figure}
		\centering
		\includegraphics[width=1\columnwidth]{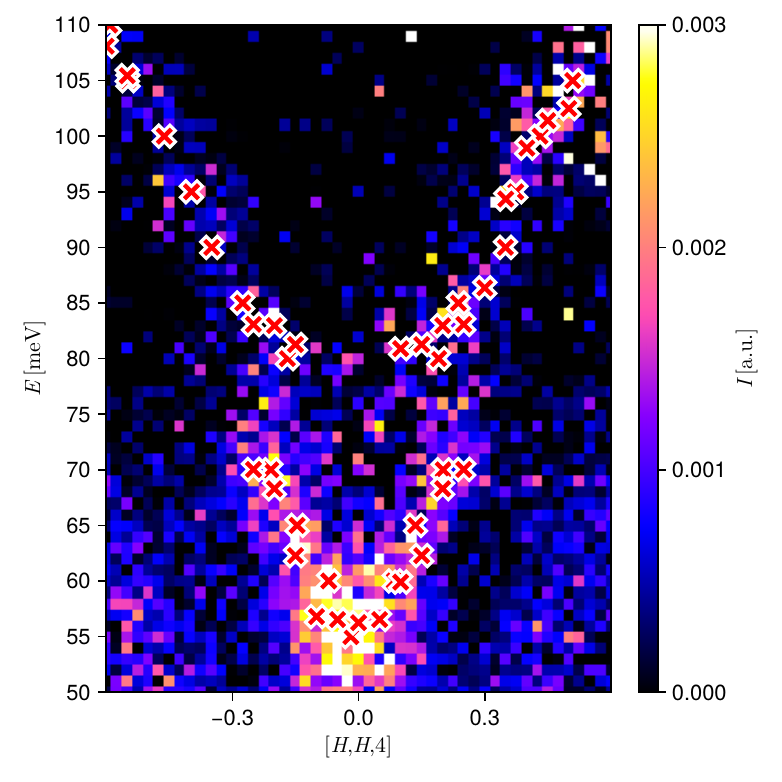}
		\caption{Evidence for anti-crossing of magnons near 75\,meV.  Gaussian-fitted dispersion peaks are overlaid and shown as crosses. A discontinuity in the dispersion curve is visible around $H =0.25$\,r.l.u. Data have been integrated $\pm 0.1$\,r.l.u perpendicular to the slice direction.}
		\label{fig:anticrossing}
	\end{figure}

	To explore the upper mode further, we performed measurements on the MAPS spectrometer with an incident energy $E_i = 250$\,meV, which facilitates access to energy transfers in the momentum region of interest up to $\sim  120$\,meV. The upper mode disperses beyond 100\,meV and has a bandwidth similar to that of the lower mode (Fig.~\ref{fig:inelastic} $e$).  The scattering intensity for both modes decreases quickly as the magnetic zone boundary is approached, complicating the task of tracking the $Q-E$ dispersion. Of particular interest is the region 70\,meV $\leq E \leq$ 80\,meV, where the spin-wave velocity appears to change abruptly near $H=\pm0.25$\,r.l.u. The data between 70\,meV $\leq E \leq$ 80\,meV are plotted in closer detail in Fig.~\ref{fig:anticrossing}, where dispersion peaks obtained from Gaussian fits to one-dimensional constant-$\mathbf{Q}$ and constant-energy cuts of symmetry-equivalent positions have been overlaid to improve statistics.  The upper mode's slope distinctly changes around 75~meV. This suggests a splitting that we will term an ``avoided crossing" throughout this paper, as it is reminiscent of such a phenomenon.
	
	Our neutron inelastic scattering measurements can be summarized as follows: (i) Two primary spin-wave branches with similar energy bandwidths are observed, separated by a gap of $\sim 55$~meV; (ii) the spin-wave intensities decrease rapidly as the zone boundary is approached; (iii) the lower mode exhibits a small energy gap of $\sim 3$~meV; (iv) near $\sim$ 75\,meV, the upper mode appears to split at an avoided level crossing.  In the following section, we discuss and reconcile these experimental observations with theoretical calculations using linear spin-wave theory.
	
	\section{Theory}
	\label{sect:theory}
	
	Given that CoFe$_2$O$_4$ has two different magnetic ions and two inequivalent crystallographic sites, it is tempting to suggest that the upper and lower modes originate from different single-ion ground states like those observed in disordered magnets.~\cite{Buyers71:27,Buyers72:5,Birgeneau75:8} Fe$^{3+}$ is spin-orbitally inactive in both tetragonal and octahedral environments and so might be expected to give rise to an almost gapless mode,~\cite{Lane21:104} whereas Co$^{2+}$ is spin-orbitally active in the octahedral environment and often exhibits a large single-ion anisotropy resulting in an energetic gap in the neutron response.~\cite{Sarte20:100} But the spinel crystal structure does not permit a clean decoupling of Co$^{2+}$ and Fe$^{3+}$ sublattices. Moreover, the similar dispersion of both modes suggests that they arise from a similar exchange-coupling mechanism, which further contradicts an interpretation of decoupled Co$^{2+}$ and Fe$^{3+}$ networks.  It has been reported that CoFe$_2$O$_4$ forms structural domains with differing strain,~\cite{Abes16:93,Yang08:77} which might give rise to magnetocrystalline anisotropy and perhaps explain the presence of energetically gapped modes. However, the strain reported in the high-resolution x-ray study of single crystal CoFe$_2$O$_4$ is $\lesssim 10^{-3}$,~\cite{Abes16:93} which is unlikely to generate a gap of this magnitude.~\cite{Schron12:86}  For example, the tetragonal distortion in CoF$_{2}$~\cite{Baur71:27} results in energetic gaps of $\sim$ 5-20 meV.~\cite{Cowley73:6}  Finally, the gap to the upper mode is approximately twice as large as the spin-orbit gap typically found in other Co$^{2+}$ compounds, which is $\sim$ 20 to 30\,meV.~\cite{Sarte18:98,Sarte20:100}

	In light of these factors, we propose that the origin of the large inter-mode gap is not due to the single-ion physics of each magnetic ion but rather stems from a strong magnetic exchange and the spinel crystal structure. In the following subsections, we model our neutron data to support this claim by using linear spin-wave theory in three steps.  First, we fit an effective linear spin-wave model to the MERLIN and MAPS data.  Second, we consider a model with no anisotropy to understand the origin of the large energy gap separating the upper and lower modes.  Third, we apply a super-cell model to try to isolate the effects of disorder on the magnetic excitations.   
	
	\subsection{Linear spin-wave model}
	\label{sect:model}
	
	To characterize the magnetic excitations plotted in Figs.~\ref{fig:inelastic} and \ref{fig:anticrossing}, we consider a Heisenberg spin Hamiltonian with single-ion anisotropy. Due to the small magnitude of the tetragonal distortion (Sect.~\ref{sect:structure}) we assume an average  cubic $Fd\overline{3}m$ structure in our analysis. The only symmetry-allowed quadratic anisotropy lies on the octahedral $B$ site of the spinel lattice, which for a perfect inverse spinel is half occupied by Fe$^{3+}$ and Co$^{2+}$ cations. The threefold $[111]$ axis and mirror plane along $[\overline{1}01]$ restricts the anisotropy axis to be along the three-fold axis. We express this in the crystallographic reference frame with $\hat{z} \parallel c$ using Stevens operators~\cite{Hutchings64:16} ($\mathcal{O}_{l}^{m}$) that are quadratic in spin operators and therefore invariant under time-reversal symmetry:
		
	\begin{widetext}
	\begin{equation}
		\hat{\mathcal{H}}=J_{1}\sum_{\langle i\in B,j\in B\rangle}\mathbf{\hat{S}}_{i,B}\cdot \mathbf{\hat{S}}_{j,B}+J_{2}\sum_{\langle i\in A,j\in B\rangle}\mathbf{\hat{S}}_{i,A}\cdot \mathbf{\hat{S}}_{j,B}+J_{3}\sum_{\langle i\in A,j\in A\rangle}\mathbf{\hat{S}}_{i,A}\cdot \mathbf{\hat{S}}_{j,A}-D\sum_{i\in B}(\mathcal{O}_{2}^{-2} + 2\mathcal{O}_{2}^{-1}+2\mathcal{O}_{2}^{1}). \nonumber
	\end{equation}
		\end{widetext}
		
	\noindent The angular brackets denote summations over nearest-neighbor bonds. The Heisenberg exchange constants are labeled from $J_{1-3}$ in order of bond length, and $D>0$ represents an easy-axis anisotropy.
	
		\begin{figure} 
		\centering
		\includegraphics[width=0.75\columnwidth]{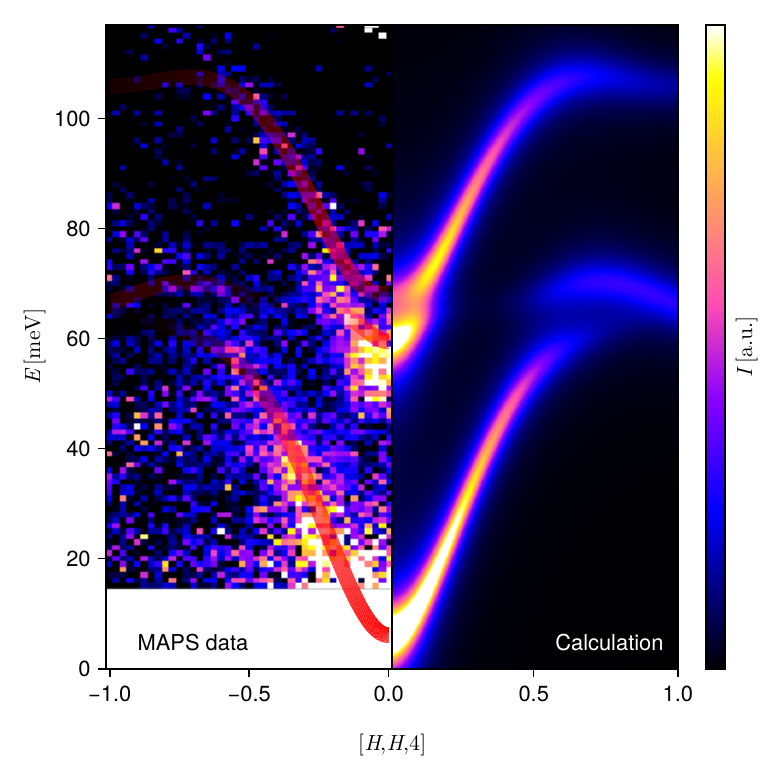}
		\caption{Comparison of the fitted model convolved with a Lorentzian kernel with a full width at half maximum (FWHM) of 8\,meV (right) with the MAPS data (left). Measured data have been integrated $\pm 0.1$\,r.l.u perpendicular to the slice direction. Over-plotted in red is the dispersion of the model with an opacity proportional to the dynamical structure factor.}
		\label{fig:fit}
	\end{figure}

	We construct an effective, averaged, linear spin-wave theory model with the exchange constants absorbing any renormalization due to anomalies in the scattering from cation site mixing. We use conventional $SU(2)$ linear spin-wave theory where only a single flavor of Holstein-Primakoff bosons is introduced per sublattice.~\cite{Toth15:27} To account for the average moment of each site, the length of the $SU(2)$ spin coherent state on site $A$ was fixed to be $5/2$ while the length on site $B$ was rescaled to be $S_B$ (see Sect.~\ref{sect:spin_coherent} for details). Such an approach is an approximation and treats the disorder on average (see Sect.~\ref{sect:disorder} for large-box calculations which provide additional justification for this simplified approach). The parameter $S_B$ was fixed by the ratios of the ordered moments observed on site $A$ and site $B$ reported above in our neutron diffraction results. The calculations were performed using the Sunny.jl software.~\cite{Dahlbom25:preprint,Zhang21:104}
	
	Our minimal linear spin-wave model is compared to the data in Fig.~\ref{fig:fit}. The fitted values are shown in Table \ref{Table:Parameters} (see Appendix~\ref{sect:fitting} for further details).
	
	\begin{table*}[h]
		\centering
		\caption{Fitted linear spin-wave theory parameters for the AB$_{2}$O$_{4}$ magnetic Hamiltonian presented in the main text fitted to the MERLIN and MAPS single crystal CoFe$_{2}$O$_{4}$ data.}
		\label{Table:Parameters}
		\begin{tabular*}{\columnwidth}{@{\extracolsep{\fill}}cc}
			\hline
			$J_{1}$ & $B-B$ site exchange: 0.43 $\pm$ 0.20 meV  \\
			$J_{2}$ & $A-B$ site exchange: 4.70 $\pm$ 0.18 meV  \\
			$J_{3}$ & $A-A$ site exchange: 2.06 $\pm$ 0.81 meV \\
			$D$ & Trigonal anisotropy: 1.66 $\pm$ 0.09 meV.\\
			\hline
		\end{tabular*}
	\end{table*}

	We stress that the parameters quoted here are phenomenological in the sense that they absorb the ``average" effect of the site disorder discussed above.  Therefore, they do not directly represent the microscopic exchange constants, which likely form some distribution centered around the values given in Table \ref{Table:Parameters} (see Sect.~\ref{sect:disorder} for further discussion).

	\subsection{Effective $S=1$ model}
	
	Having established the energy scales governing the spin dynamics in CoFe$_{2}$O$_{4}$, we identify the physical origin of the large splitting between the upper and lower modes by considering the simpler case with no anisotropy and a dominant antiferromagnetic coupling between the $A$ and the $B$ sublattices. The ground state is found by anti-aligning the spins on each sublattice and assuming isotropy.  We choose the spins to order along the $\hat{z}$-axis and map the Hamiltonian onto an effective spin model of unit length, $\mathbf{\hat{S}}_{\mu}=\alpha_\mu\mathbf{\hat{T}}_\mu$ (see Appendix~\ref{sect:spin_exchange} for details):

\begin{widetext}	
	\begin{equation}
		\mathcal{H}^{S=1}=\frac{1}{2}\tilde{J}_{AB}\sum_{i,j \in \{A,B\}}\mathbf{\hat{T}}_{i}\cdot \mathbf{\hat{T}}_{j} +\frac{1}{2}\tilde{J}_{AA}\sum_{i,j \in \{A\}}\mathbf{\hat{T}}_{i}\cdot \mathbf{\hat{T}}_{j} + \frac{1}{2}\tilde{J}_{BB}\sum_{i,j \in \{B\}}\mathbf{\hat{T}}_{i}\cdot \mathbf{\hat{T}}_{j}+\sum_{i\in \mu}h_{\mathrm{ferri}}^{\mu}\hat{T}_{i}^{z}. \nonumber
	\end{equation}
	\end{widetext}
	
	\noindent The effective couplings absorb the magnitude of the spin on each site so that $\sqrt{\alpha_\mu \alpha_\nu} J_{\mu\nu} = \tilde{J}_{\mu\nu}$.  The  factor of $\frac{1}{2}$ is included to avoid double counting. 
	
	This magnetic Hamiltonian differs from the previous one by the presence of a site-dependent molecular field $h_{\mathrm{ferri}}^{\mu}$ ($\mu = A, B$).  It arises from the static ferrimagnetism of CoFe$_2$O$_4$ because there is an imbalance between the number of $A$ and $B$ sublattice spins that are anti-aligned. This molecular field breaks time-reversal symmetry and generates an effective anisotropy on the individual magnetic sites. This molecular field also Zeeman-splits the doubly degenerate bands that would otherwise exist in an ordered antiferromagnet with two sublattices, giving rise to the experimentally observed upper and lower modes. The molecular field can be expressed in terms of the ratio of the spins on the $A$ and $B$ sites $S_{A}/S_B = \gamma$ and the coordination number $z_\mu$ as
	
	\begin{gather}
		h_{\mathrm{ferri}}^{A} = \tilde{J}_{AB}z_A (1-\sqrt{1/\gamma}), \nonumber \\ 
		h_{\mathrm{ferri}}^{B} =\tilde{J}_{AB}z_B (1-\sqrt{\gamma}). \nonumber
		\label{eq:magnetic_field}
	\end{gather}
	
	\noindent The internal molecular field can therefore be tuned by adjusting the types of cations and the degree of mixing between the spinel $A$ and $B$ sites.
	
	\begin{figure*}
		\centering
		\includegraphics[width=1\textwidth]{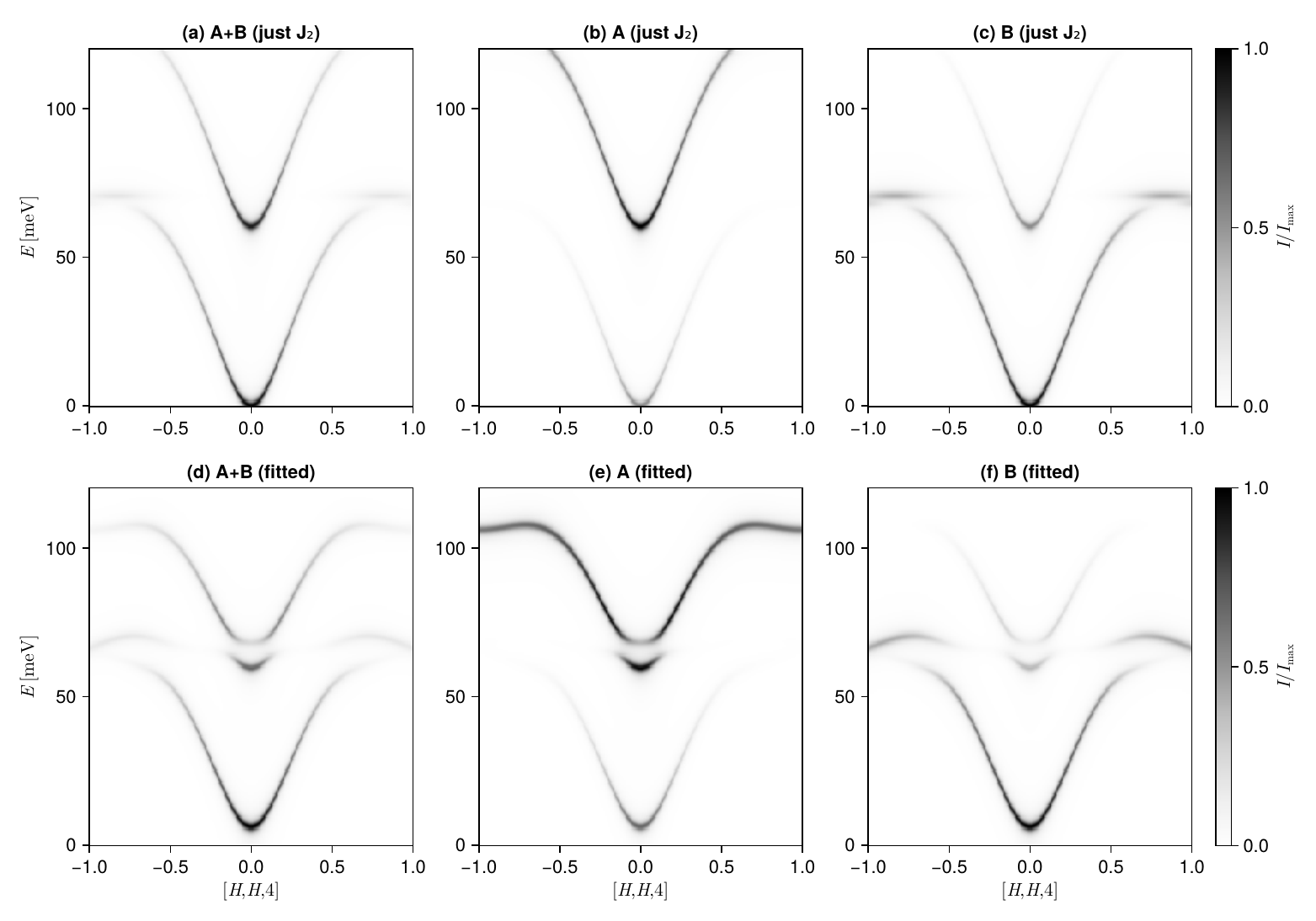}
		\caption{Trace of the partial dynamical structure factor. (a-c) show the trace for a simplified model with only the dominant $J_2$ exchange. Panels (b) and (c) show the result with the contributions only from the $A$ and $B$ sites respectively. (d-f) show the same calculations with the full fitted model.}
		\label{fig:sublattice}
	\end{figure*}
	
	This analysis finds that energy difference between the upper and lower modes is intrinsic to the ferrimagnetic structure; it does not rely on the fact that $S_A \neq S_B$, only that the molecular mean field differs on each sublattice. The spinel structure exhibits a nominal 1:2 ratio of $A$ sites to $B$ sites, and the local coordination of each site differs, as there are twelve $J_{AB}$ bonds per $A$ site ($z_A = 12$) and six per $B$ site ($z_B = 6$). The gap between the two modes at the zone center is then given by the difference between the molecular field contribution from $J_{AB}$ on each sublattice
	
	\begin{equation}
		\Delta_{S=1} = 12\tilde{J}_{AB}\langle S_B\rangle -6\tilde{J}_{AB}\langle S_A\rangle, \nonumber
	\end{equation}
	
	\noindent which from the neutron spectroscopy data is $\sim$ 60 meV.  To emphasize this point, Fig.~\ref{fig:sublattice} plots the sublattice character of each mode by calculating the trace of the dynamical spin structure factor and setting the form factor to zero for each of the sites in turn. The result is a measure of the contributions from each sublattice to the two modes. The upper panel shows a simplified model where only $J_2$ is non-zero. The $A$ sublattice (Fig.~\ref{fig:sublattice} $b$) contributes more to the upper mode, whereas the $B$ sublattice primarily contributes to the lower mode (Fig.~\ref{fig:sublattice} $c$). This pattern is reproduced with the fitted values (Sect.~\ref{sect:model}) in the lower panels (Fig.~\ref{fig:sublattice} $d-f$).
	
\begin{figure}
		\centering
		\includegraphics[width=1.0\columnwidth]{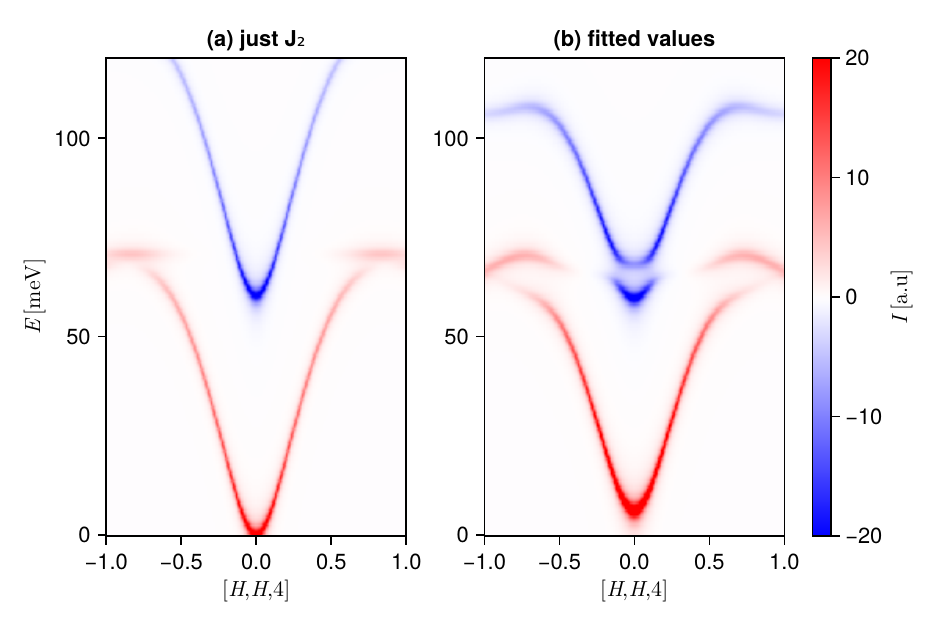}
		\caption{Chiral component of the dynamical spin structure factor for (a) a simplified model with only $J_2 \neq 0$, and (b) the fitted values.}
		\label{fig:chirality}
	\end{figure}

	The difference between the sublattice characters of the lower and upper modes has important ramifications for the nature of the magnetic excitations. At the mean-field level, one can view the upper mode as originating predominantly from $A$-sublattice moments Larmor-precessing in the local molecular field provided by the $B$ sublattice (and vice versa). Since the molecular field points in the opposite direction for each sublattice, the chiralities of the lower and upper magnon modes have opposite signs (Fig.~\ref{fig:chirality}). In the contrasting case of an antiferromagnet, the two modes are degenerate and the net chirality vanishes. Evidence of differing chiralities has been observed in Y$_{3}$Fe$_{5}$O$_{12}$ using polarized neutrons.~\cite{Nambu20:125}  This compound also displays a splitting of the degeneracy of the antiferromagnetic bands. Renewed interest in chiral spin waves was recently prompted by the emergence of altermagnetism as a new paradigm in magnetism.~\cite{Smejkal22:12,Mazin22:12,Bai24:34} In this context, the degeneracy is split by anisotropic magnetoelastic interactions,~\cite{Smejkal23:131,Liu24:133} where the product of parity and time-reversal symmetry is broken.~\cite{Cheong24:9,Smejkal23:131,Consoli25:134}  
	
	\subsection{Effects of site disorder}
	\label{sect:disorder}
	
	In this section we address the effects of cation site-disorder on the low-temperature magnetic excitations. To address such effects in the past, spin-wave calculations were performed using $T$-matrices.~\cite{Brenig92:43,Chernyshev02:65,Cowley72:44}  However, this method is not valid for large concentrations of disorder, as in CoFe$_{2}$O$_{4}$. Another approach is to perform spin-dynamic calculations on a large supercell, but this requires more fitting parameters in order to reflect the exchange parameters~\cite{Lane25:37} of each ionic species at each of the possible sites. To simplify matters, we assume that the effects of disorder can be treated on average. This assumption can be justified by noting that the scattering potential associated with defects (for example, Fe$^{3+}$ cations on the Co$^{2+}$-rich, $B$-site sublattice) is weak insofar as the spins on each sublattice have a similar magnitude ($S=5/2$ vs $3/2$).  Furthermore, first-principles calculations predict a similar exchange magnitude.~\cite{Ansari20:102} We note that the spin excitations we measured in CoFe$_{2}$O$_{4}$ differ significantly from those in systems with spin vacancies, which, in combination with disorder, give rise to additional local modes and continua.~\cite{Chernyshev02:65,Dao24:preprint,Lane25:37} We see no such excitations in our data.
	
	To explore the role of disorder, we apply linear spin-wave theory to large unit cells ($>1500$ spins) with various realizations of magnetic disorder. We first initialize a unit cell of size $(4,4,4)$ with $S=1$ spins and renormalize the spin-coherent states on each site according to a distribution that mimics the population of $S=5/2$ and $S=3/2$ spins on the $A$ and $B$ sites for a perfect inverse spinel. We then calculate the dynamical spin structure factor using the Kernel Polynomial Method for linear spin-wave theory (KPM-SWT).~\cite{Lane24:17}
	
\begin{figure*}
		\centering
		\includegraphics[width=1\textwidth]{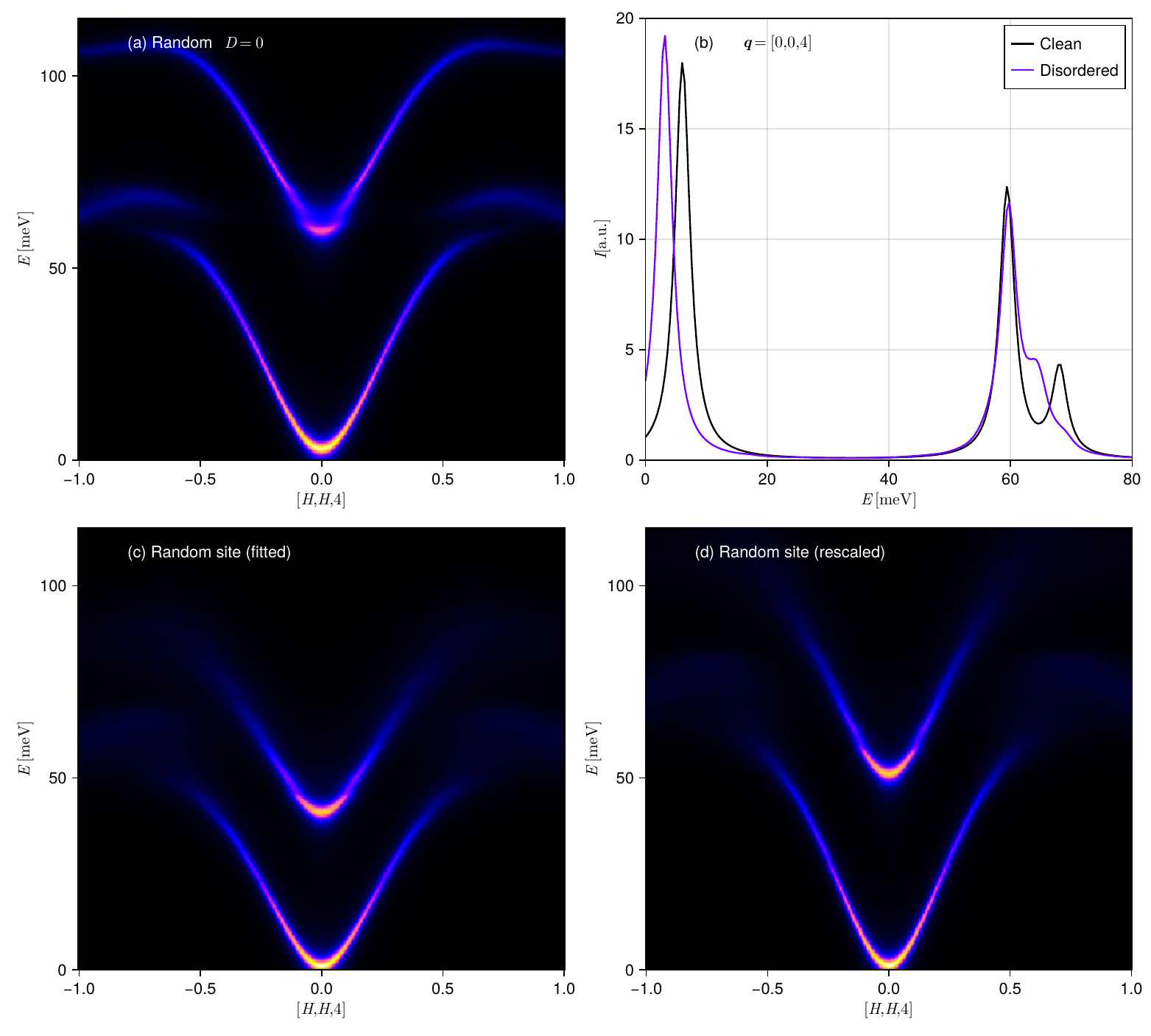}
		\caption{KPM-SWT calculation of the site-disordered model for \ce{CoFe2O4}. (a) Neutron scattering spectra using the fitted values from the model in Sect.~\ref{sect:model} with 50 \% of the $B$ sites having zero anisotropy. (b) Constant-$\mathbf{Q}$ cut through the spectra in (a) at $\mathbf{Q}=(0,0,4)$ showing a reduction in the low-energy anisotropy gap in the disordered model and a smearing of the upper-mode splitting. (c) Random-site model with spin-coherent states scaled according to the distribution of Co$^{2+} $ and Fe$^{3+}$, and (d) spectra with the exchange constants rescaled to better match the experimental bandwidth.}
		\label{fig:disorder}
	\end{figure*}

	The effects of anisotropy disorder is considered first in Fig.~\ref{fig:disorder} $(a)$ by keeping the fitted values from before and then scaling $S_A = 5/2$ and $S_B = \frac{M_B}{M_A}S_A$, with the anisotropy on half of the $B$ sites set to zero. The region surrounding the high-energy anti-crossing shows a broadening, and the two distinct upper modes are no longer resolvable. This is reflected in a constant-$\mathbf{Q}$ cut (Fig.~\ref{fig:disorder} $b$) through our calculated neutron response, where the two distinct modes are seen to merge. Notably, the peak position of the upper mode does not shift significantly, however the anisotropy gap of the lower mode (of primarily $B$ sublattice character) decreases on the inclusion of anisotropy disorder. We emphasize that while disorder introduces renormalization effects, it does not account for the large energy difference between the upper and lower modes.
	
	We next consider the case where the $A$ and $B$ sites are occupied by $S=5/2$ and $S=3/2$ cations according to our neutron powder diffraction measurements (0.62:0.38 on the $A$ site and 0.24:0.76 on the $B$ site). Fig.~\ref{fig:disorder} $(c)$ shows the calculated spectrum using the fitted values from our simple spin-wave model (Sect.~\ref{sect:model}). The smearing of the anticrossing is less severe than that in Fig.~\ref{fig:disorder} $(a)$, however the site disorder leads to a variation in the exchange strengths across bonds that couples spins of different length. This causes long-wavelength excitations near the zone center to remain well defined while shorter-wavelength excitations become progressively weaker in intensity as the wave vector $q$ approaches the zone boundary.  This response is reminiscent of models of exchange disorder.~\cite{Lane25:37,Stock05:71,Carlson04:70}  In Fig.~\ref{fig:disorder} $(d)$, we rescale the exchange constants to account for the different magnetic moments used in the simplified spin-wave model (Sect.~\ref{sect:model}) compared with the distribution of site occupancy used in this calculation.
	
	It is important to state the limitations of this analysis. First, there is known to be significant sample variation in the cation-site occupancy, which means that the values extracted from our powder diffraction data may differ considerably from those for the single crystal sample used in the neutron inelastic scattering measurements. Secondly, it is expected that the exchange constants should depend on the spin species on each site. This has not been included in our KPM-SWT calculations. Third, the effects of spin-phonon coupling have not been considered in our calculations.  Nonetheless, the model presented here reproduces the main features of our data extremely well.
	
	\subsection{Anisotropy}
	\label{sect:anisotropy}
	
\begin{figure*}
		\centering
		\includegraphics[width=1.7\columnwidth]{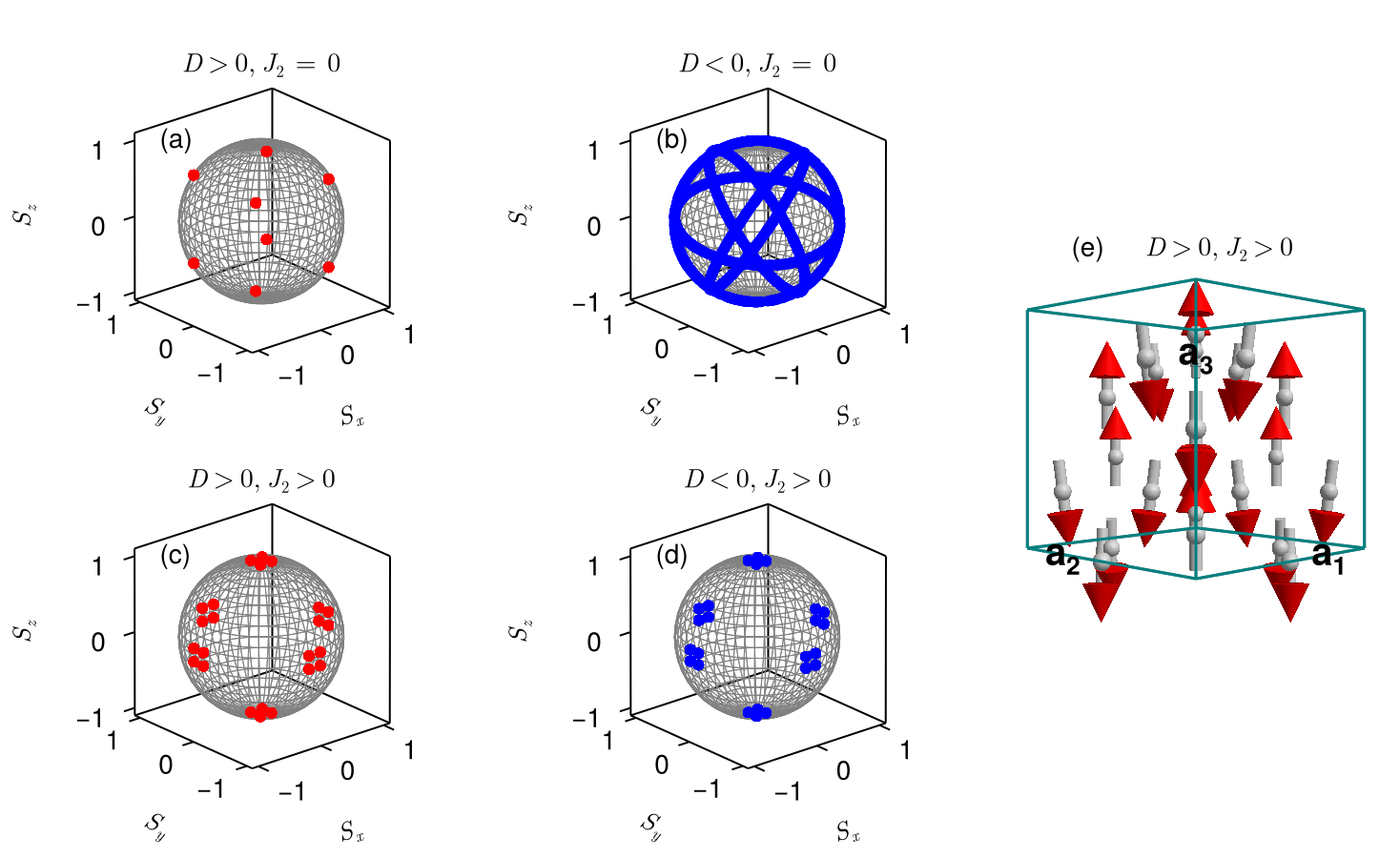}
		\caption{Ground state magnetic structure of a simplified model of \ce{CoFe2O4}. The top row shows the direction of the $B$-site moments in the magnetic ground state for a simplified model with $J_2=4, \:D=\pm2$. In both the case of easy-axis ($D>0$) and easy-plane ($D<0$) anisotropy, the ground state is collinear with alignment along the cubic axis (see right panel for full magnetic structure). Lower panels show the ground state with $J_2=0$ indicating the locations of the easy axes and easy planes.}
		\label{fig:anisotropy}
	\end{figure*}

	The anisotropy considered in the spin-wave model of Sect.~\ref{sect:model} is a uniaxial (quadratic in spin operators) anisotropy and is motivated by the high-temperature trigonal symmetry of the local Co$^{2+}$ environment on the $B$ sites. We expect this to be an easy-axis anisotropy ($D>0$) based on a point-charge calculation. The rotation of the three-fold [111] axis within the unit cell according to the $Fd\overline{3}m$ space group gives rise to four different local easy axes on the $B$ sites. In the absence of exchange coupling, the Co$^{2+}$ spins will point along these easy-axis directions if $D>0$.  Fig.~\ref{fig:anisotropy} $(a)$ depicts the directions of the [111] axes as points on the surface of the sphere. If the sign of the anisotropy is swapped, then one has an easy-plane anisotropy ($D<0$), and the spins will lie in the plane defined by a normal oriented along the [111] direction, as indicated by the blue lines in Fig.~\ref{fig:anisotropy} $(b)$.  The analysis of our neutron inelastic scattering data reveals the presence of a strong antiferromagnetic coupling $J_2$ between the $A$ and $B$ sublattices (Table \ref{Table:Parameters}). This enforces the collinear alignment of the spins within each sublattice, which couple indirectly through the spins on the $A$ sublattice (even in the absence of $J_1$ and $J_3$, the intra-sublattice bonds). In the absence of anisotropy, there is a continuous spin-rotational symmetry that is spontaneously broken, meaning the magnetization can point along any direction. 
	
	The inclusion of a non-negligible, easy-axis anisotropy $D$ and a strong antiferromagnetic inter-sublattice coupling $J_2$ leads to a competition between the local anisotropy, which favors alignment along the local $[111]$ axes, and the isotropic exchange, which favors a collinear structure. The resulting ground state is a collinear alignment of spins on the $A$ and $B$ sublattices along the crystallographic [100]-type lattice vectors. The $<$100$>$ directions lie at the center of the faces of the cube whose vertices are the [111] easy axes. A small canting of the spins from the $<$100$>$ directions is seen in our calculations, however such canting would not be resolvable in our powder diffraction measurements.  But, as suggested by Teillet \textit{et al},~\cite{Teillet93:123} this may explain the observed variation of the refined magnetic moments of each sublattice from their theoretical values. This is further supported by M{\"o}ssbauer studies of the Zn-diluted series Co$_{1-x}$Zn$_x$Fe$_2$O$_4$.~\cite{Pettit71:4}
	
	Within our model, there is a degeneracy between states with spins aligned along each of the crystallographic lattice vectors, mimicking a cubic single-ion anisotropy. It is interesting that a similar ground state is found in the case of easy-plane anisotropy (Fig.~\ref{fig:anisotropy} $d$), highlighting the fact that this is a consequence of the competition between the local anisotropy directions, which rotate with the crystal symmetry, and the isotropic exchange, which favors global coalignment. In the absence of a tetragonal distortion that further breaks the degeneracy, this would suggest the presence of magnetic domains, with the spins aligned along each of the cubic axes. This situation is ideal for the emergence of magnetostriction, where small strains within tetragonal domains can pick out one of these axes. The combined action of the [111] trigonal anisotropy and the strong antiferromagnetic exchange constant $J_2$ make the canting of these spins away from the chosen [100] axis unfavorable, leading to the switching of these structural domains rather than spin canting in applied field. 
	
	\section{Discussion}
	\label{sect:discussion}
	
	Our results have direct implications for the favorable device-relevant properties of CoFe$_2$O$_4$.  The strong antiferromagnetic exchange interaction (see $J_2$ in Table~\ref{Table:Parameters}) between the spinel $A$ and $B$ sublattices, being the largest energy scale in the Hamiltonian, is responsible for the high magnetic-ordering temperature, which is critical to device applications. It is also responsible for the large splitting between the upper and lower magnon branches, a feature that has been reported in other ferrimagnets such as Fe$_3$O$_4$,~\cite{McQueeney07:99} Tb$_3$Fe$_5$O$_{12}$,~\cite{Kawamoto24:124} and Y$_3$Fe$_5$O$_{12}$.~\cite{Princep17:2}  Our finding of an avoided crossing in the upper magnon mode and a small gap in the lower mode indicate the presence of an anisotropy that is created by local distortions of the oxygen octahedra surrounding the magnetic Co$^{2+}$ and Fe$^{2+}$ ions.  It is the combination of this anisotropy and the dominant magnetic exchange interaction, which generates very large and mismatched molecular fields on the distinct magnetic $A$ and $B$ sites, that locks the magnetization direction within each structural domain and enhances the magnetostrictive response of CoFe$_2$O$_4$.  We discuss these results and their implications in more detail below.
	
	There are two crystalline anisotropies to consider in CoFe$_{2}$O$_{4}$. The first arises from the local trigonal distortions of the oxygen octahedra that are present in the high-temperature $Fd\overline{3}m$ cubic phase.  These have been observed in other ferrite spinels, including FeMn$_2$O$_4$.~\cite{Zhang23:35}  The second comes from the tetragonal distortion that coincides with the magnetic Curie temperature $T_c$.  Given the weakness of the tetragonal distortion,~\cite{Abes16:93} we have assumed that the trigonal distortions of the oxygen octahedra provide the dominant crystalline anisotropy.  Our assumption is consistent with single crystal studies of CoFe$_2$O$_4$ that demonstrate that the magnetoelastic domains can be switched by a field of only 0.4\,T.~\cite{Abes16:93}  This equates to a small energy scale of $\sim$ 0.03\,meV (1\,T $\sim$ 0.06\,meV), which is roughly two orders of magnitude smaller than the $\sim$ 3\,meV anisotropy gap we observe in our neutron data (see Fig.~\ref{fig:inelastic} a).  It is for these reasons that we attribute the 3\,meV gap to a high-temperature trigonal anisotropy in our spin-wave calculations.   
	
	Our high-resolution x-ray diffraction data on a powder sample of CoFe$_2$O$_4$ reveal a clear anomaly in the temperature dependence of the lattice constant at $T_c$ that, in combination with our magnetic symmetry analysis, is consistent with a tetragonal distortion. Yet we detect no splitting of the nuclear Bragg peaks that should accompany a tetragonal distortion.  All peaks are resolution-limited, and this places an upper bound on the strain of $1 - c/a \le 10^{-4}$.  As previous synchrotron studies of single crystal CoFe$_2$O$_4$ see clear evidence of tetragonal domains with strains $1 - c/a \sim 10^{-3}$,~\cite{Abes16:93} i.e. roughly one order of magnitude larger than our upper bound, we conclude that finite-size effects are significant in CoFe$_2$O$_4$.  We believe this is important because the weak tetragonal distortion at $T_c$ leads to field-tunable structural domains, and there is evidence that the magnitude of this distortion is closely linked to the strength of the magnetostriction. This would explain why single crystals of CoFe$_2$O$_4$ exhibit magnetostriction values that are roughly 10 times greater than those measured in powders.  In fact, comparisons with other inverse spinels reveal that weaker magnetostriction correlates with smaller tetragonal distortions.  A notable example is Fe$_3$O$_4$, which remains cubic until the low-temperature Verwey transition~\cite{Pan76:34} and exhibits a magnetostriction coefficient that is an order of magnitude smaller than that of CoFe$_2$O$_4$.~\cite{Bozorth55:99}
	
	The observation of simultaneous structural and magnetic transitions in CoFe$_{2}$O$_{4}$ suggests that spin-orbit coupling ($\sim \vec{L} \cdot \vec{S}$) is playing an important role because it provides a mechanism by which the spin magnetism can couple to the crystal electric field.  As pointed out above, a Co$^{2+}$ cation in an octahedral crystal field has an orbital degeneracy in the $|t_{2g}\rangle$ channel, and the Jahn-Teller theorem states that the crystal should distort to lift this degeneracy.  Structural and magnetic transitions often occur simultaneously when spin-orbit coupling is present, as happens in CoO and other Co$^{2+}$-based compounds.~\cite{Jauch01:64}  Our postulate that the approximate orbital degeneracy drives the transition at $T_c \sim 780$\,K gains further credence when looking at the analogous inverse spinel CuFe$_{2}$O$_{4}$,~\cite{Levinstein65:36,Kulkarni79:14,Balagurov13:58} where the octahedrally coordinated Jahn-Teller active ion Cu$^{2+}$ has a degeneracy in the $|e_{g}\rangle$ orbital manifold.  Ions in such cases are known to be strongly Jahn-Teller active, and CuFe$_2$O$_4$ does undergo a large structural transition.  We note that the large molecular fields present at low temperatures, which introduce an energy scale comparable to the magnon bandwidths, exceeds the spin-orbit-coupling energy scale.  Therefore, we expect the low-temperature orbital magnetism to be effectively quenched.  This is consistent with our neutron diffraction analysis, which concluded that the low-temperature magnetically ordered state is dominated by spin magnetism.
	
	A framework for understanding single-ion magnetostriction has been discussed by Callen and Callen.~\cite{Callen63:129} Slonczewski applied a similar theory to cobalt-substituted iron ferrite and found a large magnetostrictive effect due to the presence of spin-orbit coupling from the Co$^{2+}$ ions.~\cite{Slonczewski61:122}  We note that first-principles calculations comparing CoFe$_{2}$O$_{4}$ to NiFe$_{2}$O$_{4}$~\cite{Fritsch12:86} report considerably stronger magnetostriction in the Co-variant.  This is consistent with the fact that Ni$^{2+}$ does not have an orbital degree of freedom in a dominant octahedral crystalline electric field environment.~\cite{Wallington15:92,Sarte18:98,Sarte18:98}
	
	Once magnetic order forms in CoFe$_2$O$_4$, the strong preference for the net moment to point along $<$100$>$ effectively locks the magnetization direction of each domain. The strong directional anisotropy imposed by the large molecular field creates a huge energy scale that cannot be overcome by modest applied fields. Therefore, as the net moment of the domain is locked in place and forbidden from reorienting in field, it is energetically favorable for the structural domains to reorient, rather than for the spins to rotate coherently independent of the structure.  The strength of the easy-axis locking can be tuned by varying the effective internal ferrimagnetic molecular field, which promotes collinear alignment, and the single-ion anisotropy, which attempts to cant the spins towards their local trigonal axes. Both of these parameters are sensitive to site occupation and likely contribute to the tunability and sample variation of \ce{CoFe2O4}.
	
	Several results from our study elucidate the microscopic basis of the favorable magnetostrictive properties of CoFe$_{2}$O$_{4}$.  The first is the large exchange coupling and the presence of two different magnetic cations; these generate the large magnon band splitting and the large magnetic Curie temperature, making CoFe$_{2}$O$_{4}$ amenable to room temperature applications.  The second is the presence of orbitally active Co$^{2+}$ ions that produce a small tetragonal distortion that is easily switchable in modest applied magnetic fields.  The third is the presence of a high-temperature trigonal anisotropy that provides the required directional dependence for magnetostriction. While anisotropy has been implicated as an important parameter in magnetostriction in many previous studies, our study points to the strong role that differing molecular fields play, which are characterized by the magnon band splitting observed in the excitations.
	
	The presence of the large magnon band splitting has implications beyond magnetostriction. Aside from shedding light on the magnetoelastic properties of \ce{CoFe2O4}, our neutron scattering measurements combined with theory suggest that the band-split magnetic modes have different chiralities. One consequence of this is the expectation of the existence of a spin Seebeck effect~\cite{Kato04:306,Sinova15:87} in insulator-metal junctions formed by placing a conductor in contact with \ce{CoFe2O4} in the presence of a thermal gradient, causing a spin current.  Such effects are also expected for band splitting in altermagnets, where the product of parity and time-reversal symmetry is broken.~\cite{Liu24:133}  This has been discussed in other ferrimagnets such as \ce{Tb3Fe5O12} ~\cite{Mori25:111,Kawamoto24:124} and \ce{Y3Fe5O12}.~\cite{Barker16:117} In agreement with our results, thin film samples of \ce{CoFe2O4} have been shown to exhibit a longitudinal Spin Seebeck effect.~\cite{Guo16:108} The interplay of the Spin Seebeck effect and magnetoelastic properties of \ce{CoFe2O4} may make for interesting avenues in device design and future work.  This highlights possible further applications of mixed inverse spinels as a result of strong band splitting in the magnetic excitations.
	
	\section{Conclusion}
	\label{sect:conclusion}
	We have presented a comprehensive series of x-ray diffraction, neutron diffraction, and neutron inelastic scattering measurements on CoFe$_2$O$_4$. A kink in the temperature-dependence of the lattice parameter that coincides with the onset of long-range magnetic order at $T_c \sim 780$\,K confirms the presence of magnetoelastic coupling. X-ray powder diffraction measurements observe no splitting of Bragg peaks down to 1.6\,K and place a bound on the structural distortion of $1 - c/a \le 10^{-4}$, which is at least one order of magnitude smaller than the tetragonal distortions reported in single-crystal specimens.~\cite{Abes16:93}  This indicates that finite-size effects are significant and helps to explain why powder samples of CoFe$_2$O$_4$ exhibit weaker saturation magnetostriction values than do single crystals.  The magnetic response features two strongly dispersive magnon branches offset by  $\sim 60$\,meV. Evidence of an easy-axis anisotropy is manifested by an avoided crossing between the upper magnon branch and a flat mode with zero intensity along with a small gap of $\sim 3$\,meV in the lower magnon branch. Using linear spin-wave theory and applying the Kernal Polynomial Method (KPM-SWT), we have constructed a simplified model of CoFe$_2$O$_4$ that captures the essential physics and discussed the consequences of magnetic disorder on the spin Hamiltonian.  We show that the large magnon band splitting arises from an extremely strong ferrimagnetic molecular field that is generated by an imbalance between the anti-aligned spins on the spinel $A$ and $B$ sites. The combined effects of this field and the easy-axis, single-ion anisotropy lock the spin direction within each structural domain, thereby making the switching of structural domains in an applied magnetic field much more favorable energetically than a global spin reorientation.  This mechanism is the basis for the excellent magnetostrictive properties in CoFe$_2$O$_4$, and it illustrates a means by which the magnetostrictive character in this and other related compounds may be enhanced.
	
	\appendix
	\begin{widetext}
	\section{Rescaling of the spin coherent states}
	\label{sect:spin_coherent}
	The gap between the two modes depends on the molecular mean field felt on one sublattice by the other. The disordered occupancy of both lattices makes a full description of this difficult, however we make the approximation that this effect can be captured, on average, by rescaling the spin magnitudes on each site according to their total moment (and hence occupancy). Linear spin wave theory describes the dynamics of $SU(2)$ spin coherent states, which, in the adjoint representation can be described as a column vector of three real spin components, $S^{\alpha}$, for $\alpha=x,y,z$, normalized to $S$. To account for the average occupancy of each sublattice, we take $SU(N)$ coherent states normalized to unity, $\ket{\hat{\Omega}}$ and renormalize them to the average spin value of each sublattice
	\begin{equation}
		\ket{\Omega_{A/B}}=S_{A/B}\ket{\hat{\Omega}}.
	\end{equation}
	\noindent The classical limit of the Hamiltonian with respect to the scaled coherent states is then compared to that of the $SU(2)$ coherent state with $|\Omega_{A/B}|=S_{A/B}$, represented by $\ket{\Omega_{A/B}}$. The correspondence is trivial for terms linear in $\hat{S}^{\alpha}_j$, however additional corrections are required for onsite anisotropies. By the orthogonality theorem, terms of the same order will have the same proportionality factor. We therefore determine this by considering the $\mathcal{O}_{2}^0$ Stevens operator and compare the classical limit with respect to each of the coherent states.

	\begin{subequations}
		\begin{gather}
			S_{A/B}^2\bra{\hat{\Omega}}\mathcal{O}_2^0\ket{\hat{\Omega}}=S_{A/B}^2\bra{\hat{\Omega}}\left(3(\hat{S}^{z})^{2}-\mathbf{S}^2\right)\ket{\hat{\Omega}}=S_{A/B}^2\\
			\bra{\Omega_{A/B}}\mathcal{O}_2^0\ket{\Omega_{A/B}}=\bra{\Omega_{A/B}}\left(3(\hat{S}^{z})^{2}-\mathbf{S}^2\right)\ket{\Omega_{A/B}}=3S_{A/B}^2 - S_{A/B}(S_{A/B}+1)
		\end{gather}
	\end{subequations}

	\noindent To ensure that the quadratic anisotropies in the rescaled model are equivalent to those in the original, we introduce the constant of proportionality, $c$
	\begin{equation}
		\begin{split}
			cS_{A/B}^2 =& 3S_{A/B}^2 - S_{A/B}(S_{A/B}+1)\\
			c =& 2-\frac{1}{S_{A/B}}.
		\end{split}
	\end{equation}
	The anisotropies of the rescaled model are therefore related to those of the true model by, $D_{S=1} = cD$.
	
	\section{Scaling of the exchanges}
	\label{sect:spin_exchange}
	In the main text, a toy model of unit spin length was introduced to qualitatively explain the splitting between the optical and the acoustic mode. In this section we derive the rescaling factors required for this mapping. We begin with the Holstein-Primakoff transformation in the harmonic approximation
	\begin{align}
		\label{eq:HPtrans}
		\begin{gathered}
			\tilde{S}_{i,A}^{+} = \sqrt{2S_A}\hat{a}_i  \\
			\tilde{S}_{i,A}^{-} = \sqrt{2S_A}\hat{a}^{\dagger}_i\\
			\tilde{S}_{i,A}^{z} =S_A-\hat{a}^{\dagger}_i\hat{a}_i
		\end{gathered}
		&&
		\begin{gathered}
			\tilde{S}_{i,B}^{+} = \sqrt{2S_B}\hat{b}_i   \\
			\tilde{S}_{i,B}^{-} = \sqrt{2S_B}\hat{b}^{\dagger}_i\\
			\tilde{S}_{i,B}^{z} =S_B-\hat{b}^{\dagger}_i\hat{b}_i
		\end{gathered}
	\end{align}
	\noindent where $\hat{a}$ and $\hat{b}$ are the magnon operators on each sublattice. We first consider the inter-sublattice Hamiltonian
	\begin{equation}
		\hat{\mathcal{H}}_{AB}=\frac{1}{2}\sum_{i,j \in \{A,B\}} \mathcal{H}_{AB}(ij)=\frac{1}{2}J_{AB}\sum_{i,j \in \{A,B\}}\mathbf{\hat{S}}_i \cdot \mathbf{\hat{S}}_j. 
	\end{equation}
	
	The spins on each sublattice are antialigned with one another and so we rotate the spins on B sublattice relative to A

	\begin{equation}
		\mathcal{H}_{AB}(ij) = J_{AB}\left(\frac{1}{2}\left(\tilde{S}_{i,A}^{+}\tilde{S}_{j,B}^{+}+\tilde{S}_{i,A}^{-}\tilde{S}_{j,B}^{-}\right)-\tilde{S}_{i,A}^{z}\tilde{S}_{j,B}^{z}\right).
		\label{eq:ABbond}
	\end{equation}

	\noindent Inserting Eq.~\ref{eq:HPtrans} into Eq.~\ref{eq:ABbond} we find

	\begin{equation}
		\mathcal{H}_{AB}(ij)=J_{AB}\sqrt{S_AS_B}(\hat{a}_i\hat{b}_j+\hat{a}_i^\dagger\hat{b}_j^\dagger)+J_{AB}(S_A \hat{b}_j^\dagger \hat{b}_j+S_B \hat{a}_i^\dagger \hat{a}_i).
	\end{equation}

	\noindent Let us adopt the scaled exchange parameter $\tilde{J}_{AB}=J_{AB}\sqrt{S_AS_B}$
	\begin{equation}
		\mathcal{H}_{AB}(ij)=\tilde{J}_{AB}(\hat{a}_i\hat{b}_j+\hat{a}_i^\dagger\hat{b}_j^\dagger)+\tilde{J}_{AB}(\sqrt{\frac{S_A}{S_B}} \hat{b}_j^\dagger \hat{b}_j+\sqrt{\frac{S_B}{S_A}} \hat{a}_i^\dagger \hat{a}_i).
	\end{equation}
	\noindent The inter-sublattice exchange term is then

	\begin{equation}
		\hat{\mathcal{H}}_{AB}=\frac{1}{2}\sum_{i,j \in \{A,B\}} \mathcal{H}_{AB}(ij)=\frac{1}{2}\tilde{J}_{AB}\sum_{i,j \in \{A,B\}}(\hat{a}_i\hat{b}_j+\hat{a}_i^\dagger\hat{b}_j^\dagger)+\tilde{J}_{AB}\sum_{i}\left(z_{B}\sqrt{\frac{S_A}{S_B}} \hat{b}_j^\dagger \hat{b}_j+z_{A}\sqrt{\frac{S_B}{S_A}} \hat{a}_i^\dagger \hat{a}_i \right).
		\label{eq:AB-HP}
	\end{equation}

	\noindent If the upper mode is associated with the A sublattice and the lower associated with the B sublattice, then the difference in the gap at the zone center is 
	\begin{equation}
		\tilde{J}_{AB}z_A  \sqrt{\frac{S_B}{S_A}} - \tilde{J}_{AB}z_B \sqrt{\frac{S_A}{S_B}} = J_{AB}z_{A}S_B-J_{AB}z_A S_{A} = J_{AB}(12S_B - 6S_A).
	\end{equation}
	\noindent A similar procedure can be undertaken for the A-A and B-B bonds
	\begin{equation}
		\hat{\mathcal{H}}_{\mu\mu}=\frac{1}{2}\sum_{i,j \in \{\mu,\mu\}} \mathcal{H}_{\mu\mu}(ij)=\frac{1}{2}\tilde{J}_{\mu\mu}\sum_{i,j \in \{\mu,\mu\}}(\hat{\mu}^{\dagger}_i\hat{\mu}_j+\hat{\mu}_i\hat{\mu}_j^\dagger)-z_{\mu}\tilde{J}_{\mu\mu}\sum_{i} \hat{\mu}_i^\dagger \hat{\mu}_i 
	\end{equation}
	\noindent with $\tilde{J}_{\mu\mu} = S_{\mu}J_{\mu\mu}$. The construction demonstrated above, where the spin magnitudes are used to rescale the exchange parameters allows us to perform linear spin wave theory in terms of $S=1$ operators on both sublattices and account for the difference in spin magnitude in terms of an effective magnetic field applied to the $A$/$B$ site along the magnetization direction, chosen to be $+\hat{z}$ for $A$ and $+\hat{z}$ for $B$, which modifies the splitting between the dispersion curves. To do this we compare Eqn.~\ref{eq:AB-HP} with the expression found by performing a Holstein-Primakoff transformation on the $S=1$ model
	\begin{equation}
		\mathcal{H}_{AB}^{S=1}=\frac{1}{2}\tilde{J}_{AB}\sum_{i,j \in \{A,B\}}\mathbf{\hat{T}}_{i}\cdot \mathbf{\hat{T}}_{j} +\sum_{i\in \mu}h_{\mathrm{ferri}}^{\mu}\hat{T}_{i}^{z}
	\end{equation}
	\noindent namely 
	\begin{equation}
		\hat{\mathcal{H}}_{AB}=\frac{1}{2}\tilde{J}_{AB}\sum_{i,j \in \{A,B\}}(\hat{a}_i\hat{b}_j+\hat{a}_i^\dagger\hat{b}_j^\dagger)+\tilde{J}_{AB}\sum_{i}\left(z_{B} \hat{b}_i^\dagger \hat{b}_i+z_{A}\hat{a}_i^\dagger \hat{a}_i \right)-\sum_i\left( h^{A}_{\mathrm{ferri}} \hat{a}_i^\dagger \hat{a}_i-h^{B}_{\mathrm{ferri}} \hat{b}_i^\dagger \hat{b}_i\right),
	\end{equation}
	\noindent from which one finds
	\begin{gather}
		h_{\mathrm{ferri}}^{A} = \tilde{J}_{AB}z_A \left(1-\sqrt{\frac{S_{B}}{S_A}}\right) = \tilde{J}_{AB}z_A (1-\sqrt{1/\gamma})\\
		h_{\mathrm{ferri}}^{B} = \tilde{J}_{AB}z_B \left(1-\sqrt{\frac{S_{A}}{S_B}}\right)=\tilde{J}_{AB}z_B (1-\sqrt{\gamma}).
	\end{gather}
	
	\noindent The effective $S=1$ model is then
	\begin{equation}
		\mathcal{H}^{S=1}=\frac{1}{2}\tilde{J}_{AB}\sum_{i,j \in \{A,B\}}\mathbf{\hat{T}}_{i}\cdot \mathbf{\hat{T}}_{j} +\frac{1}{2}\tilde{J}_{AA}\sum_{i,j \in \{A\}}\mathbf{\hat{T}}_{i}\cdot \mathbf{\hat{T}}_{j} + \frac{1}{2}\tilde{J}_{BB}\sum_{i,j \in \{B\}}\mathbf{\hat{T}}_{i}\cdot \mathbf{\hat{T}}_{j}+\sum_{i\in \mu}h_{\mathrm{ferri}}^{\mu}\hat{T}_{i}^{z}.
	\end{equation}
	
	\section{Fitting of experimental data}
	\label{sect:fitting}
	One dimensional cuts were taken through both the MAPS and MERLIN data sets and fitted with Gaussian peaks to extract the dispersion relation of CoFe$_2$O$_4$. Three primary modes were identified, a lower mode extending up to $\sim 60$~meV, and two upper modes exhibiting a level crossing at around $\sim 70$~meV. Care was taken to avoid spurious signal which follows a detector trajectory at high energy. A $\chi^2$ loss function was used to fit the experimental data
	
	\begin{equation}
		\chi^2 =\sum_{i} \frac{(\epsilon_i-\hbar\omega_i)^2}{\sigma_i^2}  
	\end{equation}
	\noindent where $\epsilon_i$ are the calculated energy eigenvalues, $\hbar\omega_i$ are the measured dispersion peaks, $\sigma$ are the experimental uncertainties. The experimental error was approximated by the energy resolution of the instrument at the appropriate energy transfer. The optimization was performed using the Optim.jl package.~\cite{Mogensen18:3} At each iteration, the classical ground state was minimized before performing LSWT and calculating the loss function. First, 1000 iterations of the Nelder Mead algorithm was performed. The optimized values were then offset by a small random number before 200 iterations of particle swarm optimization were performed. Finally, 500 iterations of Nelder Mead were performed to find the loss function minimum. This process was repeated $81$ times. 71 of these interactions converged to the same minimum. Errorbars on the fitted parameters were estimated by taking the square root of the diagonal values of the inverse of the Hessian of the loss function. The Hessian was estimated by means of the finite difference method.
	
	\acknowledgments
	H.L acknowledges financial support from the Royal Commission for the Exhibition of 1851. We would like to thank R. Rae and C. Kirk for their help in using the X-ray Smartlab Rigaku facility, and A. Ali for helpful discussions. We thank Anatoly Balbashov for the provision of single crystals. Funding is acknowledged from the EPSRC, STFC, and Royal Society of Edinburgh (RSE). Experiments at the ISIS Neutron and Muon Source were supported by beamtime allocations RB2010607 and RB2220529 from the Science and Technology Facilities Council. The data are available at \url{https://doi.org/10.5286/ISIS.E.RB2010607-1} and \url{https://doi.org/10.5286/ISIS.E.RB2220529-1} respectively.
	
		\end{widetext}
	

\begin{thebibliography}{120}%
		\makeatletter
		\providecommand \@ifxundefined [1]{%
			\@ifx{#1\undefined}
		}%
		\providecommand \@ifnum [1]{%
			\ifnum #1\expandafter \@firstoftwo
			\else \expandafter \@secondoftwo
			\fi
		}%
		\providecommand \@ifx [1]{%
			\ifx #1\expandafter \@firstoftwo
			\else \expandafter \@secondoftwo
			\fi
		}%
		\providecommand \natexlab [1]{#1}%
		\providecommand \enquote  [1]{``#1''}%
		\providecommand \bibnamefont  [1]{#1}%
		\providecommand \bibfnamefont [1]{#1}%
		\providecommand \citenamefont [1]{#1}%
		\providecommand \href@noop [0]{\@secondoftwo}%
		\providecommand \href [0]{\begingroup \@sanitize@url \@href}%
		\providecommand \@href[1]{\@@startlink{#1}\@@href}%
		\providecommand \@@href[1]{\endgroup#1\@@endlink}%
		\providecommand \@sanitize@url [0]{\catcode `\\12\catcode `\$12\catcode
			`\&12\catcode `\#12\catcode `\^12\catcode `\_12\catcode `\%12\relax}%
		\providecommand \@@startlink[1]{}%
		\providecommand \@@endlink[0]{}%
		\providecommand \url  [0]{\begingroup\@sanitize@url \@url }%
		\providecommand \@url [1]{\endgroup\@href {#1}{\urlprefix }}%
		\providecommand \urlprefix  [0]{URL }%
		\providecommand \Eprint [0]{\href }%
		\providecommand \doibase [0]{https://doi.org/}%
		\providecommand \selectlanguage [0]{\@gobble}%
		\providecommand \bibinfo  [0]{\@secondoftwo}%
		\providecommand \bibfield  [0]{\@secondoftwo}%
		\providecommand \translation [1]{[#1]}%
		\providecommand \BibitemOpen [0]{}%
		\providecommand \bibitemStop [0]{}%
		\providecommand \bibitemNoStop [0]{.\EOS\space}%
		\providecommand \EOS [0]{\spacefactor3000\relax}%
		\providecommand \BibitemShut  [1]{\csname bibitem#1\endcsname}%
		\let\auto@bib@innerbib\@empty
		\bibitem [{\citenamefont {Saiz}\ \emph {et~al.}(2022)\citenamefont {Saiz},
			\citenamefont {Luis}, \citenamefont {Lasheras}, \citenamefont {Arriortua},\
			and\ \citenamefont {Lopes}}]{Saiz22:7}%
		\BibitemOpen
		\bibfield  {author} {\bibinfo {author} {\bibfnamefont {P.~G.}\ \bibnamefont
				{Saiz}}, \bibinfo {author} {\bibfnamefont {R.~F.}\ \bibnamefont {Luis}},
			\bibinfo {author} {\bibfnamefont {A.}~\bibnamefont {Lasheras}}, \bibinfo
			{author} {\bibfnamefont {M.~I.}\ \bibnamefont {Arriortua}},\ and\ \bibinfo
			{author} {\bibfnamefont {A.~C.}\ \bibnamefont {Lopes}},\ }\href
		{https://doi.org/10.1021/acssensors.2c00032} {\bibfield  {journal} {\bibinfo
				{journal} {ACS Sens.}\ }\textbf {\bibinfo {volume} {7}},\ \bibinfo {pages}
			{1248} (\bibinfo {year} {2022})}\BibitemShut {NoStop}%
		\bibitem [{\citenamefont {Zaeimbashi}\ \emph {et~al.}(2021)\citenamefont
			{Zaeimbashi}, \citenamefont {Nasrollahpour}, \citenamefont {Khalifa},
			\citenamefont {Romano}, \citenamefont {Liang}, \citenamefont {Chen},
			\citenamefont {Sun}, \citenamefont {Matyushov}, \citenamefont {Lin},
			\citenamefont {Dong}, \citenamefont {Xu}, \citenamefont {Mittal},
			\citenamefont {Martos-Repath}, \citenamefont {Jha}, \citenamefont
			{Mirchandani}, \citenamefont {Das}, \citenamefont {Onabajo}, \citenamefont
			{Shrivastava}, \citenamefont {Cash},\ and\ \citenamefont
			{Sun}}]{Zaeimbashi21:12}%
		\BibitemOpen
		\bibfield  {author} {\bibinfo {author} {\bibfnamefont {M.}~\bibnamefont
				{Zaeimbashi}}, \bibinfo {author} {\bibfnamefont {M.}~\bibnamefont
				{Nasrollahpour}}, \bibinfo {author} {\bibfnamefont {A.}~\bibnamefont
				{Khalifa}}, \bibinfo {author} {\bibfnamefont {A.}~\bibnamefont {Romano}},
			\bibinfo {author} {\bibfnamefont {X.}~\bibnamefont {Liang}}, \bibinfo
			{author} {\bibfnamefont {H.}~\bibnamefont {Chen}}, \bibinfo {author}
			{\bibfnamefont {N.}~\bibnamefont {Sun}}, \bibinfo {author} {\bibfnamefont
				{A.}~\bibnamefont {Matyushov}}, \bibinfo {author} {\bibfnamefont
				{H.}~\bibnamefont {Lin}}, \bibinfo {author} {\bibfnamefont {C.}~\bibnamefont
				{Dong}}, \bibinfo {author} {\bibfnamefont {Z.}~\bibnamefont {Xu}}, \bibinfo
			{author} {\bibfnamefont {A.}~\bibnamefont {Mittal}}, \bibinfo {author}
			{\bibfnamefont {I.}~\bibnamefont {Martos-Repath}}, \bibinfo {author}
			{\bibfnamefont {G.}~\bibnamefont {Jha}}, \bibinfo {author} {\bibfnamefont
				{N.}~\bibnamefont {Mirchandani}}, \bibinfo {author} {\bibfnamefont
				{D.}~\bibnamefont {Das}}, \bibinfo {author} {\bibfnamefont {M.}~\bibnamefont
				{Onabajo}}, \bibinfo {author} {\bibfnamefont {A.}~\bibnamefont
				{Shrivastava}}, \bibinfo {author} {\bibfnamefont {S.}~\bibnamefont {Cash}},\
			and\ \bibinfo {author} {\bibfnamefont {N.~X.}\ \bibnamefont {Sun}},\ }\href
		{https://doi.org/10.1038/s41467-021-23256-z} {\bibfield  {journal} {\bibinfo
				{journal} {Nat. Commun.}\ }\textbf {\bibinfo {volume} {12}},\ \bibinfo
			{pages} {3141} (\bibinfo {year} {2021})}\BibitemShut {NoStop}%
		\bibitem [{\citenamefont {\ifmmode~\check{S}\else \v{S}\fi{}mejkal}\ \emph
			{et~al.}(2022{\natexlab{a}})\citenamefont {\ifmmode~\check{S}\else
				\v{S}\fi{}mejkal}, \citenamefont {Sinova},\ and\ \citenamefont
			{Jungwirth}}]{Smejkal22:12}%
		\BibitemOpen
		\bibfield  {author} {\bibinfo {author} {\bibfnamefont {L.}~\bibnamefont
				{\ifmmode~\check{S}\else \v{S}\fi{}mejkal}}, \bibinfo {author} {\bibfnamefont
				{J.}~\bibnamefont {Sinova}},\ and\ \bibinfo {author} {\bibfnamefont
				{T.}~\bibnamefont {Jungwirth}},\ }\href
		{https://doi.org/10.1103/PhysRevX.12.040501} {\bibfield  {journal} {\bibinfo
				{journal} {Phys. Rev. X}\ }\textbf {\bibinfo {volume} {12}},\ \bibinfo
			{pages} {040501} (\bibinfo {year} {2022}{\natexlab{a}})}\BibitemShut
		{NoStop}%
		\bibitem [{\citenamefont {\ifmmode~\check{S}\else \v{S}\fi{}mejkal}\ \emph
			{et~al.}(2022{\natexlab{b}})\citenamefont {\ifmmode~\check{S}\else
				\v{S}\fi{}mejkal}, \citenamefont {Sinova},\ and\ \citenamefont
			{Jungwirth}}]{Smejkal22:12_2}%
		\BibitemOpen
		\bibfield  {author} {\bibinfo {author} {\bibfnamefont {L.}~\bibnamefont
				{\ifmmode~\check{S}\else \v{S}\fi{}mejkal}}, \bibinfo {author} {\bibfnamefont
				{J.}~\bibnamefont {Sinova}},\ and\ \bibinfo {author} {\bibfnamefont
				{T.}~\bibnamefont {Jungwirth}},\ }\href
		{https://doi.org/10.1103/PhysRevX.12.031042} {\bibfield  {journal} {\bibinfo
				{journal} {Phys. Rev. X}\ }\textbf {\bibinfo {volume} {12}},\ \bibinfo
			{pages} {031042} (\bibinfo {year} {2022}{\natexlab{b}})}\BibitemShut
		{NoStop}%
		\bibitem [{\citenamefont {Cheong}\ and\ \citenamefont
			{Huang}(2024)}]{Cheong24:9}%
		\BibitemOpen
		\bibfield  {author} {\bibinfo {author} {\bibfnamefont {S.-W.}\ \bibnamefont
				{Cheong}}\ and\ \bibinfo {author} {\bibfnamefont {F.-T.}\ \bibnamefont
				{Huang}},\ }\href {https://doi.org/10.1038/s41535-024-00626-6} {\bibfield
			{journal} {\bibinfo  {journal} {npj Quantum Mater.}\ }\textbf {\bibinfo
				{volume} {9}},\ \bibinfo {pages} {13} (\bibinfo {year} {2024})}\BibitemShut
		{NoStop}%
		\bibitem [{\citenamefont {Cheong}\ and\ \citenamefont
			{Huang}(2025)}]{Cheong25:10}%
		\BibitemOpen
		\bibfield  {author} {\bibinfo {author} {\bibfnamefont {S.~W.}\ \bibnamefont
				{Cheong}}\ and\ \bibinfo {author} {\bibfnamefont {F.~T.}\ \bibnamefont
				{Huang}},\ }\href {https://doi.org/10.1038/s41535-025-00756-5} {\bibfield
			{journal} {\bibinfo  {journal} {npj Quantum Mater.}\ }\textbf {\bibinfo
				{volume} {10}},\ \bibinfo {pages} {38} (\bibinfo {year} {2025})}\BibitemShut
		{NoStop}%
		\bibitem [{\citenamefont {Yuan}\ \emph {et~al.}(2020)\citenamefont {Yuan},
			\citenamefont {Wang}, \citenamefont {Luo}, \citenamefont {Rashba},\ and\
			\citenamefont {Zunger}}]{Yuan20:102}%
		\BibitemOpen
		\bibfield  {author} {\bibinfo {author} {\bibfnamefont {L.-D.}\ \bibnamefont
				{Yuan}}, \bibinfo {author} {\bibfnamefont {Z.}~\bibnamefont {Wang}}, \bibinfo
			{author} {\bibfnamefont {J.-W.}\ \bibnamefont {Luo}}, \bibinfo {author}
			{\bibfnamefont {E.~I.}\ \bibnamefont {Rashba}},\ and\ \bibinfo {author}
			{\bibfnamefont {A.}~\bibnamefont {Zunger}},\ }\bibfield  {title} {\bibinfo
			{title} {Giant momentum-dependent spin splitting in centrosymmetric low-$z$
				antiferromagnets},\ }\href {https://doi.org/10.1103/PhysRevB.102.014422}
		{\bibfield  {journal} {\bibinfo  {journal} {Phys. Rev. B}\ }\textbf {\bibinfo
				{volume} {102}},\ \bibinfo {pages} {014422} (\bibinfo {year}
			{2020})}\BibitemShut {NoStop}%
		\bibitem [{\citenamefont {Yuan}\ \emph {et~al.}(2021)\citenamefont {Yuan},
			\citenamefont {Wang}, \citenamefont {Luo},\ and\ \citenamefont
			{Zunger}}]{Yuan21:5}%
		\BibitemOpen
		\bibfield  {author} {\bibinfo {author} {\bibfnamefont {L.-D.}\ \bibnamefont
				{Yuan}}, \bibinfo {author} {\bibfnamefont {Z.}~\bibnamefont {Wang}}, \bibinfo
			{author} {\bibfnamefont {J.-W.}\ \bibnamefont {Luo}},\ and\ \bibinfo {author}
			{\bibfnamefont {A.}~\bibnamefont {Zunger}},\ }\bibfield  {title} {\bibinfo
			{title} {Prediction of low-z collinear and noncollinear antiferromagnetic
				compounds having momentum-dependent spin splitting even without spin-orbit
				coupling},\ }\href {https://doi.org/10.1103/PhysRevMaterials.5.014409}
		{\bibfield  {journal} {\bibinfo  {journal} {Phys. Rev. Mater.}\ }\textbf
			{\bibinfo {volume} {5}},\ \bibinfo {pages} {014409} (\bibinfo {year}
			{2021})}\BibitemShut {NoStop}%
		\bibitem [{\citenamefont {Yershov}\ \emph {et~al.}(2024)\citenamefont
			{Yershov}, \citenamefont {Kravchuk}, \citenamefont {Daghofer},\ and\
			\citenamefont {van~den Brink}}]{Yershov24:110}%
		\BibitemOpen
		\bibfield  {author} {\bibinfo {author} {\bibfnamefont {K.~V.}\ \bibnamefont
				{Yershov}}, \bibinfo {author} {\bibfnamefont {V.~P.}\ \bibnamefont
				{Kravchuk}}, \bibinfo {author} {\bibfnamefont {M.}~\bibnamefont {Daghofer}},\
			and\ \bibinfo {author} {\bibfnamefont {J.}~\bibnamefont {van~den Brink}},\
		}\bibfield  {title} {\bibinfo {title} {Fluctuation-induced piezomagnetism in
				local moment altermagnets},\ }\href
		{https://doi.org/10.1103/PhysRevB.110.144421} {\bibfield  {journal} {\bibinfo
				{journal} {Phys. Rev. B}\ }\textbf {\bibinfo {volume} {110}},\ \bibinfo
			{pages} {144421} (\bibinfo {year} {2024})}\BibitemShut {NoStop}%
		\bibitem [{\citenamefont {Aoyama}\ and\ \citenamefont
			{Ohgushi}(2024)}]{Aoyama24:8}%
		\BibitemOpen
		\bibfield  {author} {\bibinfo {author} {\bibfnamefont {T.}~\bibnamefont
				{Aoyama}}\ and\ \bibinfo {author} {\bibfnamefont {K.}~\bibnamefont
				{Ohgushi}},\ }\href {https://doi.org/10.1103/PhysRevMaterials.8.L041402}
		{\bibfield  {journal} {\bibinfo  {journal} {Phys. Rev. Mater.}\ }\textbf
			{\bibinfo {volume} {8}},\ \bibinfo {pages} {L041402} (\bibinfo {year}
			{2024})}\BibitemShut {NoStop}%
		\bibitem [{\citenamefont {Ma}\ \emph {et~al.}(2021)\citenamefont {Ma},
			\citenamefont {Hu}, \citenamefont {Liu}, \citenamefont {Yao}, \citenamefont
			{Jia},\ and\ \citenamefont {Liu}}]{Ma21:12}%
		\BibitemOpen
		\bibfield  {author} {\bibinfo {author} {\bibfnamefont {H.~Y.}\ \bibnamefont
				{Ma}}, \bibinfo {author} {\bibfnamefont {M.}~\bibnamefont {Hu}}, \bibinfo
			{author} {\bibfnamefont {J.}~\bibnamefont {Liu}}, \bibinfo {author}
			{\bibfnamefont {W.}~\bibnamefont {Yao}}, \bibinfo {author} {\bibfnamefont
				{J.~F.}\ \bibnamefont {Jia}},\ and\ \bibinfo {author} {\bibfnamefont
				{J.}~\bibnamefont {Liu}},\ }\href
		{https://doi.org/10.1038/s41467-021-23127-7} {\bibfield  {journal} {\bibinfo
				{journal} {Nat. Commun.}\ }\textbf {\bibinfo {volume} {12}},\ \bibinfo
			{pages} {2846} (\bibinfo {year} {2021})}\BibitemShut {NoStop}%
		\bibitem [{\citenamefont {Callen}\ and\ \citenamefont
			{Callen}(1965)}]{Callen65:139}%
		\BibitemOpen
		\bibfield  {author} {\bibinfo {author} {\bibfnamefont {E.}~\bibnamefont
				{Callen}}\ and\ \bibinfo {author} {\bibfnamefont {H.~B.}\ \bibnamefont
				{Callen}},\ }\href {https://doi.org/10.1103/PhysRev.139.A455} {\bibfield
			{journal} {\bibinfo  {journal} {Phys. Rev.}\ }\textbf {\bibinfo {volume}
				{139}},\ \bibinfo {pages} {A455} (\bibinfo {year} {1965})}\BibitemShut
		{NoStop}%
		\bibitem [{\citenamefont {Callen}\ and\ \citenamefont
			{Callen}(1963)}]{Callen63:129}%
		\BibitemOpen
		\bibfield  {author} {\bibinfo {author} {\bibfnamefont {E.~R.}\ \bibnamefont
				{Callen}}\ and\ \bibinfo {author} {\bibfnamefont {H.~B.}\ \bibnamefont
				{Callen}},\ }\href {https://doi.org/10.1103/PhysRev.129.578} {\bibfield
			{journal} {\bibinfo  {journal} {Phys. Rev.}\ }\textbf {\bibinfo {volume}
				{129}},\ \bibinfo {pages} {578} (\bibinfo {year} {1963})}\BibitemShut
		{NoStop}%
		\bibitem [{\citenamefont {Rajagopal}\ and\ \citenamefont
			{Sinha}(2021)}]{Rajagopal21:92}%
		\BibitemOpen
		\bibfield  {author} {\bibinfo {author} {\bibfnamefont {M.~C.}\ \bibnamefont
				{Rajagopal}}\ and\ \bibinfo {author} {\bibfnamefont {S.}~\bibnamefont
				{Sinha}},\ }\href {https://doi.org/10.1063/5.0035296} {\bibfield  {journal}
			{\bibinfo  {journal} {Rev. Sci. Instrum.}\ }\textbf {\bibinfo {volume}
				{92}},\ \bibinfo {pages} {014901} (\bibinfo {year} {2021})}\BibitemShut
		{NoStop}%
		\bibitem [{\citenamefont {Herzer}(2003)}]{Herzer03:254}%
		\BibitemOpen
		\bibfield  {author} {\bibinfo {author} {\bibfnamefont {G.}~\bibnamefont
				{Herzer}},\ }\href
		{https://doi.org/https://doi.org/10.1016/S0304-8853(02)00930-7} {\bibfield
			{journal} {\bibinfo  {journal} {J. Magn. Magn. Mater.}\ }\textbf {\bibinfo
				{volume} {254-255}},\ \bibinfo {pages} {598} (\bibinfo {year}
			{2003})}\BibitemShut {NoStop}%
		\bibitem [{\citenamefont {Li}\ \emph {et~al.}(2010)\citenamefont {Li},
			\citenamefont {Simonian},\ and\ \citenamefont {Chin}}]{Li10_2:19}%
		\BibitemOpen
		\bibfield  {author} {\bibinfo {author} {\bibfnamefont {S.}~\bibnamefont
				{Li}}, \bibinfo {author} {\bibfnamefont {A.}~\bibnamefont {Simonian}},\ and\
			\bibinfo {author} {\bibfnamefont {B.~A.}\ \bibnamefont {Chin}},\ }\bibfield
		{title} {\bibinfo {title} {Sensors for agriculture and the food industry},\
		}\href {https://doi.org/10.1149/2.F05104if} {\bibfield  {journal} {\bibinfo
				{journal} {The Electrochemical Society Interface}\ }\textbf {\bibinfo
				{volume} {19}},\ \bibinfo {pages} {41} (\bibinfo {year} {2010})}\BibitemShut
		{NoStop}%
		\bibitem [{\citenamefont {Langlois}\ \emph {et~al.}(2016)\citenamefont
			{Langlois}, \citenamefont {Pillars}, \citenamefont {Monson}, \citenamefont
			{Arrington}, \citenamefont {Finnegan}, \citenamefont {St.~John},\ and\
			\citenamefont {Smartt}}]{Langlois16:tech}%
		\BibitemOpen
		\bibfield  {author} {\bibinfo {author} {\bibfnamefont {E.~D.}\ \bibnamefont
				{Langlois}}, \bibinfo {author} {\bibfnamefont {J.~R.}\ \bibnamefont
				{Pillars}}, \bibinfo {author} {\bibfnamefont {T.~C.}\ \bibnamefont {Monson}},
			\bibinfo {author} {\bibfnamefont {C.~L.}\ \bibnamefont {Arrington}}, \bibinfo
			{author} {\bibfnamefont {P.~S.}\ \bibnamefont {Finnegan}}, \bibinfo {author}
			{\bibfnamefont {C.~R.}\ \bibnamefont {St.~John}},\ and\ \bibinfo {author}
			{\bibfnamefont {H.~A.}\ \bibnamefont {Smartt}},\ }\href
		{https://doi.org/10.2172/1562661} {\emph {\bibinfo {title} {Magnetic Smart
					Tags (MaST) for Arms Control and Treaty Verification}}},\ \bibinfo {type}
		{Tech. Rep.}\ (\bibinfo  {institution} {Sandia National Lab. (SNL-NM),
			Albuquerque, NM (United States); Sandia National Lab. (SNL-CA), Livermore, CA
			(United States)},\ \bibinfo {year} {2016})\BibitemShut {NoStop}%
		\bibitem [{\citenamefont {Fletcher}(2002)}]{Fletcher02:thesis}%
		\BibitemOpen
		\bibfield  {author} {\bibinfo {author} {\bibfnamefont {R.}~\bibnamefont
				{Fletcher}},\ }\emph {\bibinfo {title} {An Analysis of Example}},\ \href@noop
		{} {\bibinfo {type} {Phd thesis}},\ \bibinfo  {school} {Massachusetts
			Institute of Technology}, \bibinfo {address} {Cambridge, MA} (\bibinfo {year}
		{2002}),\ \bibinfo {note} {available at
			\url{http://dspace.mit.edu/handle/1721.1/7582}}\BibitemShut {NoStop}%
		\bibitem [{\citenamefont {Olabi}\ and\ \citenamefont
			{Grunwald}(2008)}]{Olabi06:29}%
		\BibitemOpen
		\bibfield  {author} {\bibinfo {author} {\bibfnamefont {A.}~\bibnamefont
				{Olabi}}\ and\ \bibinfo {author} {\bibfnamefont {A.}~\bibnamefont
				{Grunwald}},\ }\href
		{https://doi.org/https://doi.org/10.1016/j.matdes.2006.12.016} {\bibfield
			{journal} {\bibinfo  {journal} {Materials \& Design}\ }\textbf {\bibinfo
				{volume} {29}},\ \bibinfo {pages} {469} (\bibinfo {year} {2008})}\BibitemShut
		{NoStop}%
		\bibitem [{\citenamefont {Dhilsha}\ \emph {et~al.}(2005)\citenamefont
			{Dhilsha}, \citenamefont {Rajeshwari},\ and\ \citenamefont
			{Rajendran}}]{Dhilsha05:55}%
		\BibitemOpen
		\bibfield  {author} {\bibinfo {author} {\bibfnamefont {R.}~\bibnamefont
				{Dhilsha}}, \bibinfo {author} {\bibfnamefont {P.~M.}\ \bibnamefont
				{Rajeshwari}},\ and\ \bibinfo {author} {\bibfnamefont {V.}~\bibnamefont
				{Rajendran}},\ }\href@noop {} {\bibfield  {journal} {\bibinfo  {journal}
				{Defence science journal.}\ }\textbf {\bibinfo {volume} {55}} (\bibinfo
			{year} {2005})}\BibitemShut {NoStop}%
		\bibitem [{\citenamefont {Ekreem}\ \emph {et~al.}(2007)\citenamefont {Ekreem},
			\citenamefont {Olabi}, \citenamefont {Prescott}, \citenamefont {Rafferty},\
			and\ \citenamefont {Hashmi}}]{Ekreem07:191}%
		\BibitemOpen
		\bibfield  {author} {\bibinfo {author} {\bibfnamefont {N.}~\bibnamefont
				{Ekreem}}, \bibinfo {author} {\bibfnamefont {A.}~\bibnamefont {Olabi}},
			\bibinfo {author} {\bibfnamefont {T.}~\bibnamefont {Prescott}}, \bibinfo
			{author} {\bibfnamefont {A.}~\bibnamefont {Rafferty}},\ and\ \bibinfo
			{author} {\bibfnamefont {M.}~\bibnamefont {Hashmi}},\ }\href
		{https://doi.org/https://doi.org/10.1016/j.jmatprotec.2007.03.064} {\bibfield
			{journal} {\bibinfo  {journal} {J. Mater. Process. Technol.}\ }\textbf
			{\bibinfo {volume} {191}},\ \bibinfo {pages} {96} (\bibinfo {year} {2007})},\
		\bibinfo {note} {advances in Materials and Processing Technologies, July 30th
			- August 3rd 2006, Las Vegas, Nevada}\BibitemShut {NoStop}%
		\bibitem [{\citenamefont {Hathaway}\ and\ \citenamefont
			{Clark}(1993)}]{Hathaway93:18}%
		\BibitemOpen
		\bibfield  {author} {\bibinfo {author} {\bibfnamefont {K.~B.}\ \bibnamefont
				{Hathaway}}\ and\ \bibinfo {author} {\bibfnamefont {A.~E.}\ \bibnamefont
				{Clark}},\ }\href {https://doi.org/https://doi.org/10.1557/S0883769400037337}
		{\bibfield  {journal} {\bibinfo  {journal} {MRS Bulletin}\ }\textbf {\bibinfo
				{volume} {18}},\ \bibinfo {pages} {34} (\bibinfo {year} {1993})}\BibitemShut
		{NoStop}%
		\bibitem [{\citenamefont {McCorkle}(1923)}]{McCorkle23:22}%
		\BibitemOpen
		\bibfield  {author} {\bibinfo {author} {\bibfnamefont {P.}~\bibnamefont
				{McCorkle}},\ }\href {https://doi.org/10.1103/PhysRev.22.271} {\bibfield
			{journal} {\bibinfo  {journal} {Phys. Rev.}\ }\textbf {\bibinfo {volume}
				{22}},\ \bibinfo {pages} {271} (\bibinfo {year} {1923})}\BibitemShut
		{NoStop}%
		\bibitem [{\citenamefont {Lee}\ and\ \citenamefont {Asgar}(1971)}]{Lee71:326}%
		\BibitemOpen
		\bibfield  {author} {\bibinfo {author} {\bibfnamefont {E.~W.}\ \bibnamefont
				{Lee}}\ and\ \bibinfo {author} {\bibfnamefont {M.~A.}\ \bibnamefont
				{Asgar}},\ }\bibfield  {title} {\bibinfo {title} {Magnetostriction of
				nickel},\ }\href {https://doi.org/10.1098/rspa.1971.0192} {\bibfield
			{journal} {\bibinfo  {journal} {Proc. R. Soc. Lond. A}\ }\textbf {\bibinfo
				{volume} {326}},\ \bibinfo {pages} {73} (\bibinfo {year} {1971})}\BibitemShut
		{NoStop}%
		\bibitem [{\citenamefont {du~Tremolet de Lacheisserie~nd
				R.~M.~Monterroso}(1983)}]{Lacheisserie83:31}%
		\BibitemOpen
		\bibfield  {author} {\bibinfo {author} {\bibfnamefont {E.}~\bibnamefont
				{du~Tremolet de Lacheisserie~nd R.~M.~Monterroso}},\ }\bibfield  {title}
		{\bibinfo {title} {Magnetostriction of iron},\ }\href
		{https://doi.org/10.1016/0304-8853(83)90704-7} {\bibfield  {journal}
			{\bibinfo  {journal} {J. Magn. Magn. Mater.}\ }\textbf {\bibinfo {volume}
				{31-34}},\ \bibinfo {pages} {837} (\bibinfo {year} {1983})}\BibitemShut
		{NoStop}%
		\bibitem [{\citenamefont {Clark}\ \emph {et~al.}(1976)\citenamefont {Clark},
			\citenamefont {Cullen}, \citenamefont {McMasters},\ and\ \citenamefont
			{Callen}}]{Clark76:29}%
		\BibitemOpen
		\bibfield  {author} {\bibinfo {author} {\bibfnamefont {A.~E.}\ \bibnamefont
				{Clark}}, \bibinfo {author} {\bibfnamefont {J.~R.}\ \bibnamefont {Cullen}},
			\bibinfo {author} {\bibfnamefont {O.~D.}\ \bibnamefont {McMasters}},\ and\
			\bibinfo {author} {\bibfnamefont {E.~R.}\ \bibnamefont {Callen}},\ }\bibfield
		{title} {\bibinfo {title} {Rhombohedral distortion in highly
				magnetostrictive laves phase compounds},\ }\href
		{https://doi.org/10.1063/1.30580} {\bibfield  {journal} {\bibinfo  {journal}
				{AIP Conference Proceedings}\ }\textbf {\bibinfo {volume} {29}},\ \bibinfo
			{pages} {192} (\bibinfo {year} {1976})},\ \Eprint
		{https://arxiv.org/abs/https://pubs.aip.org/aip/acp/article-pdf/29/1/192/12097629/192\_1\_online.pdf}
		{https://pubs.aip.org/aip/acp/article-pdf/29/1/192/12097629/192\_1\_online.pdf}
		\BibitemShut {NoStop}%
		\bibitem [{\citenamefont {Jiles}(1994)}]{Jiles94:27}%
		\BibitemOpen
		\bibfield  {author} {\bibinfo {author} {\bibfnamefont {D.~C.}\ \bibnamefont
				{Jiles}},\ }\href {https://doi.org/10.1088/0022-3727/27/1/001} {\bibfield
			{journal} {\bibinfo  {journal} {J. Phys. D: Appl. Phys.}\ }\textbf {\bibinfo
				{volume} {27}},\ \bibinfo {pages} {1} (\bibinfo {year} {1994})}\BibitemShut
		{NoStop}%
		\bibitem [{\citenamefont {Ishikawa}\ and\ \citenamefont
			{Syono}(1971)}]{Ishikawa71:26}%
		\BibitemOpen
		\bibfield  {author} {\bibinfo {author} {\bibfnamefont {Y.}~\bibnamefont
				{Ishikawa}}\ and\ \bibinfo {author} {\bibfnamefont {Y.}~\bibnamefont
				{Syono}},\ }\href {https://doi.org/10.1103/PhysRevLett.26.1335} {\bibfield
			{journal} {\bibinfo  {journal} {Phys. Rev. Lett.}\ }\textbf {\bibinfo
				{volume} {26}},\ \bibinfo {pages} {1335} (\bibinfo {year}
			{1971})}\BibitemShut {NoStop}%
		\bibitem [{\citenamefont {Kittel}(1949)}]{Kittel49:21}%
		\BibitemOpen
		\bibfield  {author} {\bibinfo {author} {\bibfnamefont {C.}~\bibnamefont
				{Kittel}},\ }\href {https://doi.org/10.1103/RevModPhys.21.541} {\bibfield
			{journal} {\bibinfo  {journal} {Rev. Mod. Phys.}\ }\textbf {\bibinfo {volume}
				{21}},\ \bibinfo {pages} {541} (\bibinfo {year} {1949})}\BibitemShut
		{NoStop}%
		\bibitem [{\citenamefont {Landau}\ and\ \citenamefont
			{Lifshitz}(1977)}]{Landau:book}%
		\BibitemOpen
		\bibfield  {author} {\bibinfo {author} {\bibfnamefont {L.~D.}\ \bibnamefont
				{Landau}}\ and\ \bibinfo {author} {\bibfnamefont {E.~M.}\ \bibnamefont
				{Lifshitz}},\ }\href@noop {} {\emph {\bibinfo {title} {Quantum Mechanics
					(Non-relativistic Theory)}}}\ (\bibinfo  {publisher} {Elsevier Science},\
		\bibinfo {address} {Oxford},\ \bibinfo {year} {1977})\BibitemShut {NoStop}%
		\bibitem [{\citenamefont {Turtelli}\ \emph {et~al.}(2014)\citenamefont
			{Turtelli}, \citenamefont {Kriegisch}, \citenamefont {Atif},\ and\
			\citenamefont {Grössinger}}]{Turtelli14:60}%
		\BibitemOpen
		\bibfield  {author} {\bibinfo {author} {\bibfnamefont {R.~S.}\ \bibnamefont
				{Turtelli}}, \bibinfo {author} {\bibfnamefont {M.}~\bibnamefont {Kriegisch}},
			\bibinfo {author} {\bibfnamefont {M.}~\bibnamefont {Atif}},\ and\ \bibinfo
			{author} {\bibfnamefont {R.}~\bibnamefont {Grössinger}},\ }\href
		{https://doi.org/10.1088/1757-899X/60/1/012020} {\bibfield  {journal}
			{\bibinfo  {journal} {IOP Conf. Ser.: Mater. Sci. Eng}\ }\textbf {\bibinfo
				{volume} {60}},\ \bibinfo {pages} {012020} (\bibinfo {year}
			{2014})}\BibitemShut {NoStop}%
		\bibitem [{\citenamefont {McCallum}\ \emph {et~al.}(2001)\citenamefont
			{McCallum}, \citenamefont {Dennis}, \citenamefont {Jiles}, \citenamefont
			{Snyder},\ and\ \citenamefont {Chen}}]{McCallum01:27}%
		\BibitemOpen
		\bibfield  {author} {\bibinfo {author} {\bibfnamefont {R.~W.}\ \bibnamefont
				{McCallum}}, \bibinfo {author} {\bibfnamefont {K.~W.}\ \bibnamefont
				{Dennis}}, \bibinfo {author} {\bibfnamefont {D.~C.}\ \bibnamefont {Jiles}},
			\bibinfo {author} {\bibfnamefont {J.~E.}\ \bibnamefont {Snyder}},\ and\
			\bibinfo {author} {\bibfnamefont {Y.~H.}\ \bibnamefont {Chen}},\ }\href
		{https://doi.org/10.1063/1.1365598} {\bibfield  {journal} {\bibinfo
				{journal} {Low Temp. Phys.}\ }\textbf {\bibinfo {volume} {27}},\ \bibinfo
			{pages} {266} (\bibinfo {year} {2001})}\BibitemShut {NoStop}%
		\bibitem [{\citenamefont {Bozorth}\ \emph {et~al.}(1955)\citenamefont
			{Bozorth}, \citenamefont {Tilden},\ and\ \citenamefont
			{Williams}}]{Bozorth55:99}%
		\BibitemOpen
		\bibfield  {author} {\bibinfo {author} {\bibfnamefont {R.~M.}\ \bibnamefont
				{Bozorth}}, \bibinfo {author} {\bibfnamefont {E.~F.}\ \bibnamefont
				{Tilden}},\ and\ \bibinfo {author} {\bibfnamefont {A.~J.}\ \bibnamefont
				{Williams}},\ }\href {https://doi.org/10.1103/PhysRev.99.1788} {\bibfield
			{journal} {\bibinfo  {journal} {Phys. Rev.}\ }\textbf {\bibinfo {volume}
				{99}},\ \bibinfo {pages} {1788} (\bibinfo {year} {1955})}\BibitemShut
		{NoStop}%
		\bibitem [{\citenamefont {Abes}\ \emph {et~al.}(2016)\citenamefont {Abes},
			\citenamefont {Koops}, \citenamefont {Hrkac}, \citenamefont {McCord},
			\citenamefont {Urs}, \citenamefont {Wolff}, \citenamefont {Kienle},
			\citenamefont {Ren}, \citenamefont {Bouchenoire}, \citenamefont {Murphy},\
			and\ \citenamefont {Magnussen}}]{Abes16:93}%
		\BibitemOpen
		\bibfield  {author} {\bibinfo {author} {\bibfnamefont {M.}~\bibnamefont
				{Abes}}, \bibinfo {author} {\bibfnamefont {C.~T.}\ \bibnamefont {Koops}},
			\bibinfo {author} {\bibfnamefont {S.~B.}\ \bibnamefont {Hrkac}}, \bibinfo
			{author} {\bibfnamefont {J.}~\bibnamefont {McCord}}, \bibinfo {author}
			{\bibfnamefont {N.~O.}\ \bibnamefont {Urs}}, \bibinfo {author} {\bibfnamefont
				{N.}~\bibnamefont {Wolff}}, \bibinfo {author} {\bibfnamefont
				{L.}~\bibnamefont {Kienle}}, \bibinfo {author} {\bibfnamefont {W.~J.}\
				\bibnamefont {Ren}}, \bibinfo {author} {\bibfnamefont {L.}~\bibnamefont
				{Bouchenoire}}, \bibinfo {author} {\bibfnamefont {B.~M.}\ \bibnamefont
				{Murphy}},\ and\ \bibinfo {author} {\bibfnamefont {O.~M.}\ \bibnamefont
				{Magnussen}},\ }\href {https://doi.org/10.1103/PhysRevB.93.195427} {\bibfield
			{journal} {\bibinfo  {journal} {Phys. Rev. B}\ }\textbf {\bibinfo {volume}
				{93}},\ \bibinfo {pages} {195427} (\bibinfo {year} {2016})}\BibitemShut
		{NoStop}%
		\bibitem [{\citenamefont {Yang}\ and\ \citenamefont {Ren}(2008)}]{Yang08:77}%
		\BibitemOpen
		\bibfield  {author} {\bibinfo {author} {\bibfnamefont {S.}~\bibnamefont
				{Yang}}\ and\ \bibinfo {author} {\bibfnamefont {X.}~\bibnamefont {Ren}},\
		}\href {https://doi.org/10.1103/PhysRevB.77.014407} {\bibfield  {journal}
			{\bibinfo  {journal} {Phys. Rev. B}\ }\textbf {\bibinfo {volume} {77}},\
			\bibinfo {pages} {014407} (\bibinfo {year} {2008})}\BibitemShut {NoStop}%
		\bibitem [{\citenamefont {Song}\ and\ \citenamefont
			{Zhang}(2004)}]{Song04:126}%
		\BibitemOpen
		\bibfield  {author} {\bibinfo {author} {\bibfnamefont {Q.}~\bibnamefont
				{Song}}\ and\ \bibinfo {author} {\bibfnamefont {Z.~J.}\ \bibnamefont
				{Zhang}},\ }\href {https://doi.org/10.1021/ja049931r} {\bibfield  {journal}
			{\bibinfo  {journal} {J. Am. Chem. Soc.}\ }\textbf {\bibinfo {volume}
				{126}},\ \bibinfo {pages} {6164} (\bibinfo {year} {2004})}\BibitemShut
		{NoStop}%
		\bibitem [{\citenamefont {Bewley}\ \emph {et~al.}(2006)\citenamefont {Bewley},
			\citenamefont {Eccleston}, \citenamefont {McEwen}, \citenamefont {Hayden},
			\citenamefont {Dove}, \citenamefont {Bennington}, \citenamefont {Treadgold},\
			and\ \citenamefont {Coleman}}]{Bewley06:385}%
		\BibitemOpen
		\bibfield  {author} {\bibinfo {author} {\bibfnamefont {R.}~\bibnamefont
				{Bewley}}, \bibinfo {author} {\bibfnamefont {R.}~\bibnamefont {Eccleston}},
			\bibinfo {author} {\bibfnamefont {K.}~\bibnamefont {McEwen}}, \bibinfo
			{author} {\bibfnamefont {S.}~\bibnamefont {Hayden}}, \bibinfo {author}
			{\bibfnamefont {M.}~\bibnamefont {Dove}}, \bibinfo {author} {\bibfnamefont
				{S.}~\bibnamefont {Bennington}}, \bibinfo {author} {\bibfnamefont
				{J.}~\bibnamefont {Treadgold}},\ and\ \bibinfo {author} {\bibfnamefont
				{R.}~\bibnamefont {Coleman}},\ }\href
		{https://doi.org/https://doi.org/10.1016/j.physb.2006.05.328} {\bibfield
			{journal} {\bibinfo  {journal} {Physica B: Condensed Matter}\ }\textbf
			{\bibinfo {volume} {385-386}},\ \bibinfo {pages} {1029} (\bibinfo {year}
			{2006})}\BibitemShut {NoStop}%
		\bibitem [{\citenamefont {Ewings}\ \emph {et~al.}(2019)\citenamefont {Ewings},
			\citenamefont {Stewart}, \citenamefont {Perring}, \citenamefont {Bewley},
			\citenamefont {Le}, \citenamefont {Raspino}, \citenamefont {Pooley},
			\citenamefont {Škoro}, \citenamefont {Waller}, \citenamefont {Zacek},
			\citenamefont {Smith},\ and\ \citenamefont {Riehl-Shaw}}]{Ewings19:90}%
		\BibitemOpen
		\bibfield  {author} {\bibinfo {author} {\bibfnamefont {R.~A.}\ \bibnamefont
				{Ewings}}, \bibinfo {author} {\bibfnamefont {J.~R.}\ \bibnamefont {Stewart}},
			\bibinfo {author} {\bibfnamefont {T.~G.}\ \bibnamefont {Perring}}, \bibinfo
			{author} {\bibfnamefont {R.~I.}\ \bibnamefont {Bewley}}, \bibinfo {author}
			{\bibfnamefont {M.~D.}\ \bibnamefont {Le}}, \bibinfo {author} {\bibfnamefont
				{D.}~\bibnamefont {Raspino}}, \bibinfo {author} {\bibfnamefont {D.~E.}\
				\bibnamefont {Pooley}}, \bibinfo {author} {\bibfnamefont {G.}~\bibnamefont
				{Škoro}}, \bibinfo {author} {\bibfnamefont {S.~P.}\ \bibnamefont {Waller}},
			\bibinfo {author} {\bibfnamefont {D.}~\bibnamefont {Zacek}}, \bibinfo
			{author} {\bibfnamefont {C.~A.}\ \bibnamefont {Smith}},\ and\ \bibinfo
			{author} {\bibfnamefont {R.~C.}\ \bibnamefont {Riehl-Shaw}},\ }\href
		{https://doi.org/10.1063/1.5086255} {\bibfield  {journal} {\bibinfo
				{journal} {Rev. Sci. Instrum.}\ }\textbf {\bibinfo {volume} {90}},\ \bibinfo
			{pages} {035110} (\bibinfo {year} {2019})}\BibitemShut {NoStop}%
		\bibitem [{\citenamefont {Arnold}\ \emph {et~al.}(2014)\citenamefont {Arnold},
			\citenamefont {Bilheux}, \citenamefont {Borreguero}, \citenamefont {Buts},
			\citenamefont {Campbell}, \citenamefont {Chapon}, \citenamefont {Doucet},
			\citenamefont {Draper}, \citenamefont {{Ferraz Leal}}, \citenamefont {Gigg},
			\citenamefont {Lynch}, \citenamefont {Markvardsen}, \citenamefont
			{Mikkelson}, \citenamefont {Mikkelson}, \citenamefont {Miller}, \citenamefont
			{Palmen}, \citenamefont {Parker}, \citenamefont {Passos}, \citenamefont
			{Perring}, \citenamefont {Peterson}, \citenamefont {Ren}, \citenamefont
			{Reuter}, \citenamefont {Savici}, \citenamefont {Taylor}, \citenamefont
			{Taylor}, \citenamefont {Tolchenov}, \citenamefont {Zhou},\ and\
			\citenamefont {Zikovsky}}]{Arnold14:764}%
		\BibitemOpen
		\bibfield  {author} {\bibinfo {author} {\bibfnamefont {O.}~\bibnamefont
				{Arnold}}, \bibinfo {author} {\bibfnamefont {J.}~\bibnamefont {Bilheux}},
			\bibinfo {author} {\bibfnamefont {J.}~\bibnamefont {Borreguero}}, \bibinfo
			{author} {\bibfnamefont {A.}~\bibnamefont {Buts}}, \bibinfo {author}
			{\bibfnamefont {S.}~\bibnamefont {Campbell}}, \bibinfo {author}
			{\bibfnamefont {L.}~\bibnamefont {Chapon}}, \bibinfo {author} {\bibfnamefont
				{M.}~\bibnamefont {Doucet}}, \bibinfo {author} {\bibfnamefont
				{N.}~\bibnamefont {Draper}}, \bibinfo {author} {\bibfnamefont
				{R.}~\bibnamefont {{Ferraz Leal}}}, \bibinfo {author} {\bibfnamefont
				{M.}~\bibnamefont {Gigg}}, \bibinfo {author} {\bibfnamefont {V.}~\bibnamefont
				{Lynch}}, \bibinfo {author} {\bibfnamefont {A.}~\bibnamefont {Markvardsen}},
			\bibinfo {author} {\bibfnamefont {D.}~\bibnamefont {Mikkelson}}, \bibinfo
			{author} {\bibfnamefont {R.}~\bibnamefont {Mikkelson}}, \bibinfo {author}
			{\bibfnamefont {R.}~\bibnamefont {Miller}}, \bibinfo {author} {\bibfnamefont
				{K.}~\bibnamefont {Palmen}}, \bibinfo {author} {\bibfnamefont
				{P.}~\bibnamefont {Parker}}, \bibinfo {author} {\bibfnamefont
				{G.}~\bibnamefont {Passos}}, \bibinfo {author} {\bibfnamefont
				{T.}~\bibnamefont {Perring}}, \bibinfo {author} {\bibfnamefont
				{P.}~\bibnamefont {Peterson}}, \bibinfo {author} {\bibfnamefont
				{S.}~\bibnamefont {Ren}}, \bibinfo {author} {\bibfnamefont {M.}~\bibnamefont
				{Reuter}}, \bibinfo {author} {\bibfnamefont {A.}~\bibnamefont {Savici}},
			\bibinfo {author} {\bibfnamefont {J.}~\bibnamefont {Taylor}}, \bibinfo
			{author} {\bibfnamefont {R.}~\bibnamefont {Taylor}}, \bibinfo {author}
			{\bibfnamefont {R.}~\bibnamefont {Tolchenov}}, \bibinfo {author}
			{\bibfnamefont {W.}~\bibnamefont {Zhou}},\ and\ \bibinfo {author}
			{\bibfnamefont {J.}~\bibnamefont {Zikovsky}},\ }\href
		{https://doi.org/https://doi.org/10.1016/j.nima.2014.07.029} {\bibfield
			{journal} {\bibinfo  {journal} {Nucl. Instrum. Methods Phys. Res. A}\
			}\textbf {\bibinfo {volume} {764}},\ \bibinfo {pages} {156} (\bibinfo {year}
			{2014})}\BibitemShut {NoStop}%
		\bibitem [{\citenamefont {Ewings}\ \emph {et~al.}(2016)\citenamefont {Ewings},
			\citenamefont {Buts}, \citenamefont {Le}, \citenamefont {{van Duijn}},
			\citenamefont {Bustinduy},\ and\ \citenamefont {Perring}}]{Ewings16:834}%
		\BibitemOpen
		\bibfield  {author} {\bibinfo {author} {\bibfnamefont {R.}~\bibnamefont
				{Ewings}}, \bibinfo {author} {\bibfnamefont {A.}~\bibnamefont {Buts}},
			\bibinfo {author} {\bibfnamefont {M.}~\bibnamefont {Le}}, \bibinfo {author}
			{\bibfnamefont {J.}~\bibnamefont {{van Duijn}}}, \bibinfo {author}
			{\bibfnamefont {I.}~\bibnamefont {Bustinduy}},\ and\ \bibinfo {author}
			{\bibfnamefont {T.}~\bibnamefont {Perring}},\ }\bibfield  {title} {\bibinfo
			{title} {Horace: Software for the analysis of data from single crystal
				spectroscopy experiments at time-of-flight neutron instruments},\ }\href
		{https://doi.org/https://doi.org/10.1016/j.nima.2016.07.036} {\bibfield
			{journal} {\bibinfo  {journal} {Nucl. Instrum. Methods Phys. Res. A}\
			}\textbf {\bibinfo {volume} {834}},\ \bibinfo {pages} {132} (\bibinfo {year}
			{2016})}\BibitemShut {NoStop}%
		\bibitem [{\citenamefont {Rodr{\'\i}guez-Carvajal}\ \emph
			{et~al.}(2001)\citenamefont {Rodr{\'\i}guez-Carvajal} \emph
			{et~al.}}]{rodriguez2001}%
		\BibitemOpen
		\bibfield  {author} {\bibinfo {author} {\bibfnamefont {J.}~\bibnamefont
				{Rodr{\'\i}guez-Carvajal}} \emph {et~al.},\ }\href@noop {} {\bibfield
			{journal} {\bibinfo  {journal} {Laboratoire L{\'e}on Brillouin (CEA-CNRS):
					Saclay, France}\ } (\bibinfo {year} {2001})}\BibitemShut {NoStop}%
		\bibitem [{\citenamefont {Moyer}\ \emph {et~al.}(2011)\citenamefont {Moyer},
			\citenamefont {Vaz}, \citenamefont {Negusse}, \citenamefont {Arena},\ and\
			\citenamefont {Henrich}}]{Moyer11:83}%
		\BibitemOpen
		\bibfield  {author} {\bibinfo {author} {\bibfnamefont {J.~A.}\ \bibnamefont
				{Moyer}}, \bibinfo {author} {\bibfnamefont {C.~A.~F.}\ \bibnamefont {Vaz}},
			\bibinfo {author} {\bibfnamefont {E.}~\bibnamefont {Negusse}}, \bibinfo
			{author} {\bibfnamefont {D.~A.}\ \bibnamefont {Arena}},\ and\ \bibinfo
			{author} {\bibfnamefont {V.~E.}\ \bibnamefont {Henrich}},\ }\href
		{https://doi.org/10.1103/PhysRevB.83.035121} {\bibfield  {journal} {\bibinfo
				{journal} {Phys. Rev. B}\ }\textbf {\bibinfo {volume} {83}},\ \bibinfo
			{pages} {035121} (\bibinfo {year} {2011})}\BibitemShut {NoStop}%
		\bibitem [{\citenamefont {Sharma}\ \emph {et~al.}(2022)\citenamefont {Sharma},
			\citenamefont {Calmels}, \citenamefont {Li}, \citenamefont {Barbier},\ and\
			\citenamefont {Arras}}]{Sharma22:6}%
		\BibitemOpen
		\bibfield  {author} {\bibinfo {author} {\bibfnamefont {K.}~\bibnamefont
				{Sharma}}, \bibinfo {author} {\bibfnamefont {L.}~\bibnamefont {Calmels}},
			\bibinfo {author} {\bibfnamefont {D.}~\bibnamefont {Li}}, \bibinfo {author}
			{\bibfnamefont {A.}~\bibnamefont {Barbier}},\ and\ \bibinfo {author}
			{\bibfnamefont {R.}~\bibnamefont {Arras}},\ }\href
		{https://doi.org/10.1103/PhysRevMaterials.6.124402} {\bibfield  {journal}
			{\bibinfo  {journal} {Phys. Rev. Mater.}\ }\textbf {\bibinfo {volume} {6}},\
			\bibinfo {pages} {124402} (\bibinfo {year} {2022})}\BibitemShut {NoStop}%
		\bibitem [{\citenamefont {Sawatzky}\ \emph {et~al.}(1968)\citenamefont
			{Sawatzky}, \citenamefont {van~der Woude},\ and\ \citenamefont
			{Morrish}}]{Sawatzky68:39}%
		\BibitemOpen
		\bibfield  {author} {\bibinfo {author} {\bibfnamefont {G.~A.}\ \bibnamefont
				{Sawatzky}}, \bibinfo {author} {\bibfnamefont {F.}~\bibnamefont {van~der
					Woude}},\ and\ \bibinfo {author} {\bibfnamefont {A.~H.}\ \bibnamefont
				{Morrish}},\ }\href {https://doi.org/10.1063/1.1656224} {\bibfield  {journal}
			{\bibinfo  {journal} {J. Appl. Phys.}\ }\textbf {\bibinfo {volume} {39}},\
			\bibinfo {pages} {1204} (\bibinfo {year} {1968})}\BibitemShut {NoStop}%
		\bibitem [{\citenamefont {Teillet}\ \emph {et~al.}(1993)\citenamefont
			{Teillet}, \citenamefont {Bouree},\ and\ \citenamefont
			{Krishnan}}]{Teillet93:123}%
		\BibitemOpen
		\bibfield  {author} {\bibinfo {author} {\bibfnamefont {J.}~\bibnamefont
				{Teillet}}, \bibinfo {author} {\bibfnamefont {F.}~\bibnamefont {Bouree}},\
			and\ \bibinfo {author} {\bibfnamefont {R.}~\bibnamefont {Krishnan}},\ }\href
		{https://doi.org/https://doi.org/10.1016/0304-8853(93)90017-V} {\bibfield
			{journal} {\bibinfo  {journal} {J. Magn. Magn. Mater.}\ }\textbf {\bibinfo
				{volume} {123}},\ \bibinfo {pages} {93} (\bibinfo {year} {1993})}\BibitemShut
		{NoStop}%
		\bibitem [{\citenamefont {Tucek}\ \emph {et~al.}(2011)\citenamefont {Tucek},
			\citenamefont {Ohkoshi},\ and\ \citenamefont {Zboril}}]{Tucek11:99}%
		\BibitemOpen
		\bibfield  {author} {\bibinfo {author} {\bibfnamefont {J.}~\bibnamefont
				{Tucek}}, \bibinfo {author} {\bibfnamefont {S.}~\bibnamefont {Ohkoshi}},\
			and\ \bibinfo {author} {\bibfnamefont {R.}~\bibnamefont {Zboril}},\ }\href
		{https://doi.org/10.1063/1.3671114} {\bibfield  {journal} {\bibinfo
				{journal} {Appl. Phys. Lett.}\ }\textbf {\bibinfo {volume} {99}},\ \bibinfo
			{pages} {253108} (\bibinfo {year} {2011})}\BibitemShut {NoStop}%
		\bibitem [{\citenamefont {Reehuis}\ \emph {et~al.}(2003)\citenamefont
			{Reehuis}, \citenamefont {Krimmel}, \citenamefont {Buttgen}, \citenamefont
			{Loidl},\ and\ \citenamefont {Prokofiev}}]{Reehuis03:35}%
		\BibitemOpen
		\bibfield  {author} {\bibinfo {author} {\bibfnamefont {M.}~\bibnamefont
				{Reehuis}}, \bibinfo {author} {\bibfnamefont {A.}~\bibnamefont {Krimmel}},
			\bibinfo {author} {\bibfnamefont {N.}~\bibnamefont {Buttgen}}, \bibinfo
			{author} {\bibfnamefont {A.}~\bibnamefont {Loidl}},\ and\ \bibinfo {author}
			{\bibfnamefont {A.}~\bibnamefont {Prokofiev}},\ }\href
		{https://doi.org/10.1140/epjb/e2003-00282-4} {\bibfield  {journal} {\bibinfo
				{journal} {Eur. Phys. J. B}\ }\textbf {\bibinfo {volume} {35}},\ \bibinfo
			{pages} {311} (\bibinfo {year} {2003})}\BibitemShut {NoStop}%
		\bibitem [{\citenamefont {Zhang}\ and\ \citenamefont
			{Batista}(2021)}]{Zhang21:104}%
		\BibitemOpen
		\bibfield  {author} {\bibinfo {author} {\bibfnamefont {H.}~\bibnamefont
				{Zhang}}\ and\ \bibinfo {author} {\bibfnamefont {C.~D.}\ \bibnamefont
				{Batista}},\ }\bibfield  {title} {\bibinfo {title} {Classical spin dynamics
				based on $\mathrm{SU}(n)$ coherent states},\ }\href
		{https://doi.org/10.1103/PhysRevB.104.104409} {\bibfield  {journal} {\bibinfo
				{journal} {Phys. Rev. B}\ }\textbf {\bibinfo {volume} {104}},\ \bibinfo
			{pages} {104409} (\bibinfo {year} {2021})}\BibitemShut {NoStop}%
		\bibitem [{\citenamefont {Guratinder}\ \emph {et~al.}(2019)\citenamefont
			{Guratinder}, \citenamefont {Rau}, \citenamefont {Tsurkan}, \citenamefont
			{Ritter}, \citenamefont {Embs}, \citenamefont {Fennell}, \citenamefont
			{Walker}, \citenamefont {Medarde}, \citenamefont {Shang}, \citenamefont
			{Cervellino}, \citenamefont {R\"uegg},\ and\ \citenamefont
			{Zaharko}}]{Guratinder19:100}%
		\BibitemOpen
		\bibfield  {author} {\bibinfo {author} {\bibfnamefont {K.}~\bibnamefont
				{Guratinder}}, \bibinfo {author} {\bibfnamefont {J.~G.}\ \bibnamefont {Rau}},
			\bibinfo {author} {\bibfnamefont {V.}~\bibnamefont {Tsurkan}}, \bibinfo
			{author} {\bibfnamefont {C.}~\bibnamefont {Ritter}}, \bibinfo {author}
			{\bibfnamefont {J.}~\bibnamefont {Embs}}, \bibinfo {author} {\bibfnamefont
				{T.}~\bibnamefont {Fennell}}, \bibinfo {author} {\bibfnamefont {H.~C.}\
				\bibnamefont {Walker}}, \bibinfo {author} {\bibfnamefont {M.}~\bibnamefont
				{Medarde}}, \bibinfo {author} {\bibfnamefont {T.}~\bibnamefont {Shang}},
			\bibinfo {author} {\bibfnamefont {A.}~\bibnamefont {Cervellino}}, \bibinfo
			{author} {\bibfnamefont {C.}~\bibnamefont {R\"uegg}},\ and\ \bibinfo {author}
			{\bibfnamefont {O.}~\bibnamefont {Zaharko}},\ }\href
		{https://doi.org/10.1103/PhysRevB.100.094420} {\bibfield  {journal} {\bibinfo
				{journal} {Phys. Rev. B}\ }\textbf {\bibinfo {volume} {100}},\ \bibinfo
			{pages} {094420} (\bibinfo {year} {2019})}\BibitemShut {NoStop}%
		\bibitem [{\citenamefont {Wills}(2000)}]{Wills00:276}%
		\BibitemOpen
		\bibfield  {author} {\bibinfo {author} {\bibfnamefont {A.}~\bibnamefont
				{Wills}},\ }\bibfield  {title} {\bibinfo {title} {A new protocol for the
				determination of magnetic structures using simulated annealing and
				representational analysis (sarah)},\ }\href
		{https://doi.org/https://doi.org/10.1016/S0921-4526(99)01722-6} {\bibfield
			{journal} {\bibinfo  {journal} {Physica B: Condensed Matter}\ }\textbf
			{\bibinfo {volume} {276-278}},\ \bibinfo {pages} {680} (\bibinfo {year}
			{2000})}\BibitemShut {NoStop}%
		\bibitem [{\citenamefont {Wills}(2009)}]{Wills09:book}%
		\BibitemOpen
		\bibfield  {author} {\bibinfo {author} {\bibfnamefont {A.~S.}\ \bibnamefont
				{Wills}},\ }\bibinfo {title} {Indexing magnetic structures and
			crystallographic distortions from powder diffraction: Brillouin zone
			indexing},\ in\ \href {https://doi.org/doi:10.1524/9783486992588-011} {\emph
			{\bibinfo {booktitle} {Eleventh European Powder Diffraction Conference}}}\
		(\bibinfo  {publisher} {Oldenbourg Wissenschaftsverlag},\ \bibinfo {address}
		{München},\ \bibinfo {year} {2009})\ pp.\ \bibinfo {pages}
		{39--44}\BibitemShut {NoStop}%
		\bibitem [{\citenamefont {Abragam}\ and\ \citenamefont
			{Bleaney}(1986)}]{Abragam:book}%
		\BibitemOpen
		\bibfield  {author} {\bibinfo {author} {\bibfnamefont {A.}~\bibnamefont
				{Abragam}}\ and\ \bibinfo {author} {\bibfnamefont {B.}~\bibnamefont
				{Bleaney}},\ }\href@noop {} {\emph {\bibinfo {title} {Electron paramagnetic
					resonance of transition ions}}}\ (\bibinfo  {publisher} {Dover
			Publications},\ \bibinfo {address} {New York},\ \bibinfo {year}
		{1986})\BibitemShut {NoStop}%
		\bibitem [{\citenamefont {Cowley}\ \emph {et~al.}(2013)\citenamefont {Cowley},
			\citenamefont {Buyers}, \citenamefont {Stock}, \citenamefont {Yamani},
			\citenamefont {Frost}, \citenamefont {Taylor},\ and\ \citenamefont
			{Prabhakaran}}]{Cowley13:88}%
		\BibitemOpen
		\bibfield  {author} {\bibinfo {author} {\bibfnamefont {R.~A.}\ \bibnamefont
				{Cowley}}, \bibinfo {author} {\bibfnamefont {W.~J.~L.}\ \bibnamefont
				{Buyers}}, \bibinfo {author} {\bibfnamefont {C.}~\bibnamefont {Stock}},
			\bibinfo {author} {\bibfnamefont {Z.}~\bibnamefont {Yamani}}, \bibinfo
			{author} {\bibfnamefont {C.}~\bibnamefont {Frost}}, \bibinfo {author}
			{\bibfnamefont {J.~W.}\ \bibnamefont {Taylor}},\ and\ \bibinfo {author}
			{\bibfnamefont {D.}~\bibnamefont {Prabhakaran}},\ }\href
		{https://doi.org/10.1103/PhysRevB.88.205117} {\bibfield  {journal} {\bibinfo
				{journal} {Phys. Rev. B}\ }\textbf {\bibinfo {volume} {88}},\ \bibinfo
			{pages} {205117} (\bibinfo {year} {2013})}\BibitemShut {NoStop}%
		\bibitem [{\citenamefont {Sarte}\ \emph {et~al.}(2019)\citenamefont {Sarte},
			\citenamefont {Songvilay}, \citenamefont {Pachoud}, \citenamefont {Ewings},
			\citenamefont {Frost}, \citenamefont {Prabhakaran}, \citenamefont {Hong},
			\citenamefont {Browne}, \citenamefont {Yamani}, \citenamefont {Attfield},
			\citenamefont {Rodriguez}, \citenamefont {Wilson},\ and\ \citenamefont
			{Stock}}]{Sarte20:100}%
		\BibitemOpen
		\bibfield  {author} {\bibinfo {author} {\bibfnamefont {P.~M.}\ \bibnamefont
				{Sarte}}, \bibinfo {author} {\bibfnamefont {M.}~\bibnamefont {Songvilay}},
			\bibinfo {author} {\bibfnamefont {E.}~\bibnamefont {Pachoud}}, \bibinfo
			{author} {\bibfnamefont {R.~A.}\ \bibnamefont {Ewings}}, \bibinfo {author}
			{\bibfnamefont {C.~D.}\ \bibnamefont {Frost}}, \bibinfo {author}
			{\bibfnamefont {D.}~\bibnamefont {Prabhakaran}}, \bibinfo {author}
			{\bibfnamefont {K.~H.}\ \bibnamefont {Hong}}, \bibinfo {author}
			{\bibfnamefont {A.~J.}\ \bibnamefont {Browne}}, \bibinfo {author}
			{\bibfnamefont {Z.}~\bibnamefont {Yamani}}, \bibinfo {author} {\bibfnamefont
				{J.~P.}\ \bibnamefont {Attfield}}, \bibinfo {author} {\bibfnamefont {E.~E.}\
				\bibnamefont {Rodriguez}}, \bibinfo {author} {\bibfnamefont {S.~D.}\
				\bibnamefont {Wilson}},\ and\ \bibinfo {author} {\bibfnamefont
				{C.}~\bibnamefont {Stock}},\ }\href
		{https://doi.org/10.1103/PhysRevB.100.075143} {\bibfield  {journal} {\bibinfo
				{journal} {Phys. Rev. B}\ }\textbf {\bibinfo {volume} {100}},\ \bibinfo
			{pages} {075143} (\bibinfo {year} {2019})}\BibitemShut {NoStop}%
		\bibitem [{\citenamefont {Slonczewski}(1961{\natexlab{a}})}]{Slonczewski61:32}%
		\BibitemOpen
		\bibfield  {author} {\bibinfo {author} {\bibfnamefont {J.~C.}\ \bibnamefont
				{Slonczewski}},\ }\href {https://doi.org/10.1063/1.2000425} {\bibfield
			{journal} {\bibinfo  {journal} {J. Appl. Phys.}\ }\textbf {\bibinfo {volume}
				{32}},\ \bibinfo {pages} {253S} (\bibinfo {year}
			{1961}{\natexlab{a}})}\BibitemShut {NoStop}%
		\bibitem [{\citenamefont {Jahn}\ and\ \citenamefont
			{Teller}(1937)}]{Jahn37:161}%
		\BibitemOpen
		\bibfield  {author} {\bibinfo {author} {\bibfnamefont {H.~A.}\ \bibnamefont
				{Jahn}}\ and\ \bibinfo {author} {\bibfnamefont {E.}~\bibnamefont {Teller}},\
		}\href {https://doi.org/10.1098/rspa.1937.0142} {\bibfield  {journal}
			{\bibinfo  {journal} {Proc. R. Soc. A}\ }\textbf {\bibinfo {volume} {161}},\
			\bibinfo {pages} {220} (\bibinfo {year} {1937})}\BibitemShut {NoStop}%
		\bibitem [{\citenamefont {Gehring}\ and\ \citenamefont
			{Gehring}(1975)}]{Gehring75:38}%
		\BibitemOpen
		\bibfield  {author} {\bibinfo {author} {\bibfnamefont {G.~A.}\ \bibnamefont
				{Gehring}}\ and\ \bibinfo {author} {\bibfnamefont {K.~A.}\ \bibnamefont
				{Gehring}},\ }\href {https://doi.org/10.1088/0034-4885/38/1/001} {\bibfield
			{journal} {\bibinfo  {journal} {Rep. Prog. Phys.}\ }\textbf {\bibinfo
				{volume} {38}},\ \bibinfo {pages} {1} (\bibinfo {year} {1975})}\BibitemShut
		{NoStop}%
		\bibitem [{\citenamefont {Baral}\ \emph {et~al.}(2023)\citenamefont {Baral},
			\citenamefont {Abeykoon}, \citenamefont {Campbell},\ and\ \citenamefont
			{Frandsen}}]{Baral23:33}%
		\BibitemOpen
		\bibfield  {author} {\bibinfo {author} {\bibfnamefont {R.}~\bibnamefont
				{Baral}}, \bibinfo {author} {\bibfnamefont {A.~M.}\ \bibnamefont {Abeykoon}},
			\bibinfo {author} {\bibfnamefont {B.~J.}\ \bibnamefont {Campbell}},\ and\
			\bibinfo {author} {\bibfnamefont {B.~A.}\ \bibnamefont {Frandsen}},\
		}\bibfield  {title} {\bibinfo {title} {Giant spontaneous magnetostriction in
				mnte driven by a novel magnetostructural coupling mechanism},\ }\href
		{https://doi.org/https://doi.org/10.1002/adfm.202305247} {\bibfield
			{journal} {\bibinfo  {journal} {Advanced Functional Materials}\ }\textbf
			{\bibinfo {volume} {33}},\ \bibinfo {pages} {2305247} (\bibinfo {year}
			{2023})}\BibitemShut {NoStop}%
		\bibitem [{\citenamefont {Petricek}\ \emph {et~al.}(2014)\citenamefont
			{Petricek}, \citenamefont {Dusek},\ and\ \citenamefont
			{Palatinus}}]{Petricek14:229}%
		\BibitemOpen
		\bibfield  {author} {\bibinfo {author} {\bibfnamefont {V.}~\bibnamefont
				{Petricek}}, \bibinfo {author} {\bibfnamefont {M.}~\bibnamefont {Dusek}},\
			and\ \bibinfo {author} {\bibfnamefont {L.}~\bibnamefont {Palatinus}},\ }\href
		{https://doi.org/10.1515/zkri-2014-1737} {\bibfield  {journal} {\bibinfo
				{journal} {Z. Kristallogr.}\ }\textbf {\bibinfo {volume} {229(5)}},\ \bibinfo
			{pages} {345} (\bibinfo {year} {2014})}\BibitemShut {NoStop}%
		\bibitem [{\citenamefont {Tchernyshyov}(2004)}]{Tchern03:93}%
		\BibitemOpen
		\bibfield  {author} {\bibinfo {author} {\bibfnamefont {O.}~\bibnamefont
				{Tchernyshyov}},\ }\href {https://doi.org/10.1103/PhysRevLett.93.157206}
		{\bibfield  {journal} {\bibinfo  {journal} {Phys. Rev. Lett.}\ }\textbf
			{\bibinfo {volume} {93}},\ \bibinfo {pages} {157206} (\bibinfo {year}
			{2004})}\BibitemShut {NoStop}%
		\bibitem [{\citenamefont {Lane}\ \emph {et~al.}(2023)\citenamefont {Lane},
			\citenamefont {Sarte}, \citenamefont {Guratinder}, \citenamefont
			{Arevalo-Lopez}, \citenamefont {Perry}, \citenamefont {Hunter}, \citenamefont
			{Weber}, \citenamefont {Roessli}, \citenamefont {Stunault}, \citenamefont
			{Su}, \citenamefont {Ewings}, \citenamefont {Wilson}, \citenamefont {B\"oni},
			\citenamefont {Attfield},\ and\ \citenamefont {Stock}}]{Lane23:5}%
		\BibitemOpen
		\bibfield  {author} {\bibinfo {author} {\bibfnamefont {H.}~\bibnamefont
				{Lane}}, \bibinfo {author} {\bibfnamefont {P.~M.}\ \bibnamefont {Sarte}},
			\bibinfo {author} {\bibfnamefont {K.}~\bibnamefont {Guratinder}}, \bibinfo
			{author} {\bibfnamefont {A.~M.}\ \bibnamefont {Arevalo-Lopez}}, \bibinfo
			{author} {\bibfnamefont {R.~S.}\ \bibnamefont {Perry}}, \bibinfo {author}
			{\bibfnamefont {E.~C.}\ \bibnamefont {Hunter}}, \bibinfo {author}
			{\bibfnamefont {T.}~\bibnamefont {Weber}}, \bibinfo {author} {\bibfnamefont
				{B.}~\bibnamefont {Roessli}}, \bibinfo {author} {\bibfnamefont
				{A.}~\bibnamefont {Stunault}}, \bibinfo {author} {\bibfnamefont
				{Y.}~\bibnamefont {Su}}, \bibinfo {author} {\bibfnamefont {R.~A.}\
				\bibnamefont {Ewings}}, \bibinfo {author} {\bibfnamefont {S.~D.}\
				\bibnamefont {Wilson}}, \bibinfo {author} {\bibfnamefont {P.}~\bibnamefont
				{B\"oni}}, \bibinfo {author} {\bibfnamefont {J.~P.}\ \bibnamefont
				{Attfield}},\ and\ \bibinfo {author} {\bibfnamefont {C.}~\bibnamefont
				{Stock}},\ }\href {https://doi.org/10.1103/PhysRevResearch.5.043146}
		{\bibfield  {journal} {\bibinfo  {journal} {Phys. Rev. Res.}\ }\textbf
			{\bibinfo {volume} {5}},\ \bibinfo {pages} {043146} (\bibinfo {year}
			{2023})}\BibitemShut {NoStop}%
		\bibitem [{\citenamefont {Wolin}\ \emph {et~al.}(2018)\citenamefont {Wolin},
			\citenamefont {Wang}, \citenamefont {Naibert}, \citenamefont {Gleason},
			\citenamefont {MacDougall}, \citenamefont {Zhou}, \citenamefont {Cooper},\
			and\ \citenamefont {Budakian}}]{Wolin18:2}%
		\BibitemOpen
		\bibfield  {author} {\bibinfo {author} {\bibfnamefont {B.}~\bibnamefont
				{Wolin}}, \bibinfo {author} {\bibfnamefont {X.}~\bibnamefont {Wang}},
			\bibinfo {author} {\bibfnamefont {T.}~\bibnamefont {Naibert}}, \bibinfo
			{author} {\bibfnamefont {S.~L.}\ \bibnamefont {Gleason}}, \bibinfo {author}
			{\bibfnamefont {G.~J.}\ \bibnamefont {MacDougall}}, \bibinfo {author}
			{\bibfnamefont {H.~D.}\ \bibnamefont {Zhou}}, \bibinfo {author}
			{\bibfnamefont {S.~L.}\ \bibnamefont {Cooper}},\ and\ \bibinfo {author}
			{\bibfnamefont {R.}~\bibnamefont {Budakian}},\ }\href
		{https://doi.org/10.1103/PhysRevMaterials.2.064407} {\bibfield  {journal}
			{\bibinfo  {journal} {Phys. Rev. Mater.}\ }\textbf {\bibinfo {volume} {2}},\
			\bibinfo {pages} {064407} (\bibinfo {year} {2018})}\BibitemShut {NoStop}%
		\bibitem [{\citenamefont {Tsurkan}\ \emph {et~al.}(2021)\citenamefont
			{Tsurkan}, \citenamefont {von Nidda}, \citenamefont {Desenofer},\ and\
			\citenamefont {nd~A.~Loidl}}]{Tsurkan21:926}%
		\BibitemOpen
		\bibfield  {author} {\bibinfo {author} {\bibfnamefont {V.}~\bibnamefont
				{Tsurkan}}, \bibinfo {author} {\bibfnamefont {H.~A.~K.}\ \bibnamefont {von
					Nidda}}, \bibinfo {author} {\bibfnamefont {J.}~\bibnamefont {Desenofer}},\
			and\ \bibinfo {author} {\bibfnamefont {P.~L.}\ \bibnamefont {nd~A.~Loidl}},\
		}\href {https://doi.org/10.1016/j.physrep.2021.04.002} {\bibfield  {journal}
			{\bibinfo  {journal} {Phys. Rep.}\ }\textbf {\bibinfo {volume} {926}},\
			\bibinfo {pages} {1} (\bibinfo {year} {2021})}\BibitemShut {NoStop}%
		\bibitem [{\citenamefont {Radaelli}(2005)}]{Radaelli05:7}%
		\BibitemOpen
		\bibfield  {author} {\bibinfo {author} {\bibfnamefont {P.~G.}\ \bibnamefont
				{Radaelli}},\ }\href {https://doi.org/10.1088/1367-2630/7/1/053} {\bibfield
			{journal} {\bibinfo  {journal} {New. J. Phys.}\ }\textbf {\bibinfo {volume}
				{7}},\ \bibinfo {pages} {53} (\bibinfo {year} {2005})}\BibitemShut {NoStop}%
		\bibitem [{\citenamefont {Fritsch}\ and\ \citenamefont
			{Ederer}(2012)}]{Fritsch12:86}%
		\BibitemOpen
		\bibfield  {author} {\bibinfo {author} {\bibfnamefont {D.}~\bibnamefont
				{Fritsch}}\ and\ \bibinfo {author} {\bibfnamefont {C.}~\bibnamefont
				{Ederer}},\ }\href {https://doi.org/10.1103/PhysRevB.86.014406} {\bibfield
			{journal} {\bibinfo  {journal} {Phys. Rev. B}\ }\textbf {\bibinfo {volume}
				{86}},\ \bibinfo {pages} {014406} (\bibinfo {year} {2012})}\BibitemShut
		{NoStop}%
		\bibitem [{\citenamefont {Xu}\ \emph {et~al.}(2003)\citenamefont {Xu},
			\citenamefont {Zhong}, \citenamefont {Bing}, \citenamefont {Ye},
			\citenamefont {Stock},\ and\ \citenamefont {Shirane}}]{Xu03:67}%
		\BibitemOpen
		\bibfield  {author} {\bibinfo {author} {\bibfnamefont {G.}~\bibnamefont
				{Xu}}, \bibinfo {author} {\bibfnamefont {Z.}~\bibnamefont {Zhong}}, \bibinfo
			{author} {\bibfnamefont {Y.}~\bibnamefont {Bing}}, \bibinfo {author}
			{\bibfnamefont {Z.-G.}\ \bibnamefont {Ye}}, \bibinfo {author} {\bibfnamefont
				{C.}~\bibnamefont {Stock}},\ and\ \bibinfo {author} {\bibfnamefont
				{G.}~\bibnamefont {Shirane}},\ }\href
		{https://doi.org/10.1103/PhysRevB.67.104102} {\bibfield  {journal} {\bibinfo
				{journal} {Phys. Rev. B}\ }\textbf {\bibinfo {volume} {67}},\ \bibinfo
			{pages} {104102} (\bibinfo {year} {2003})}\BibitemShut {NoStop}%
		\bibitem [{\citenamefont {Xu}\ \emph {et~al.}(2004)\citenamefont {Xu},
			\citenamefont {Zhong}, \citenamefont {Bing}, \citenamefont {Ye},
			\citenamefont {Stock},\ and\ \citenamefont {Shirane}}]{Xu04:70}%
		\BibitemOpen
		\bibfield  {author} {\bibinfo {author} {\bibfnamefont {G.}~\bibnamefont
				{Xu}}, \bibinfo {author} {\bibfnamefont {Z.}~\bibnamefont {Zhong}}, \bibinfo
			{author} {\bibfnamefont {Y.}~\bibnamefont {Bing}}, \bibinfo {author}
			{\bibfnamefont {Z.-G.}\ \bibnamefont {Ye}}, \bibinfo {author} {\bibfnamefont
				{C.}~\bibnamefont {Stock}},\ and\ \bibinfo {author} {\bibfnamefont
				{G.}~\bibnamefont {Shirane}},\ }\href
		{https://doi.org/10.1103/PhysRevB.70.064107} {\bibfield  {journal} {\bibinfo
				{journal} {Phys. Rev. B}\ }\textbf {\bibinfo {volume} {70}},\ \bibinfo
			{pages} {064107} (\bibinfo {year} {2004})}\BibitemShut {NoStop}%
		\bibitem [{\citenamefont {Conlon}\ \emph {et~al.}(2004)\citenamefont {Conlon},
			\citenamefont {Luo}, \citenamefont {Viehland}, \citenamefont {Li},
			\citenamefont {Whan}, \citenamefont {Fox}, \citenamefont {Stock},\ and\
			\citenamefont {Shirane}}]{Conlon04:70}%
		\BibitemOpen
		\bibfield  {author} {\bibinfo {author} {\bibfnamefont {K.~H.}\ \bibnamefont
				{Conlon}}, \bibinfo {author} {\bibfnamefont {H.}~\bibnamefont {Luo}},
			\bibinfo {author} {\bibfnamefont {D.}~\bibnamefont {Viehland}}, \bibinfo
			{author} {\bibfnamefont {J.~F.}\ \bibnamefont {Li}}, \bibinfo {author}
			{\bibfnamefont {T.}~\bibnamefont {Whan}}, \bibinfo {author} {\bibfnamefont
				{J.~H.}\ \bibnamefont {Fox}}, \bibinfo {author} {\bibfnamefont
				{C.}~\bibnamefont {Stock}},\ and\ \bibinfo {author} {\bibfnamefont
				{G.}~\bibnamefont {Shirane}},\ }\href
		{https://doi.org/10.1103/PhysRevB.70.172204} {\bibfield  {journal} {\bibinfo
				{journal} {Phys. Rev. B}\ }\textbf {\bibinfo {volume} {70}},\ \bibinfo
			{pages} {172204} (\bibinfo {year} {2004})}\BibitemShut {NoStop}%
		\bibitem [{\citenamefont {Stock}\ \emph {et~al.}(2004)\citenamefont {Stock},
			\citenamefont {Birgeneau}, \citenamefont {Wakimoto}, \citenamefont {Gardner},
			\citenamefont {Chen}, \citenamefont {Ye},\ and\ \citenamefont
			{Shirane}}]{Stock04:69}%
		\BibitemOpen
		\bibfield  {author} {\bibinfo {author} {\bibfnamefont {C.}~\bibnamefont
				{Stock}}, \bibinfo {author} {\bibfnamefont {R.~J.}\ \bibnamefont
				{Birgeneau}}, \bibinfo {author} {\bibfnamefont {S.}~\bibnamefont {Wakimoto}},
			\bibinfo {author} {\bibfnamefont {J.~S.}\ \bibnamefont {Gardner}}, \bibinfo
			{author} {\bibfnamefont {W.}~\bibnamefont {Chen}}, \bibinfo {author}
			{\bibfnamefont {Z.-G.}\ \bibnamefont {Ye}},\ and\ \bibinfo {author}
			{\bibfnamefont {G.}~\bibnamefont {Shirane}},\ }\href
		{https://doi.org/10.1103/PhysRevB.69.094104} {\bibfield  {journal} {\bibinfo
				{journal} {Phys. Rev. B}\ }\textbf {\bibinfo {volume} {69}},\ \bibinfo
			{pages} {094104} (\bibinfo {year} {2004})}\BibitemShut {NoStop}%
		\bibitem [{\citenamefont {Gehring}\ \emph {et~al.}(2004)\citenamefont
			{Gehring}, \citenamefont {Chen}, \citenamefont {Ye},\ and\ \citenamefont
			{Shirane}}]{Gehring04:16}%
		\BibitemOpen
		\bibfield  {author} {\bibinfo {author} {\bibfnamefont {P.~M.}\ \bibnamefont
				{Gehring}}, \bibinfo {author} {\bibfnamefont {W.}~\bibnamefont {Chen}},
			\bibinfo {author} {\bibfnamefont {Z.~G.}\ \bibnamefont {Ye}},\ and\ \bibinfo
			{author} {\bibfnamefont {G.}~\bibnamefont {Shirane}},\ }\href
		{https://doi.org/10.1088/0953-8984/16/39/042} {\bibfield  {journal} {\bibinfo
				{journal} {J. Phys. Condens. Matter}\ }\textbf {\bibinfo {volume} {16}},\
			\bibinfo {pages} {7113} (\bibinfo {year} {2004})}\BibitemShut {NoStop}%
		\bibitem [{\citenamefont {Xu}\ \emph {et~al.}(2006)\citenamefont {Xu},
			\citenamefont {Gehring}, \citenamefont {Stock},\ and\ \citenamefont
			{Conlon}}]{Xu06:79}%
		\BibitemOpen
		\bibfield  {author} {\bibinfo {author} {\bibfnamefont {G.}~\bibnamefont
				{Xu}}, \bibinfo {author} {\bibfnamefont {P.~M.}\ \bibnamefont {Gehring}},
			\bibinfo {author} {\bibfnamefont {C.}~\bibnamefont {Stock}},\ and\ \bibinfo
			{author} {\bibfnamefont {K.}~\bibnamefont {Conlon}},\ }\href
		{https://doi.org/10.1080/02652030600558682} {\bibfield  {journal} {\bibinfo
				{journal} {Phase Transitions}\ }\textbf {\bibinfo {volume} {79}},\ \bibinfo
			{pages} {135} (\bibinfo {year} {2006})}\BibitemShut {NoStop}%
		\bibitem [{\citenamefont {Giles-Donovan}\ \emph {et~al.}(2024)\citenamefont
			{Giles-Donovan}, \citenamefont {Hillier}, \citenamefont {Ishida},
			\citenamefont {Hampshire}, \citenamefont {Giblin}, \citenamefont {Roessli},
			\citenamefont {Gehring}, \citenamefont {Xu}, \citenamefont {Li},
			\citenamefont {Luo}, \citenamefont {Cochran},\ and\ \citenamefont
			{Stock}}]{Donovan24:36}%
		\BibitemOpen
		\bibfield  {author} {\bibinfo {author} {\bibfnamefont {N.}~\bibnamefont
				{Giles-Donovan}}, \bibinfo {author} {\bibfnamefont {A.~D.}\ \bibnamefont
				{Hillier}}, \bibinfo {author} {\bibfnamefont {K.}~\bibnamefont {Ishida}},
			\bibinfo {author} {\bibfnamefont {B.~V.}\ \bibnamefont {Hampshire}}, \bibinfo
			{author} {\bibfnamefont {S.~R.}\ \bibnamefont {Giblin}}, \bibinfo {author}
			{\bibfnamefont {B.}~\bibnamefont {Roessli}}, \bibinfo {author} {\bibfnamefont
				{P.~M.}\ \bibnamefont {Gehring}}, \bibinfo {author} {\bibfnamefont
				{G.}~\bibnamefont {Xu}}, \bibinfo {author} {\bibfnamefont {X.}~\bibnamefont
				{Li}}, \bibinfo {author} {\bibfnamefont {H.}~\bibnamefont {Luo}}, \bibinfo
			{author} {\bibfnamefont {S.}~\bibnamefont {Cochran}},\ and\ \bibinfo {author}
			{\bibfnamefont {C.}~\bibnamefont {Stock}},\ }\href
		{https://doi.org/10.1080/02652030600558682} {\bibfield  {journal} {\bibinfo
				{journal} {J. Phys. Condens. Matter}\ }\textbf {\bibinfo {volume} {36}},\
			\bibinfo {pages} {435802} (\bibinfo {year} {2024})}\BibitemShut {NoStop}%
		\bibitem [{\citenamefont {Kumar}\ \emph {et~al.}(2020)\citenamefont {Kumar},
			\citenamefont {Kong}, \citenamefont {Sharma}, \citenamefont {Shi},
			\citenamefont {Vats}, \citenamefont {Checchia}, \citenamefont {Seidel},
			\citenamefont {Hoffman},\ and\ \citenamefont {Daniels}}]{Kumar20:8}%
		\BibitemOpen
		\bibfield  {author} {\bibinfo {author} {\bibfnamefont {N.}~\bibnamefont
				{Kumar}}, \bibinfo {author} {\bibfnamefont {S.}~\bibnamefont {Kong}},
			\bibinfo {author} {\bibfnamefont {P.}~\bibnamefont {Sharma}}, \bibinfo
			{author} {\bibfnamefont {X.}~\bibnamefont {Shi}}, \bibinfo {author}
			{\bibfnamefont {G.}~\bibnamefont {Vats}}, \bibinfo {author} {\bibfnamefont
				{S.}~\bibnamefont {Checchia}}, \bibinfo {author} {\bibfnamefont
				{J.}~\bibnamefont {Seidel}}, \bibinfo {author} {\bibfnamefont
				{M.}~\bibnamefont {Hoffman}},\ and\ \bibinfo {author} {\bibfnamefont
				{J.}~\bibnamefont {Daniels}},\ }\href {https://doi.org/10.1039/D0TC01300E}
		{\bibfield  {journal} {\bibinfo  {journal} {J. Mater. Chem. C}\ }\textbf
			{\bibinfo {volume} {8}},\ \bibinfo {pages} {7663} (\bibinfo {year}
			{2020})}\BibitemShut {NoStop}%
		\bibitem [{\citenamefont {Hill}\ \emph {et~al.}(1997)\citenamefont {Hill},
			\citenamefont {Feng}, \citenamefont {Harris}, \citenamefont {Birgeneau},
			\citenamefont {Ramirez},\ and\ \citenamefont {Cassanho}}]{Hill97:55}%
		\BibitemOpen
		\bibfield  {author} {\bibinfo {author} {\bibfnamefont {J.~P.}\ \bibnamefont
				{Hill}}, \bibinfo {author} {\bibfnamefont {Q.}~\bibnamefont {Feng}}, \bibinfo
			{author} {\bibfnamefont {Q.~J.}\ \bibnamefont {Harris}}, \bibinfo {author}
			{\bibfnamefont {R.~J.}\ \bibnamefont {Birgeneau}}, \bibinfo {author}
			{\bibfnamefont {A.~P.}\ \bibnamefont {Ramirez}},\ and\ \bibinfo {author}
			{\bibfnamefont {A.}~\bibnamefont {Cassanho}},\ }\href
		{https://doi.org/10.1103/PhysRevB.55.356} {\bibfield  {journal} {\bibinfo
				{journal} {Phys. Rev. B}\ }\textbf {\bibinfo {volume} {55}},\ \bibinfo
			{pages} {356} (\bibinfo {year} {1997})}\BibitemShut {NoStop}%
		\bibitem [{\citenamefont {Islam}\ \emph {et~al.}(2012)\citenamefont {Islam},
			\citenamefont {Wheeler}, \citenamefont {Reehuis}, \citenamefont
			{Siemensmeyer}, \citenamefont {Tovar}, \citenamefont {Klemke}, \citenamefont
			{Kiefer}, \citenamefont {Hill},\ and\ \citenamefont {Lake}}]{Islam12:85}%
		\BibitemOpen
		\bibfield  {author} {\bibinfo {author} {\bibfnamefont {A.~T. M.~N.}\
				\bibnamefont {Islam}}, \bibinfo {author} {\bibfnamefont {E.~M.}\ \bibnamefont
				{Wheeler}}, \bibinfo {author} {\bibfnamefont {M.}~\bibnamefont {Reehuis}},
			\bibinfo {author} {\bibfnamefont {K.}~\bibnamefont {Siemensmeyer}}, \bibinfo
			{author} {\bibfnamefont {M.}~\bibnamefont {Tovar}}, \bibinfo {author}
			{\bibfnamefont {B.}~\bibnamefont {Klemke}}, \bibinfo {author} {\bibfnamefont
				{K.}~\bibnamefont {Kiefer}}, \bibinfo {author} {\bibfnamefont {A.~H.}\
				\bibnamefont {Hill}},\ and\ \bibinfo {author} {\bibfnamefont
				{B.}~\bibnamefont {Lake}},\ }\href
		{https://doi.org/10.1103/PhysRevB.85.024203} {\bibfield  {journal} {\bibinfo
				{journal} {Phys. Rev. B}\ }\textbf {\bibinfo {volume} {85}},\ \bibinfo
			{pages} {024203} (\bibinfo {year} {2012})}\BibitemShut {NoStop}%
		\bibitem [{\citenamefont {Masuda}\ \emph {et~al.}(2006)\citenamefont {Masuda},
			\citenamefont {Zheludev}, \citenamefont {Manaka}, \citenamefont {Regnault},
			\citenamefont {Chung},\ and\ \citenamefont {Qiu}}]{Masuda06:96}%
		\BibitemOpen
		\bibfield  {author} {\bibinfo {author} {\bibfnamefont {T.}~\bibnamefont
				{Masuda}}, \bibinfo {author} {\bibfnamefont {A.}~\bibnamefont {Zheludev}},
			\bibinfo {author} {\bibfnamefont {H.}~\bibnamefont {Manaka}}, \bibinfo
			{author} {\bibfnamefont {L.-P.}\ \bibnamefont {Regnault}}, \bibinfo {author}
			{\bibfnamefont {J.-H.}\ \bibnamefont {Chung}},\ and\ \bibinfo {author}
			{\bibfnamefont {Y.}~\bibnamefont {Qiu}},\ }\href
		{https://doi.org/10.1103/PhysRevLett.96.047210} {\bibfield  {journal}
			{\bibinfo  {journal} {Phys. Rev. Lett.}\ }\textbf {\bibinfo {volume} {96}},\
			\bibinfo {pages} {047210} (\bibinfo {year} {2006})}\BibitemShut {NoStop}%
		\bibitem [{\citenamefont {Buyers}\ \emph {et~al.}(1971)\citenamefont {Buyers},
			\citenamefont {Holden}, \citenamefont {Svensson}, \citenamefont {Cowley},\
			and\ \citenamefont {Stevenson}}]{Buyers71:27}%
		\BibitemOpen
		\bibfield  {author} {\bibinfo {author} {\bibfnamefont {W.~J.~L.}\
				\bibnamefont {Buyers}}, \bibinfo {author} {\bibfnamefont {T.~M.}\
				\bibnamefont {Holden}}, \bibinfo {author} {\bibfnamefont {E.~C.}\
				\bibnamefont {Svensson}}, \bibinfo {author} {\bibfnamefont {R.~A.}\
				\bibnamefont {Cowley}},\ and\ \bibinfo {author} {\bibfnamefont {R.~W.~H.}\
				\bibnamefont {Stevenson}},\ }\href
		{https://doi.org/10.1103/PhysRevLett.27.1442} {\bibfield  {journal} {\bibinfo
				{journal} {Phys. Rev. Lett.}\ }\textbf {\bibinfo {volume} {27}},\ \bibinfo
			{pages} {1442} (\bibinfo {year} {1971})}\BibitemShut {NoStop}%
		\bibitem [{\citenamefont {Buyers}\ \emph {et~al.}(1972)\citenamefont {Buyers},
			\citenamefont {Pepper},\ and\ \citenamefont {Elliott}}]{Buyers72:5}%
		\BibitemOpen
		\bibfield  {author} {\bibinfo {author} {\bibfnamefont {W.~J.~L.}\
				\bibnamefont {Buyers}}, \bibinfo {author} {\bibfnamefont {D.~E.}\
				\bibnamefont {Pepper}},\ and\ \bibinfo {author} {\bibfnamefont {R.~J.}\
				\bibnamefont {Elliott}},\ }\href {https://doi.org/10.1088/0022-3719/5/18/011}
		{\bibfield  {journal} {\bibinfo  {journal} {J. Phys. C}\ }\textbf {\bibinfo
				{volume} {5}},\ \bibinfo {pages} {2611} (\bibinfo {year} {1972})}\BibitemShut
		{NoStop}%
		\bibitem [{\citenamefont {Birgeneau}\ \emph {et~al.}(1975)\citenamefont
			{Birgeneau}, \citenamefont {Walker}, \citenamefont {Guggenheim},
			\citenamefont {Als-Nielsen},\ and\ \citenamefont {Shirane}}]{Birgeneau75:8}%
		\BibitemOpen
		\bibfield  {author} {\bibinfo {author} {\bibfnamefont {R.~J.}\ \bibnamefont
				{Birgeneau}}, \bibinfo {author} {\bibfnamefont {L.~R.}\ \bibnamefont
				{Walker}}, \bibinfo {author} {\bibfnamefont {H.~J.}\ \bibnamefont
				{Guggenheim}}, \bibinfo {author} {\bibfnamefont {J.}~\bibnamefont
				{Als-Nielsen}},\ and\ \bibinfo {author} {\bibfnamefont {G.}~\bibnamefont
				{Shirane}},\ }\href {https://doi.org/10.1088/0022-3719/8/15/006} {\bibfield
			{journal} {\bibinfo  {journal} {J. Phys. C}\ }\textbf {\bibinfo {volume}
				{8}},\ \bibinfo {pages} {L327} (\bibinfo {year} {1975})}\BibitemShut
		{NoStop}%
		\bibitem [{\citenamefont {Lane}\ \emph {et~al.}(2021)\citenamefont {Lane},
			\citenamefont {Rodriguez}, \citenamefont {Walker}, \citenamefont
			{Niedermayer}, \citenamefont {Stuhr}, \citenamefont {Bewley}, \citenamefont
			{Voneshen}, \citenamefont {Green}, \citenamefont {Rodriguez-Rivera},
			\citenamefont {Fouquet}, \citenamefont {Cheong}, \citenamefont {Attfield},
			\citenamefont {Ewings},\ and\ \citenamefont {Stock}}]{Lane21:104}%
		\BibitemOpen
		\bibfield  {author} {\bibinfo {author} {\bibfnamefont {H.}~\bibnamefont
				{Lane}}, \bibinfo {author} {\bibfnamefont {E.~E.}\ \bibnamefont {Rodriguez}},
			\bibinfo {author} {\bibfnamefont {H.~C.}\ \bibnamefont {Walker}}, \bibinfo
			{author} {\bibfnamefont {C.}~\bibnamefont {Niedermayer}}, \bibinfo {author}
			{\bibfnamefont {U.}~\bibnamefont {Stuhr}}, \bibinfo {author} {\bibfnamefont
				{R.~I.}\ \bibnamefont {Bewley}}, \bibinfo {author} {\bibfnamefont {D.~J.}\
				\bibnamefont {Voneshen}}, \bibinfo {author} {\bibfnamefont {M.~A.}\
				\bibnamefont {Green}}, \bibinfo {author} {\bibfnamefont {J.~A.}\ \bibnamefont
				{Rodriguez-Rivera}}, \bibinfo {author} {\bibfnamefont {P.}~\bibnamefont
				{Fouquet}}, \bibinfo {author} {\bibfnamefont {S.-W.}\ \bibnamefont {Cheong}},
			\bibinfo {author} {\bibfnamefont {J.~P.}\ \bibnamefont {Attfield}}, \bibinfo
			{author} {\bibfnamefont {R.~A.}\ \bibnamefont {Ewings}},\ and\ \bibinfo
			{author} {\bibfnamefont {C.}~\bibnamefont {Stock}},\ }\href
		{https://doi.org/10.1103/PhysRevB.104.104404} {\bibfield  {journal} {\bibinfo
				{journal} {Phys. Rev. B}\ }\textbf {\bibinfo {volume} {104}},\ \bibinfo
			{pages} {104404} (\bibinfo {year} {2021})}\BibitemShut {NoStop}%
		\bibitem [{\citenamefont {Schr\"on}\ \emph {et~al.}(2012)\citenamefont
			{Schr\"on}, \citenamefont {R\"odl},\ and\ \citenamefont
			{Bechstedt}}]{Schron12:86}%
		\BibitemOpen
		\bibfield  {author} {\bibinfo {author} {\bibfnamefont {A.}~\bibnamefont
				{Schr\"on}}, \bibinfo {author} {\bibfnamefont {C.}~\bibnamefont {R\"odl}},\
			and\ \bibinfo {author} {\bibfnamefont {F.}~\bibnamefont {Bechstedt}},\ }\href
		{https://doi.org/10.1103/PhysRevB.86.115134} {\bibfield  {journal} {\bibinfo
				{journal} {Phys. Rev. B}\ }\textbf {\bibinfo {volume} {86}},\ \bibinfo
			{pages} {115134} (\bibinfo {year} {2012})}\BibitemShut {NoStop}%
		\bibitem [{\citenamefont {Baur}\ and\ \citenamefont {Khan}(1971)}]{Baur71:27}%
		\BibitemOpen
		\bibfield  {author} {\bibinfo {author} {\bibfnamefont {W.~H.}\ \bibnamefont
				{Baur}}\ and\ \bibinfo {author} {\bibfnamefont {A.~A.}\ \bibnamefont
				{Khan}},\ }\href {https://doi.org/10.1107/S0567740871005466} {\bibfield
			{journal} {\bibinfo  {journal} {Acta Cryst.}\ }\textbf {\bibinfo {volume}
				{B27}},\ \bibinfo {pages} {2133} (\bibinfo {year} {1971})}\BibitemShut
		{NoStop}%
		\bibitem [{\citenamefont {Cowley}\ \emph {et~al.}(1973)\citenamefont {Cowley},
			\citenamefont {Buyers}, \citenamefont {Martel},\ and\ \citenamefont
			{Stevenson}}]{Cowley73:6}%
		\BibitemOpen
		\bibfield  {author} {\bibinfo {author} {\bibfnamefont {R.~A.}\ \bibnamefont
				{Cowley}}, \bibinfo {author} {\bibfnamefont {W.~J.~L.}\ \bibnamefont
				{Buyers}}, \bibinfo {author} {\bibfnamefont {P.}~\bibnamefont {Martel}},\
			and\ \bibinfo {author} {\bibfnamefont {R.~W.~H.}\ \bibnamefont {Stevenson}},\
		}\href {https://doi.org/10.1088/0022-3719/6/20/014} {\bibfield  {journal}
			{\bibinfo  {journal} {J. Phys. C: Solid State Phys.}\ }\textbf {\bibinfo
				{volume} {6}},\ \bibinfo {pages} {2997} (\bibinfo {year} {1973})}\BibitemShut
		{NoStop}%
		\bibitem [{\citenamefont {Sarte}\ \emph {et~al.}(2018)\citenamefont {Sarte},
			\citenamefont {Cowley}, \citenamefont {Rodriguez}, \citenamefont {Pachoud},
			\citenamefont {Le}, \citenamefont {Garc\'{\i}a-Sakai}, \citenamefont
			{Taylor}, \citenamefont {Frost}, \citenamefont {Prabhakaran}, \citenamefont
			{MacEwen}, \citenamefont {Kitada}, \citenamefont {Browne}, \citenamefont
			{Songvilay}, \citenamefont {Yamani}, \citenamefont {Buyers}, \citenamefont
			{Attfield},\ and\ \citenamefont {Stock}}]{Sarte18:98}%
		\BibitemOpen
		\bibfield  {author} {\bibinfo {author} {\bibfnamefont {P.~M.}\ \bibnamefont
				{Sarte}}, \bibinfo {author} {\bibfnamefont {R.~A.}\ \bibnamefont {Cowley}},
			\bibinfo {author} {\bibfnamefont {E.~E.}\ \bibnamefont {Rodriguez}}, \bibinfo
			{author} {\bibfnamefont {E.}~\bibnamefont {Pachoud}}, \bibinfo {author}
			{\bibfnamefont {D.}~\bibnamefont {Le}}, \bibinfo {author} {\bibfnamefont
				{V.}~\bibnamefont {Garc\'{\i}a-Sakai}}, \bibinfo {author} {\bibfnamefont
				{J.~W.}\ \bibnamefont {Taylor}}, \bibinfo {author} {\bibfnamefont {C.~D.}\
				\bibnamefont {Frost}}, \bibinfo {author} {\bibfnamefont {D.}~\bibnamefont
				{Prabhakaran}}, \bibinfo {author} {\bibfnamefont {C.}~\bibnamefont
				{MacEwen}}, \bibinfo {author} {\bibfnamefont {A.}~\bibnamefont {Kitada}},
			\bibinfo {author} {\bibfnamefont {A.~J.}\ \bibnamefont {Browne}}, \bibinfo
			{author} {\bibfnamefont {M.}~\bibnamefont {Songvilay}}, \bibinfo {author}
			{\bibfnamefont {Z.}~\bibnamefont {Yamani}}, \bibinfo {author} {\bibfnamefont
				{W.~J.~L.}\ \bibnamefont {Buyers}}, \bibinfo {author} {\bibfnamefont {J.~P.}\
				\bibnamefont {Attfield}},\ and\ \bibinfo {author} {\bibfnamefont
				{C.}~\bibnamefont {Stock}},\ }\href
		{https://doi.org/10.1103/PhysRevB.98.024415} {\bibfield  {journal} {\bibinfo
				{journal} {Phys. Rev. B}\ }\textbf {\bibinfo {volume} {98}},\ \bibinfo
			{pages} {024415} (\bibinfo {year} {2018})}\BibitemShut {NoStop}%
		\bibitem [{\citenamefont {Hutchings}(1964)}]{Hutchings64:16}%
		\BibitemOpen
		\bibfield  {author} {\bibinfo {author} {\bibfnamefont {M.~T.}\ \bibnamefont
				{Hutchings}},\ }\href {https://doi.org/10.1016/S0081-1947(08)60517-2}
		{\bibfield  {journal} {\bibinfo  {journal} {Solid State Phys.}\ }\textbf
			{\bibinfo {volume} {16}},\ \bibinfo {pages} {227} (\bibinfo {year}
			{1964})}\BibitemShut {NoStop}%
		\bibitem [{\citenamefont {Toth}\ and\ \citenamefont {Lake}(2015)}]{Toth15:27}%
		\BibitemOpen
		\bibfield  {author} {\bibinfo {author} {\bibfnamefont {S.}~\bibnamefont
				{Toth}}\ and\ \bibinfo {author} {\bibfnamefont {B.}~\bibnamefont {Lake}},\
		}\href {https://doi.org/10.1088/0953-8984/27/16/166002} {\bibfield  {journal}
			{\bibinfo  {journal} {J. Phys.: Conden. Matt.}\ }\textbf {\bibinfo {volume}
				{27}},\ \bibinfo {pages} {166002} (\bibinfo {year} {2015})}\BibitemShut
		{NoStop}%
		\bibitem [{\citenamefont {Dahlbom}\ \emph {et~al.}(2025)\citenamefont
			{Dahlbom}, \citenamefont {Zhang}, \citenamefont {Miles}, \citenamefont
			{Quinn}, \citenamefont {Niraula}, \citenamefont {Thipe}, \citenamefont
			{Wilson}, \citenamefont {Matin}, \citenamefont {Mankad}, \citenamefont
			{Hahn}, \citenamefont {Pajerowski}, \citenamefont {Johnston}, \citenamefont
			{Wang}, \citenamefont {Lane}, \citenamefont {Li}, \citenamefont {Bai},
			\citenamefont {Mourigal}, \citenamefont {Batista},\ and\ \citenamefont
			{Barros}}]{Dahlbom25:preprint}%
		\BibitemOpen
		\bibfield  {author} {\bibinfo {author} {\bibfnamefont {D.}~\bibnamefont
				{Dahlbom}}, \bibinfo {author} {\bibfnamefont {H.}~\bibnamefont {Zhang}},
			\bibinfo {author} {\bibfnamefont {C.}~\bibnamefont {Miles}}, \bibinfo
			{author} {\bibfnamefont {S.}~\bibnamefont {Quinn}}, \bibinfo {author}
			{\bibfnamefont {A.}~\bibnamefont {Niraula}}, \bibinfo {author} {\bibfnamefont
				{B.}~\bibnamefont {Thipe}}, \bibinfo {author} {\bibfnamefont
				{M.}~\bibnamefont {Wilson}}, \bibinfo {author} {\bibfnamefont
				{S.}~\bibnamefont {Matin}}, \bibinfo {author} {\bibfnamefont
				{H.}~\bibnamefont {Mankad}}, \bibinfo {author} {\bibfnamefont
				{S.}~\bibnamefont {Hahn}}, \bibinfo {author} {\bibfnamefont {D.}~\bibnamefont
				{Pajerowski}}, \bibinfo {author} {\bibfnamefont {S.}~\bibnamefont
				{Johnston}}, \bibinfo {author} {\bibfnamefont {Z.}~\bibnamefont {Wang}},
			\bibinfo {author} {\bibfnamefont {H.}~\bibnamefont {Lane}}, \bibinfo {author}
			{\bibfnamefont {Y.~W.}\ \bibnamefont {Li}}, \bibinfo {author} {\bibfnamefont
				{X.}~\bibnamefont {Bai}}, \bibinfo {author} {\bibfnamefont {M.}~\bibnamefont
				{Mourigal}}, \bibinfo {author} {\bibfnamefont {C.~D.}\ \bibnamefont
				{Batista}},\ and\ \bibinfo {author} {\bibfnamefont {K.}~\bibnamefont
				{Barros}},\ }\href {https://arxiv.org/abs/2501.13095} {\bibinfo {title}
			{Sunny.jl: A julia package for spin dynamics}} (\bibinfo {year} {2025}),\
		\Eprint {https://arxiv.org/abs/2501.13095} {arXiv:2501.13095 [quant-ph]}
		\BibitemShut {NoStop}%
		\bibitem [{\citenamefont {Nambu}\ \emph {et~al.}(2020)\citenamefont {Nambu},
			\citenamefont {Barker}, \citenamefont {Okino}, \citenamefont {Kikkawa},
			\citenamefont {Shiomi}, \citenamefont {Enderle}, \citenamefont {Weber},
			\citenamefont {Winn}, \citenamefont {Graves-Brook}, \citenamefont
			{Tranquada}, \citenamefont {Ziman}, \citenamefont {Fujita}, \citenamefont
			{Bauer}, \citenamefont {Saitoh},\ and\ \citenamefont
			{Kakurai}}]{Nambu20:125}%
		\BibitemOpen
		\bibfield  {author} {\bibinfo {author} {\bibfnamefont {Y.}~\bibnamefont
				{Nambu}}, \bibinfo {author} {\bibfnamefont {J.}~\bibnamefont {Barker}},
			\bibinfo {author} {\bibfnamefont {Y.}~\bibnamefont {Okino}}, \bibinfo
			{author} {\bibfnamefont {T.}~\bibnamefont {Kikkawa}}, \bibinfo {author}
			{\bibfnamefont {Y.}~\bibnamefont {Shiomi}}, \bibinfo {author} {\bibfnamefont
				{M.}~\bibnamefont {Enderle}}, \bibinfo {author} {\bibfnamefont
				{T.}~\bibnamefont {Weber}}, \bibinfo {author} {\bibfnamefont
				{B.}~\bibnamefont {Winn}}, \bibinfo {author} {\bibfnamefont {M.}~\bibnamefont
				{Graves-Brook}}, \bibinfo {author} {\bibfnamefont {J.~M.}\ \bibnamefont
				{Tranquada}}, \bibinfo {author} {\bibfnamefont {T.}~\bibnamefont {Ziman}},
			\bibinfo {author} {\bibfnamefont {M.}~\bibnamefont {Fujita}}, \bibinfo
			{author} {\bibfnamefont {G.~E.~W.}\ \bibnamefont {Bauer}}, \bibinfo {author}
			{\bibfnamefont {E.}~\bibnamefont {Saitoh}},\ and\ \bibinfo {author}
			{\bibfnamefont {K.}~\bibnamefont {Kakurai}},\ }\href
		{https://doi.org/10.1103/PhysRevLett.125.027201} {\bibfield  {journal}
			{\bibinfo  {journal} {Phys. Rev. Lett.}\ }\textbf {\bibinfo {volume} {125}},\
			\bibinfo {pages} {027201} (\bibinfo {year} {2020})}\BibitemShut {NoStop}%
		\bibitem [{\citenamefont {Mazin}(2022)}]{Mazin22:12}%
		\BibitemOpen
		\bibfield  {author} {\bibinfo {author} {\bibfnamefont {I.}~\bibnamefont
				{Mazin}},\ }\bibfield  {title} {\bibinfo {title} {Editorial:
				Altermagnetism---a new punch line of fundamental magnetism},\ }\href
		{https://doi.org/10.1103/PhysRevX.12.040002} {\bibfield  {journal} {\bibinfo
				{journal} {Phys. Rev. X}\ }\textbf {\bibinfo {volume} {12}},\ \bibinfo
			{pages} {040002} (\bibinfo {year} {2022})}\BibitemShut {NoStop}%
		\bibitem [{\citenamefont {Bai}\ \emph {et~al.}(2024)\citenamefont {Bai},
			\citenamefont {Feng}, \citenamefont {Liu}, \citenamefont {Šmejkal},
			\citenamefont {Mokrousov},\ and\ \citenamefont {Yao}}]{Bai24:34}%
		\BibitemOpen
		\bibfield  {author} {\bibinfo {author} {\bibfnamefont {L.}~\bibnamefont
				{Bai}}, \bibinfo {author} {\bibfnamefont {W.}~\bibnamefont {Feng}}, \bibinfo
			{author} {\bibfnamefont {S.}~\bibnamefont {Liu}}, \bibinfo {author}
			{\bibfnamefont {L.}~\bibnamefont {Šmejkal}}, \bibinfo {author}
			{\bibfnamefont {Y.}~\bibnamefont {Mokrousov}},\ and\ \bibinfo {author}
			{\bibfnamefont {Y.}~\bibnamefont {Yao}},\ }\href
		{https://doi.org/https://doi.org/10.1002/adfm.202409327} {\bibfield
			{journal} {\bibinfo  {journal} {Adv. Funct. Mater.}\ }\textbf {\bibinfo
				{volume} {34}},\ \bibinfo {pages} {2409327} (\bibinfo {year}
			{2024})}\BibitemShut {NoStop}%
		\bibitem [{\citenamefont {\ifmmode~\check{S}\else \v{S}\fi{}mejkal}\ \emph
			{et~al.}(2023)\citenamefont {\ifmmode~\check{S}\else \v{S}\fi{}mejkal},
			\citenamefont {Marmodoro}, \citenamefont {Ahn}, \citenamefont
			{Gonz\'alez-Hern\'andez}, \citenamefont {Turek}, \citenamefont {Mankovsky},
			\citenamefont {Ebert}, \citenamefont {D'Souza}, \citenamefont
			{\ifmmode~\check{S}\else \v{S}\fi{}ipr}, \citenamefont {Sinova},\ and\
			\citenamefont {Jungwirth}}]{Smejkal23:131}%
		\BibitemOpen
		\bibfield  {author} {\bibinfo {author} {\bibfnamefont {L.}~\bibnamefont
				{\ifmmode~\check{S}\else \v{S}\fi{}mejkal}}, \bibinfo {author} {\bibfnamefont
				{A.}~\bibnamefont {Marmodoro}}, \bibinfo {author} {\bibfnamefont {K.-H.}\
				\bibnamefont {Ahn}}, \bibinfo {author} {\bibfnamefont {R.}~\bibnamefont
				{Gonz\'alez-Hern\'andez}}, \bibinfo {author} {\bibfnamefont {I.}~\bibnamefont
				{Turek}}, \bibinfo {author} {\bibfnamefont {S.}~\bibnamefont {Mankovsky}},
			\bibinfo {author} {\bibfnamefont {H.}~\bibnamefont {Ebert}}, \bibinfo
			{author} {\bibfnamefont {S.~W.}\ \bibnamefont {D'Souza}}, \bibinfo {author}
			{\bibfnamefont {O.~c.~v.}\ \bibnamefont {\ifmmode~\check{S}\else
					\v{S}\fi{}ipr}}, \bibinfo {author} {\bibfnamefont {J.}~\bibnamefont
				{Sinova}},\ and\ \bibinfo {author} {\bibfnamefont {T.~c.~v.}\ \bibnamefont
				{Jungwirth}},\ }\href {https://doi.org/10.1103/PhysRevLett.131.256703}
		{\bibfield  {journal} {\bibinfo  {journal} {Phys. Rev. Lett.}\ }\textbf
			{\bibinfo {volume} {131}},\ \bibinfo {pages} {256703} (\bibinfo {year}
			{2023})}\BibitemShut {NoStop}%
		\bibitem [{\citenamefont {Liu}\ \emph {et~al.}(2024)\citenamefont {Liu},
			\citenamefont {Ozeki}, \citenamefont {Asai}, \citenamefont {Itoh},\ and\
			\citenamefont {Masuda}}]{Liu24:133}%
		\BibitemOpen
		\bibfield  {author} {\bibinfo {author} {\bibfnamefont {Z.}~\bibnamefont
				{Liu}}, \bibinfo {author} {\bibfnamefont {M.}~\bibnamefont {Ozeki}}, \bibinfo
			{author} {\bibfnamefont {S.}~\bibnamefont {Asai}}, \bibinfo {author}
			{\bibfnamefont {S.}~\bibnamefont {Itoh}},\ and\ \bibinfo {author}
			{\bibfnamefont {T.}~\bibnamefont {Masuda}},\ }\href
		{https://doi.org/10.1103/PhysRevLett.133.156702} {\bibfield  {journal}
			{\bibinfo  {journal} {Phys. Rev. Lett.}\ }\textbf {\bibinfo {volume} {133}},\
			\bibinfo {pages} {156702} (\bibinfo {year} {2024})}\BibitemShut {NoStop}%
		\bibitem [{\citenamefont {C\^onsoli}\ and\ \citenamefont
			{Vojta}(2025)}]{Consoli25:134}%
		\BibitemOpen
		\bibfield  {author} {\bibinfo {author} {\bibfnamefont {P.~M.}\ \bibnamefont
				{C\^onsoli}}\ and\ \bibinfo {author} {\bibfnamefont {M.}~\bibnamefont
				{Vojta}},\ }\href {https://doi.org/10.1103/PhysRevLett.134.196701} {\bibfield
			{journal} {\bibinfo  {journal} {Phys. Rev. Lett.}\ }\textbf {\bibinfo
				{volume} {134}},\ \bibinfo {pages} {196701} (\bibinfo {year}
			{2025})}\BibitemShut {NoStop}%
		\bibitem [{\citenamefont {Brenig}\ and\ \citenamefont
			{Kampf}(1991)}]{Brenig92:43}%
		\BibitemOpen
		\bibfield  {author} {\bibinfo {author} {\bibfnamefont {W.}~\bibnamefont
				{Brenig}}\ and\ \bibinfo {author} {\bibfnamefont {A.~P.}\ \bibnamefont
				{Kampf}},\ }\href {https://doi.org/10.1103/PhysRevB.43.12914} {\bibfield
			{journal} {\bibinfo  {journal} {Phys. Rev. B}\ }\textbf {\bibinfo {volume}
				{43}},\ \bibinfo {pages} {12914} (\bibinfo {year} {1991})}\BibitemShut
		{NoStop}%
		\bibitem [{\citenamefont {Chernyshev}\ \emph {et~al.}(2002)\citenamefont
			{Chernyshev}, \citenamefont {Chen},\ and\ \citenamefont
			{Castro~Neto}}]{Chernyshev02:65}%
		\BibitemOpen
		\bibfield  {author} {\bibinfo {author} {\bibfnamefont {A.~L.}\ \bibnamefont
				{Chernyshev}}, \bibinfo {author} {\bibfnamefont {Y.~C.}\ \bibnamefont
				{Chen}},\ and\ \bibinfo {author} {\bibfnamefont {A.~H.}\ \bibnamefont
				{Castro~Neto}},\ }\href@noop {} {\bibfield  {journal} {\bibinfo  {journal}
				{Phys. Rev. B}\ }\textbf {\bibinfo {volume} {65}},\ \bibinfo {pages} {104407}
			(\bibinfo {year} {2002})}\BibitemShut {NoStop}%
		\bibitem [{\citenamefont {Cowley}\ and\ \citenamefont
			{Buyers}(1972)}]{Cowley72:44}%
		\BibitemOpen
		\bibfield  {author} {\bibinfo {author} {\bibfnamefont {R.~A.}\ \bibnamefont
				{Cowley}}\ and\ \bibinfo {author} {\bibfnamefont {W.~J.~L.}\ \bibnamefont
				{Buyers}},\ }\bibfield  {title} {\bibinfo {title} {The properties of defects
				in magnetic insulators},\ }\href {https://doi.org/10.1103/RevModPhys.44.406}
		{\bibfield  {journal} {\bibinfo  {journal} {Rev. Mod. Phys.}\ }\textbf
			{\bibinfo {volume} {44}},\ \bibinfo {pages} {406} (\bibinfo {year}
			{1972})}\BibitemShut {NoStop}%
		\bibitem [{\citenamefont {Lane}\ \emph {et~al.}(2025)\citenamefont {Lane},
			\citenamefont {Barros},\ and\ \citenamefont {Mourigal}}]{Lane25:37}%
		\BibitemOpen
		\bibfield  {author} {\bibinfo {author} {\bibfnamefont {H.}~\bibnamefont
				{Lane}}, \bibinfo {author} {\bibfnamefont {K.}~\bibnamefont {Barros}},\ and\
			\bibinfo {author} {\bibfnamefont {M.}~\bibnamefont {Mourigal}},\ }\bibfield
		{title} {\bibinfo {title} {Classical signatures of quenched and thermal
				disorder in the dynamics of correlated spin systems},\ }\href
		{https://doi.org/10.1088/1361-648X/addd55} {\bibfield  {journal} {\bibinfo
				{journal} {Journal of Physics: Condensed Matter}\ }\textbf {\bibinfo {volume}
				{37}},\ \bibinfo {pages} {265802} (\bibinfo {year} {2025})}\BibitemShut
		{NoStop}%
		\bibitem [{\citenamefont {Ansari}\ \emph {et~al.}(2020)\citenamefont {Ansari},
			\citenamefont {Kashid}, \citenamefont {Salunke}, \citenamefont {Sen},
			\citenamefont {Kolekar},\ and\ \citenamefont {Ramana}}]{Ansari20:102}%
		\BibitemOpen
		\bibfield  {author} {\bibinfo {author} {\bibfnamefont {S.~M.}\ \bibnamefont
				{Ansari}}, \bibinfo {author} {\bibfnamefont {V.}~\bibnamefont {Kashid}},
			\bibinfo {author} {\bibfnamefont {H.}~\bibnamefont {Salunke}}, \bibinfo
			{author} {\bibfnamefont {D.}~\bibnamefont {Sen}}, \bibinfo {author}
			{\bibfnamefont {Y.~D.}\ \bibnamefont {Kolekar}},\ and\ \bibinfo {author}
			{\bibfnamefont {C.~V.}\ \bibnamefont {Ramana}},\ }\href
		{https://doi.org/10.1103/PhysRevB.102.035446} {\bibfield  {journal} {\bibinfo
				{journal} {Phys. Rev. B}\ }\textbf {\bibinfo {volume} {102}},\ \bibinfo
			{pages} {035446} (\bibinfo {year} {2020})}\BibitemShut {NoStop}%
		\bibitem [{\citenamefont {Dao}\ \emph {et~al.}(2024)\citenamefont {Dao},
			\citenamefont {Shao},\ and\ \citenamefont {Sandvik}}]{Dao24:preprint}%
		\BibitemOpen
		\bibfield  {author} {\bibinfo {author} {\bibfnamefont {L.}~\bibnamefont
				{Dao}}, \bibinfo {author} {\bibfnamefont {H.}~\bibnamefont {Shao}},\ and\
			\bibinfo {author} {\bibfnamefont {A.~W.}\ \bibnamefont {Sandvik}},\ }\href
		{https://arxiv.org/abs/2408.06749} {\bibinfo {title} {Multiscale excitations
				in the diluted two-dimensional s = 1/2 heisenberg antiferromagnet}} (\bibinfo
		{year} {2024}),\ \Eprint {https://arxiv.org/abs/2408.06749} {arXiv:2408.06749
			[cond-mat.str-el]} \BibitemShut {NoStop}%
		\bibitem [{\citenamefont {Lane}\ \emph {et~al.}(2024)\citenamefont {Lane},
			\citenamefont {Zhang}, \citenamefont {Dahlbom}, \citenamefont {Quinn},
			\citenamefont {Somma}, \citenamefont {Mourigal}, \citenamefont {Batista},\
			and\ \citenamefont {Barros}}]{Lane24:17}%
		\BibitemOpen
		\bibfield  {author} {\bibinfo {author} {\bibfnamefont {H.}~\bibnamefont
				{Lane}}, \bibinfo {author} {\bibfnamefont {H.}~\bibnamefont {Zhang}},
			\bibinfo {author} {\bibfnamefont {D.}~\bibnamefont {Dahlbom}}, \bibinfo
			{author} {\bibfnamefont {S.}~\bibnamefont {Quinn}}, \bibinfo {author}
			{\bibfnamefont {R.~D.}\ \bibnamefont {Somma}}, \bibinfo {author}
			{\bibfnamefont {M.}~\bibnamefont {Mourigal}}, \bibinfo {author}
			{\bibfnamefont {C.~D.}\ \bibnamefont {Batista}},\ and\ \bibinfo {author}
			{\bibfnamefont {K.}~\bibnamefont {Barros}},\ }\bibfield  {title} {\bibinfo
			{title} {{Kernel polynomial method for linear spin wave theory}},\ }\href
		{https://doi.org/10.21468/SciPostPhys.17.5.145} {\bibfield  {journal}
			{\bibinfo  {journal} {SciPost Phys.}\ }\textbf {\bibinfo {volume} {17}},\
			\bibinfo {pages} {145} (\bibinfo {year} {2024})}\BibitemShut {NoStop}%
		\bibitem [{\citenamefont {Stock}\ \emph {et~al.}(2005)\citenamefont {Stock},
			\citenamefont {Buyers}, \citenamefont {Cowley}, \citenamefont {Clegg},
			\citenamefont {Coldea}, \citenamefont {Frost}, \citenamefont {Liang},
			\citenamefont {Peets}, \citenamefont {Bonn}, \citenamefont {Hardy},\ and\
			\citenamefont {Birgeneau}}]{Stock05:71}%
		\BibitemOpen
		\bibfield  {author} {\bibinfo {author} {\bibfnamefont {C.}~\bibnamefont
				{Stock}}, \bibinfo {author} {\bibfnamefont {W.~J.~L.}\ \bibnamefont
				{Buyers}}, \bibinfo {author} {\bibfnamefont {R.~A.}\ \bibnamefont {Cowley}},
			\bibinfo {author} {\bibfnamefont {P.~S.}\ \bibnamefont {Clegg}}, \bibinfo
			{author} {\bibfnamefont {R.}~\bibnamefont {Coldea}}, \bibinfo {author}
			{\bibfnamefont {C.~D.}\ \bibnamefont {Frost}}, \bibinfo {author}
			{\bibfnamefont {R.}~\bibnamefont {Liang}}, \bibinfo {author} {\bibfnamefont
				{D.}~\bibnamefont {Peets}}, \bibinfo {author} {\bibfnamefont
				{D.}~\bibnamefont {Bonn}}, \bibinfo {author} {\bibfnamefont {W.~N.}\
				\bibnamefont {Hardy}},\ and\ \bibinfo {author} {\bibfnamefont {R.~J.}\
				\bibnamefont {Birgeneau}},\ }\href
		{https://doi.org/10.1103/PhysRevB.71.024522} {\bibfield  {journal} {\bibinfo
				{journal} {Phys. Rev. B}\ }\textbf {\bibinfo {volume} {71}},\ \bibinfo
			{pages} {024522} (\bibinfo {year} {2005})}\BibitemShut {NoStop}%
		\bibitem [{\citenamefont {Carlson}\ \emph {et~al.}(2004)\citenamefont
			{Carlson}, \citenamefont {Yao},\ and\ \citenamefont
			{Campbell}}]{Carlson04:70}%
		\BibitemOpen
		\bibfield  {author} {\bibinfo {author} {\bibfnamefont {E.~W.}\ \bibnamefont
				{Carlson}}, \bibinfo {author} {\bibfnamefont {D.~X.}\ \bibnamefont {Yao}},\
			and\ \bibinfo {author} {\bibfnamefont {D.~K.}\ \bibnamefont {Campbell}},\
		}\href {https://doi.org/10.1103/PhysRevB.70.064505} {\bibfield  {journal}
			{\bibinfo  {journal} {Phys. Rev. B}\ }\textbf {\bibinfo {volume} {70}},\
			\bibinfo {pages} {064505} (\bibinfo {year} {2004})}\BibitemShut {NoStop}%
		\bibitem [{\citenamefont {Petitt}\ and\ \citenamefont
			{Forester}(1971)}]{Pettit71:4}%
		\BibitemOpen
		\bibfield  {author} {\bibinfo {author} {\bibfnamefont {G.~A.}\ \bibnamefont
				{Petitt}}\ and\ \bibinfo {author} {\bibfnamefont {D.~W.}\ \bibnamefont
				{Forester}},\ }\href {https://doi.org/10.1103/PhysRevB.4.3912} {\bibfield
			{journal} {\bibinfo  {journal} {Phys. Rev. B}\ }\textbf {\bibinfo {volume}
				{4}},\ \bibinfo {pages} {3912} (\bibinfo {year} {1971})}\BibitemShut
		{NoStop}%
		\bibitem [{\citenamefont {McQueeney}\ \emph {et~al.}(2007)\citenamefont
			{McQueeney}, \citenamefont {Yethiraj}, \citenamefont {Chang}, \citenamefont
			{Montfrooij}, \citenamefont {Perring}, \citenamefont {Honig},\ and\
			\citenamefont {Metcalf}}]{McQueeney07:99}%
		\BibitemOpen
		\bibfield  {author} {\bibinfo {author} {\bibfnamefont {R.~J.}\ \bibnamefont
				{McQueeney}}, \bibinfo {author} {\bibfnamefont {M.}~\bibnamefont {Yethiraj}},
			\bibinfo {author} {\bibfnamefont {S.}~\bibnamefont {Chang}}, \bibinfo
			{author} {\bibfnamefont {W.}~\bibnamefont {Montfrooij}}, \bibinfo {author}
			{\bibfnamefont {T.~G.}\ \bibnamefont {Perring}}, \bibinfo {author}
			{\bibfnamefont {J.~M.}\ \bibnamefont {Honig}},\ and\ \bibinfo {author}
			{\bibfnamefont {P.}~\bibnamefont {Metcalf}},\ }\href
		{https://doi.org/10.1103/PhysRevLett.99.246401} {\bibfield  {journal}
			{\bibinfo  {journal} {Phys. Rev. Lett.}\ }\textbf {\bibinfo {volume} {99}},\
			\bibinfo {pages} {246401} (\bibinfo {year} {2007})}\BibitemShut {NoStop}%
		\bibitem [{\citenamefont {Kawamoto}\ \emph {et~al.}(2024)\citenamefont
			{Kawamoto}, \citenamefont {Kikkawa}, \citenamefont {Kawamata}, \citenamefont
			{Umemoto}, \citenamefont {Manning}, \citenamefont {Rule}, \citenamefont
			{Ikeuchi}, \citenamefont {Kamazawa}, \citenamefont {Fujita}, \citenamefont
			{Saitoh}, \citenamefont {Kakurai},\ and\ \citenamefont
			{Nambu}}]{Kawamoto24:124}%
		\BibitemOpen
		\bibfield  {author} {\bibinfo {author} {\bibfnamefont {Y.}~\bibnamefont
				{Kawamoto}}, \bibinfo {author} {\bibfnamefont {T.}~\bibnamefont {Kikkawa}},
			\bibinfo {author} {\bibfnamefont {M.}~\bibnamefont {Kawamata}}, \bibinfo
			{author} {\bibfnamefont {Y.}~\bibnamefont {Umemoto}}, \bibinfo {author}
			{\bibfnamefont {A.~G.}\ \bibnamefont {Manning}}, \bibinfo {author}
			{\bibfnamefont {K.~C.}\ \bibnamefont {Rule}}, \bibinfo {author}
			{\bibfnamefont {K.}~\bibnamefont {Ikeuchi}}, \bibinfo {author} {\bibfnamefont
				{K.}~\bibnamefont {Kamazawa}}, \bibinfo {author} {\bibfnamefont
				{M.}~\bibnamefont {Fujita}}, \bibinfo {author} {\bibfnamefont
				{E.}~\bibnamefont {Saitoh}}, \bibinfo {author} {\bibfnamefont
				{K.}~\bibnamefont {Kakurai}},\ and\ \bibinfo {author} {\bibfnamefont
				{Y.}~\bibnamefont {Nambu}},\ }\href {https://doi.org/10.1063/5.0197831}
		{\bibfield  {journal} {\bibinfo  {journal} {Appl. Phys. Lett.}\ }\textbf
			{\bibinfo {volume} {124}},\ \bibinfo {pages} {132406} (\bibinfo {year}
			{2024})}\BibitemShut {NoStop}%
		\bibitem [{\citenamefont {Princep}\ \emph {et~al.}(2017)\citenamefont
			{Princep}, \citenamefont {Ewings}, \citenamefont {Ward}, \citenamefont {Toh},
			\citenamefont {Dubs}, \citenamefont {Prabhakaran},\ and\ \citenamefont
			{Boothroyd}}]{Princep17:2}%
		\BibitemOpen
		\bibfield  {author} {\bibinfo {author} {\bibfnamefont {A.~J.}\ \bibnamefont
				{Princep}}, \bibinfo {author} {\bibfnamefont {R.~A.}\ \bibnamefont {Ewings}},
			\bibinfo {author} {\bibfnamefont {S.}~\bibnamefont {Ward}}, \bibinfo {author}
			{\bibfnamefont {S.}~\bibnamefont {Toh}}, \bibinfo {author} {\bibfnamefont
				{C.}~\bibnamefont {Dubs}}, \bibinfo {author} {\bibfnamefont {D.}~\bibnamefont
				{Prabhakaran}},\ and\ \bibinfo {author} {\bibfnamefont {A.~T.}\ \bibnamefont
				{Boothroyd}},\ }\href {https://doi.org/10.1038/s41535-017-0067-y} {\bibfield
			{journal} {\bibinfo  {journal} {npj Quantum Mater.}\ }\textbf {\bibinfo
				{volume} {2}},\ \bibinfo {pages} {63} (\bibinfo {year} {2017})}\BibitemShut
		{NoStop}%
		\bibitem [{\citenamefont {Zhang}\ \emph {et~al.}(2023)\citenamefont {Zhang},
			\citenamefont {Tian}, \citenamefont {Nepal}, \citenamefont {Huq},
			\citenamefont {Nagler}, \citenamefont {DiTusa},\ and\ \citenamefont
			{Jin}}]{Zhang23:35}%
		\BibitemOpen
		\bibfield  {author} {\bibinfo {author} {\bibfnamefont {Q.}~\bibnamefont
				{Zhang}}, \bibinfo {author} {\bibfnamefont {W.}~\bibnamefont {Tian}},
			\bibinfo {author} {\bibfnamefont {R.}~\bibnamefont {Nepal}}, \bibinfo
			{author} {\bibfnamefont {A.}~\bibnamefont {Huq}}, \bibinfo {author}
			{\bibfnamefont {S.}~\bibnamefont {Nagler}}, \bibinfo {author} {\bibfnamefont
				{J.~F.}\ \bibnamefont {DiTusa}},\ and\ \bibinfo {author} {\bibfnamefont
				{R.}~\bibnamefont {Jin}},\ }\bibfield  {title} {\bibinfo {title} {Polyhedral
				distortions and unusual magnetic order in spinel femn2o4},\ }\href
		{https://doi.org/10.1021/acs.chemmater.2c03182} {\bibfield  {journal}
			{\bibinfo  {journal} {Chem. Mater.}\ }\textbf {\bibinfo {volume} {35}},\
			\bibinfo {pages} {2330} (\bibinfo {year} {2023})}\BibitemShut {NoStop}%
		\bibitem [{\citenamefont {Pan}\ and\ \citenamefont {Evans}(1976)}]{Pan76:34}%
		\BibitemOpen
		\bibfield  {author} {\bibinfo {author} {\bibfnamefont {L.-S.}\ \bibnamefont
				{Pan}}\ and\ \bibinfo {author} {\bibfnamefont {B.~J.}\ \bibnamefont
				{Evans}},\ }\href {https://doi.org/10.1063/1.2946059} {\bibfield  {journal}
			{\bibinfo  {journal} {AIP Conf. Proc.}\ }\textbf {\bibinfo {volume} {34}},\
			\bibinfo {pages} {181} (\bibinfo {year} {1976})}\BibitemShut {NoStop}%
		\bibitem [{\citenamefont {Jauch}\ \emph {et~al.}(2001)\citenamefont {Jauch},
			\citenamefont {Reehuis}, \citenamefont {Bleif}, \citenamefont {Kubanek},\
			and\ \citenamefont {Pattison}}]{Jauch01:64}%
		\BibitemOpen
		\bibfield  {author} {\bibinfo {author} {\bibfnamefont {W.}~\bibnamefont
				{Jauch}}, \bibinfo {author} {\bibfnamefont {M.}~\bibnamefont {Reehuis}},
			\bibinfo {author} {\bibfnamefont {H.~J.}\ \bibnamefont {Bleif}}, \bibinfo
			{author} {\bibfnamefont {F.}~\bibnamefont {Kubanek}},\ and\ \bibinfo {author}
			{\bibfnamefont {P.}~\bibnamefont {Pattison}},\ }\href
		{https://doi.org/10.1103/PhysRevB.64.052102} {\bibfield  {journal} {\bibinfo
				{journal} {Phys. Rev. B}\ }\textbf {\bibinfo {volume} {64}},\ \bibinfo
			{pages} {052102} (\bibinfo {year} {2001})}\BibitemShut {NoStop}%
		\bibitem [{\citenamefont {Levinstein}\ \emph {et~al.}(1965)\citenamefont
			{Levinstein}, \citenamefont {Schnettler},\ and\ \citenamefont
			{Gyorgy}}]{Levinstein65:36}%
		\BibitemOpen
		\bibfield  {author} {\bibinfo {author} {\bibfnamefont {H.~J.}\ \bibnamefont
				{Levinstein}}, \bibinfo {author} {\bibfnamefont {F.~J.}\ \bibnamefont
				{Schnettler}},\ and\ \bibinfo {author} {\bibfnamefont {E.~M.}\ \bibnamefont
				{Gyorgy}},\ }\href {https://doi.org/10.1063/1.1714150} {\bibfield  {journal}
			{\bibinfo  {journal} {J. Appl. Phys.}\ }\textbf {\bibinfo {volume} {36}},\
			\bibinfo {pages} {1163} (\bibinfo {year} {1965})}\BibitemShut {NoStop}%
		\bibitem [{\citenamefont {Kulkarni}\ and\ \citenamefont
			{Vishwas}(1979)}]{Kulkarni79:14}%
		\BibitemOpen
		\bibfield  {author} {\bibinfo {author} {\bibfnamefont {R.~G.}\ \bibnamefont
				{Kulkarni}}\ and\ \bibinfo {author} {\bibfnamefont {U.~P.}\ \bibnamefont
				{Vishwas}},\ }\href {https://doi.org/10.1007/BF00552309} {\bibfield
			{journal} {\bibinfo  {journal} {J. Mater. Sci.}\ }\textbf {\bibinfo {volume}
				{14}},\ \bibinfo {pages} {2221} (\bibinfo {year} {1979})}\BibitemShut
		{NoStop}%
		\bibitem [{\citenamefont {Balagurov}\ \emph {et~al.}(2013)\citenamefont
			{Balagurov}, \citenamefont {Bobrikov}, \citenamefont {Maschenko},
			\citenamefont {sangaa},\ and\ \citenamefont {Simkin}}]{Balagurov13:58}%
		\BibitemOpen
		\bibfield  {author} {\bibinfo {author} {\bibfnamefont {A.~M.}\ \bibnamefont
				{Balagurov}}, \bibinfo {author} {\bibfnamefont {I.~A.}\ \bibnamefont
				{Bobrikov}}, \bibinfo {author} {\bibfnamefont {M.~S.}\ \bibnamefont
				{Maschenko}}, \bibinfo {author} {\bibfnamefont {D.}~\bibnamefont {sangaa}},\
			and\ \bibinfo {author} {\bibfnamefont {V.~G.}\ \bibnamefont {Simkin}},\
		}\href {https://doi.org/10.1134/S1063774513040044} {\bibfield  {journal}
			{\bibinfo  {journal} {Crystallogr. Rep.}\ }\textbf {\bibinfo {volume} {58}},\
			\bibinfo {pages} {710} (\bibinfo {year} {2013})}\BibitemShut {NoStop}%
		\bibitem [{\citenamefont
			{Slonczewski}(1961{\natexlab{b}})}]{Slonczewski61:122}%
		\BibitemOpen
		\bibfield  {author} {\bibinfo {author} {\bibfnamefont {J.~C.}\ \bibnamefont
				{Slonczewski}},\ }\href {https://doi.org/10.1103/PhysRev.122.1367} {\bibfield
			{journal} {\bibinfo  {journal} {Phys. Rev.}\ }\textbf {\bibinfo {volume}
				{122}},\ \bibinfo {pages} {1367} (\bibinfo {year}
			{1961}{\natexlab{b}})}\BibitemShut {NoStop}%
		\bibitem [{\citenamefont {Wallington}\ \emph {et~al.}(2015)\citenamefont
			{Wallington}, \citenamefont {Arevalo-Lopez}, \citenamefont {Taylor},
			\citenamefont {Stewart}, \citenamefont {Garcia-Sakai}, \citenamefont
			{Attfield},\ and\ \citenamefont {Stock}}]{Wallington15:92}%
		\BibitemOpen
		\bibfield  {author} {\bibinfo {author} {\bibfnamefont {F.}~\bibnamefont
				{Wallington}}, \bibinfo {author} {\bibfnamefont {A.~M.}\ \bibnamefont
				{Arevalo-Lopez}}, \bibinfo {author} {\bibfnamefont {J.~W.}\ \bibnamefont
				{Taylor}}, \bibinfo {author} {\bibfnamefont {J.~R.}\ \bibnamefont {Stewart}},
			\bibinfo {author} {\bibfnamefont {V.}~\bibnamefont {Garcia-Sakai}}, \bibinfo
			{author} {\bibfnamefont {J.~P.}\ \bibnamefont {Attfield}},\ and\ \bibinfo
			{author} {\bibfnamefont {C.}~\bibnamefont {Stock}},\ }\href
		{https://doi.org/10.1103/PhysRevB.92.125116} {\bibfield  {journal} {\bibinfo
				{journal} {Phys. Rev. B}\ }\textbf {\bibinfo {volume} {92}},\ \bibinfo
			{pages} {125116} (\bibinfo {year} {2015})}\BibitemShut {NoStop}%
		\bibitem [{\citenamefont {Kato}\ \emph {et~al.}(2004)\citenamefont {Kato},
			\citenamefont {Myers}, \citenamefont {Gossard},\ and\ \citenamefont
			{Awschalom}}]{Kato04:306}%
		\BibitemOpen
		\bibfield  {author} {\bibinfo {author} {\bibfnamefont {Y.~K.}\ \bibnamefont
				{Kato}}, \bibinfo {author} {\bibfnamefont {R.~C.}\ \bibnamefont {Myers}},
			\bibinfo {author} {\bibfnamefont {A.~C.}\ \bibnamefont {Gossard}},\ and\
			\bibinfo {author} {\bibfnamefont {D.~D.}\ \bibnamefont {Awschalom}},\ }\href
		{https://doi.org/10.1126/science.1105514} {\bibfield  {journal} {\bibinfo
				{journal} {Science}\ }\textbf {\bibinfo {volume} {306}},\ \bibinfo {pages}
			{1910} (\bibinfo {year} {2004})}\BibitemShut {NoStop}%
		\bibitem [{\citenamefont {Sinova}\ \emph {et~al.}(2015)\citenamefont {Sinova},
			\citenamefont {Valenzuela}, \citenamefont {Wunderlich}, \citenamefont
			{Back},\ and\ \citenamefont {Jungwirth}}]{Sinova15:87}%
		\BibitemOpen
		\bibfield  {author} {\bibinfo {author} {\bibfnamefont {J.}~\bibnamefont
				{Sinova}}, \bibinfo {author} {\bibfnamefont {S.~O.}\ \bibnamefont
				{Valenzuela}}, \bibinfo {author} {\bibfnamefont {J.}~\bibnamefont
				{Wunderlich}}, \bibinfo {author} {\bibfnamefont {C.~H.}\ \bibnamefont
				{Back}},\ and\ \bibinfo {author} {\bibfnamefont {T.}~\bibnamefont
				{Jungwirth}},\ }\href {https://doi.org/10.1103/RevModPhys.87.1213} {\bibfield
			{journal} {\bibinfo  {journal} {Rev. Mod. Phys.}\ }\textbf {\bibinfo
				{volume} {87}},\ \bibinfo {pages} {1213} (\bibinfo {year}
			{2015})}\BibitemShut {NoStop}%
		\bibitem [{\citenamefont {Mori}\ \emph {et~al.}(2025)\citenamefont {Mori},
			\citenamefont {Tomasello},\ and\ \citenamefont {Ziman}}]{Mori25:111}%
		\BibitemOpen
		\bibfield  {author} {\bibinfo {author} {\bibfnamefont {M.}~\bibnamefont
				{Mori}}, \bibinfo {author} {\bibfnamefont {B.}~\bibnamefont {Tomasello}},\
			and\ \bibinfo {author} {\bibfnamefont {T.}~\bibnamefont {Ziman}},\ }\href
		{https://doi.org/10.1103/PhysRevB.111.014407} {\bibfield  {journal} {\bibinfo
				{journal} {Phys. Rev. B}\ }\textbf {\bibinfo {volume} {111}},\ \bibinfo
			{pages} {014407} (\bibinfo {year} {2025})}\BibitemShut {NoStop}%
		\bibitem [{\citenamefont {Barker}\ and\ \citenamefont
			{Bauer}(2016)}]{Barker16:117}%
		\BibitemOpen
		\bibfield  {author} {\bibinfo {author} {\bibfnamefont {J.}~\bibnamefont
				{Barker}}\ and\ \bibinfo {author} {\bibfnamefont {G.~E.~W.}\ \bibnamefont
				{Bauer}},\ }\href {https://doi.org/10.1103/PhysRevLett.117.217201} {\bibfield
			{journal} {\bibinfo  {journal} {Phys. Rev. Lett.}\ }\textbf {\bibinfo
				{volume} {117}},\ \bibinfo {pages} {217201} (\bibinfo {year}
			{2016})}\BibitemShut {NoStop}%
		\bibitem [{\citenamefont {Guo}\ \emph {et~al.}(2016)\citenamefont {Guo},
			\citenamefont {Herklotz}, \citenamefont {Kehlberger}, \citenamefont {Cramer},
			\citenamefont {Jakob},\ and\ \citenamefont {Kläui}}]{Guo16:108}%
		\BibitemOpen
		\bibfield  {author} {\bibinfo {author} {\bibfnamefont {E.-J.}\ \bibnamefont
				{Guo}}, \bibinfo {author} {\bibfnamefont {A.}~\bibnamefont {Herklotz}},
			\bibinfo {author} {\bibfnamefont {A.}~\bibnamefont {Kehlberger}}, \bibinfo
			{author} {\bibfnamefont {J.}~\bibnamefont {Cramer}}, \bibinfo {author}
			{\bibfnamefont {G.}~\bibnamefont {Jakob}},\ and\ \bibinfo {author}
			{\bibfnamefont {M.}~\bibnamefont {Kläui}},\ }\href
		{https://doi.org/10.1063/1.4939625} {\bibfield  {journal} {\bibinfo
				{journal} {Appl. Phys. Lett.}\ }\textbf {\bibinfo {volume} {108}},\ \bibinfo
			{pages} {022403} (\bibinfo {year} {2016})}\BibitemShut {NoStop}%
		\bibitem [{\citenamefont {Mogensen}\ and\ \citenamefont
			{Riseth}(2018)}]{Mogensen18:3}%
		\BibitemOpen
		\bibfield  {author} {\bibinfo {author} {\bibfnamefont {P.~K.}\ \bibnamefont
				{Mogensen}}\ and\ \bibinfo {author} {\bibfnamefont {A.~N.}\ \bibnamefont
				{Riseth}},\ }\href {https://doi.org/10.21105/joss.00615} {\bibfield
			{journal} {\bibinfo  {journal} {J. Open Source Softw.}\ }\textbf {\bibinfo
				{volume} {3}},\ \bibinfo {pages} {615} (\bibinfo {year} {2018})}\BibitemShut
		{NoStop}%
	\end{thebibliography}
	
	%

\end{document}